\documentclass[
  journal=pasa,
  manuscript=research-paper, 
  year=202X,
  volume=YY,
]{cup-journal}

\usepackage[percent]{overpic}
\usepackage{amsmath}  
\usepackage[normalem]{ulem}
\usepackage{amssymb,microtype,siunitx,booktabs}
\usepackage{comment}
\usepackage{longtable}
\usepackage{xltabular}
\usepackage{makecell}
\usepackage{subfig}
\usepackage{tabularx}
\newcommand{\arcsec}{\mbox{$^{\prime\prime}$}}

\title{Poor Hot Gas Content and Rare AGN Occurrence in Low-mass Galaxies in Galaxy Clusters}

\author{Juhi Tiwari}
\affiliation{Department of Physics and Astronomy, The University of Alabama in Huntsville, 301 Sparkman Dr NW, Huntsville, AL 35899, USA}

\author{Ming Sun}
\affiliation{Department of Physics and Astronomy, The University of Alabama in Huntsville, 301 Sparkman Dr NW, Huntsville, AL 35899, USA}

\author{Paul Nulsen}
\affiliation{ICRAR, University of Western Australia, 35 Stirling Hwy, Crawley, WA 6009, Australia}
\alsoaffiliation{Center for Astrophysics | Harvard and Smithsonian, 60 Garden Street, Cambridge, MA 02138, USA}

\author{Eric Rohr}
\affiliation{Max-Planck-Institut f\"{u}r Astrophysik, Karl-Schwarzschild-Strasse 1, 85748 Garching, Germany}

\email[Juhi Tiwari and Ming Sun]
{jt0189@uah.edu; ms0071@uah.edu}

\received {dd Mmm YYYY}
\revised  {dd Mmm YYYY}
\accepted {dd Mmm YYYY}
\published{xxxx}


\begin{document}

\begin{abstract}
Hot gas content in cluster member galaxies is important for studying galaxy evolution in dense environments. While massive galaxies \mbox{($\text{M}_* > 10^{10.5}\text{--}10^{11}\text{ M}_\odot$)} in galaxy clusters are commonly found to host soft X-ray-emitting hot gas concentrated around their centres, this remains largely unexplored in lower-mass members. To test for the presence of hot gas, we present the first systematic X-ray observational study of galaxies with $\text{M}_* \leq 10^{10.5} \text{ M} _{\odot}$ in 18 nearby ($0.011 < z < 0.063$) galaxy clusters with average intracluster medium (ICM) temperatures $kT= 3 - 10$ keV, using $\sim$22 Msec of archival Chandra data. Our sample includes 735 spectroscopically confirmed dwarf galaxies ($\text{M}_*\leq10^{9.5} \text{ M}_{\odot}$) and 823 intermediate-mass galaxies ($10^{9.5}< \text{M}_*\leq10^{10.5} \text{ M}_{\odot}$). Using stacked X-ray spectra, 
we search for diffuse soft X-ray emission indicative of hot coronal gas across different stellar masses and cluster environments. We find no significant evidence of hot-gas coronae on average in either population. This is consistent with such gas being either absent, removed by environmental processes
or potentially expelled by internal feedback due to their shallow gravitational potentials. Alternatively, hot-gas coronae may be present but remain below the current X-ray detection limits. The null detection constrains the soft (0.5--2.0 keV) X-ray luminosity ($L_X$) of diffuse gas to $< 3.17\times10^{38} \text{ erg s}^{-1}$ and $< 1.93\times10^{38} \text{ erg s}^{-1}$, for the dwarf and intermediate-mass populations, respectively. Individually, 22 galaxies 
host hot-gas coronae ($kT\sim 0.3\text{ to } 1.0 \text{ keV}$, $L_X\sim \text{ a few times } 10^{39} \text{ erg s}^{-1}$). The fraction of galaxies hosting
X-ray coronae is small, $\sim$$2.4\%$ for intermediate-mass galaxies and $\sim$$0.3\%$ for dwarfs, highlighting the apparent rarity of detectable hot-gas coronae in these systems. Overall, our results reinforce the picture that low-mass galaxies in clusters are highly susceptible to environmental effects, 
with retention of hot gas governed mainly by stellar mass and further influenced by intrinsic galaxy properties, early central potential formation, and ICM pressure confinement. We also identify six luminous X-ray AGN ($L_{X_{(2-10 \text{ keV)}}}>10^{41}\text{ erg s}^{-1}$), an uncommon occurrence among less massive galaxies in nearby clusters. These include one dwarf and five intermediate-mass hosts, yielding an AGN fraction of $\sim$$0.4\%$. Two intermediate-mass galaxies host AGN with $L_X \sim 10^{43}$ erg s$^{-1}$, while the dwarf host is particularly notable for its low stellar mass of $10^{9.2}$ M$_{\odot}$ and high $L_X$ of $\sim$$2.2\times10^{42}$ erg s$^{-1}$.
\end{abstract}

\section{Introduction} \label{sec:intro}
Hierarchical structure formation theory suggests that large halos form through the merging of subhalos. While the growth of dark matter structures is well described by N-body simulations, baryonic physics limits our understanding of galaxy formation and evolution. One important baryonic process is ram-pressure stripping (RPS), which regulates the gas content of subhalos in clusters. Early galaxy formation models assumed almost instantaneous removal of hot gas once galaxies enter clusters (e.g., \citealt{kauffmann1993, cole2000}). However, $Chandra$ observations have shown that this assumption is too simplistic, as hot-gas halos are seen to survive in groups and clusters (e.g., \citealt{sun2007,jeltema2008}). Over the years, the implementations of RPS in both cosmological simulations and semi-analytic models have steadily improved (e.g., \citealt{mccarthy2008, font2008,bahe2013,steinhauser2016,jung2018,arthur2019,mostoghiu2021,rohr2023,xie2025}). While these developments alleviate some long-standing tensions between predictions and observations of the relative fraction of blue and red galaxies in clusters, current RPS prescriptions remain simplified due to limited resolution and an incomplete treatment of gas physics, and so require continued calibration against observations and detailed simulations \citep[e.g.,][]{boselli2022}.

It is now well established observationally that galaxies in dense cluster environments often retain `embedded thermal coronae', compact reservoirs of hot, X-ray emitting gas in their inner regions (e.g. \citealt{forman1985,vikhlinin2001,sun2007,sun2009}). Typically extending only a few kiloparsecs (1--6 kpc), these structures represent the hot phase of the interstellar medium (ISM). With temperatures $\sim$$10^{6}-10^{7}$ K, they often survive within cluster galaxies despite stripping and evaporation by the surrounding intracluster medium (ICM). In contrast to the more extended, multiphase circumgalactic medium (CGM) -- which can span tens to hundreds of kiloparsecs around non-cluster members -- these coronae are centrally concentrated, and often the only remaining hot gas phase in cluster galaxies that have undergone environmental processing. Their presence implies that heat conduction at the interface with the ICM is strongly suppressed (e.g., \citealt{vikhlinin2001,sun2005,sun2007, rv2017b}) and that the hot gas is confined by the thermal pressure of the surrounding ICM, which may increase the coronal density and X-ray luminosity (e.g., \citealt{vikhlinin2001,lu2011}). The temperatures and energetics of coronae are also influenced by shock heating during galaxy formation and by ongoing feedback from supernovae and active galactic nuclei (AGN). A useful way to characterize the thermal energy of the hot gas relative to the depth of the galaxy's gravitational potential is through the $\beta$ parameter, defined as the ratio of the specific stellar kinetic energy to the specific thermal energy of the gas, where `specific' denotes per unit mass. In an idealised scenario in which gravity is the dominant source of heating, $\beta$ is expected to be close to unity, while lower values indicate additional heating processes.

Embedded thermal coronae have been found to be ubiquitous in early-type galaxies, particularly those with higher stellar masses ($> 10^{10.5}-10^{11} \text{M} _{\odot}$), likely due to the deeper gravitational potential wells of these galaxies, which enable the retention of hot gas \citep{sun2007}. Additionally, these coronae are likely more prevalent in the outer regions of galaxy clusters, where the ram pressure exerted by the ICM is weaker compared to the higher pressure and denser environments found towards cluster centres (e.g., \citealt{wagner2018_a1795}). Both intrinsic factors -- such as galaxy mass, star formation rate (e.g., early-type versus late-type; see \citet{hou2021_virgo, hou2024_virgo} for similar studies in the Virgo cluster) -- and environmental factors -- such as cluster dynamical state and location within the cluster, whether the galaxy is central or a satellite -- play key roles in determining the presence and characteristics of these hot coronae. The effects of the cluster environment also depend on the time galaxies have spent within it, as prolonged exposure to the ICM can lead to progressive stripping of their hot gas (e.g., \citealt{rv2015,rohr2024}). This makes the infall history of cluster galaxies important for understanding the long term survival of their hot coronae. Phase-space diagrams of galaxies, based on their cluster-centric distances and line-of-sight velocities, are useful for studying their infall histories (e.g., \citealt{oman2013, rhee2017}).

Despite their ubiquity in certain galaxy populations, the survival of embedded thermal coronae in less massive galaxies ($\text{M}_*<10^{10.5} \text{ M}_{\odot}$) in clusters remains poorly constrained observationally. Studies investigating X-ray-emitting gas in cluster galaxies are limited, with existing efforts primarily focused on massive systems -- leaving the dwarf and intermediate-mass populations largely underexplored. Detecting coronae in less massive galaxies is challenging due to their presumably faint, soft X-ray emission, which is difficult to distinguish from the much brighter surrounding ICM. Moreover, these galaxies possess shallower gravitational potentials, likely making them more vulnerable to environmental processes like ram-pressure stripping and tidal interactions that can efficiently remove their hot gas reservoirs. Dwarf galaxies ($\text{M}_*\lesssim10^{9.0}-10^{9.5} \text{ M}_{\odot}$) are a particularly compelling population: compared to massive galaxies, they are more fragile, more responsive to their environment, and more likely to undergo morphological transformation and rapid quenching due to gas loss. Their vulnerability to the environmental effects makes them good tracers of the strength and effectiveness of environmental processes operating in clusters. However, despite the abundance of these galaxies in clusters, observational studies specifically targeting the X-ray gas properties of low-mass galaxies in clusters remain rare. This gap restricts our knowledge of how environmental processes affect hot gas retention and consequently govern the evolution of cluster galaxies across a range of stellar masses. To address this gap, we investigate the potential presence or absence of X-ray thermal coronae in dwarf ($\text{M}_*\leq10^{9.5} \text{ M}_{\odot}$ as defined in this work) and intermediate-mass member galaxies ($10^{9.5}< \text{M}_*\leq10^{10.5} \text{ M}_{\odot}$) of nearby galaxy clusters with average temperatures $> 3$ keV. We use the rich legacy of $Chandra$ X-ray observations collected over more than two decades of the mission (through 2024) to conduct this analysis, aiming to shed light on the environmental and intrinsic factors that may influence the existence of hot coronae in these less massive systems.

In addition to diffuse thermal emission, we investigate the luminous X-ray AGN population in our dwarf and intermediate-mass galaxy samples. Although X-ray AGN have been extensively studied in many nearby clusters, previous surveys have largely focused on optically luminous ($M_R < -20 \text{ mag}$) or massive galaxies. For massive cluster members, X-ray AGN fractions of $\sim$$3-5\%$ (for $L_X>10^{41} \text{ erg s}^{-1}$; e.g., \citealt{martini2007,sun2007, sivakoff2008}) have typically been reported with AGN incidence increasing towards more luminous host galaxies. Several such studies also find that luminous X-ray AGN in massive galaxies are less common in rich clusters than in galaxy groups or the field (e.g., \citealt{arnold2009}). By contrast, comparable X-ray studies of AGN in dwarf and intermediate-mass galaxies have been carried out in a smaller number of cluster samples (e.g., \citealt{gallo2008,gallo2010,antlia2023,seydasen2023,hou2024}). The occurrence and properties of luminous X-ray AGN in lower-mass cluster galaxies therefore remain less well established. This mass range is particularly important because less massive galaxies are more susceptible to environmental gas removal, which may restrict the supply available for black-hole accretion. Characterising luminous AGN in these systems therefore extends studies of black-hole activity across a broader range of host-galaxy masses and can establish whether substantial nuclear accretion can occur even in low-mass galaxies residing in massive cluster environments. 

In Section \ref{sec:galaxy sample and data analyses}, we describe the data sample and analyses. The results and discussion are presented in Section \ref{sec:results} and Section \ref{sec:discussion}, respectively, and our conclusions are summarised in Section 5. Throughout this work, we adopt H$_{0}$=70 km s$^{-1}$ Mpc$^{-1}$, $\Omega$$_{M}$=0.3, and $\Omega$$_{\Lambda}$=0.7 for our calculations.
\newpage
\section{Sample Selection and Data Analyses}
\label{sec:galaxy sample and data analyses}
\subsection{Cluster and Galaxy Sample Selection}
In this work, we investigate the presence of X-ray thermal coronae in cluster galaxies across two stellar mass ranges: low-mass ($\text{M}_* \leq 10^{9.5} \text{ M}_{\odot}$) -- hereafter referred to as dwarf galaxies -- and intermediate-mass galaxies ($10^{9.5}< \text{M}_*\leq10^{10.5} \text{ M}_{\odot}$). The sample includes 18 nearby galaxy clusters ($0.011 < z < 0.063$), with ICM  temperatures of $3 - 10$ keV. The clusters were selected based on the availability of $Chandra$ ACIS archival observations obtained through 2024, making this the first study to leverage the large publicly available $Chandra$ legacy dataset  to examine hot gas in less massive cluster galaxies. The full galaxy cluster sample is listed in Table \ref{tab:clustersample}.

\begin{figure*}
 \centering
	\includegraphics[width=1.0\linewidth]{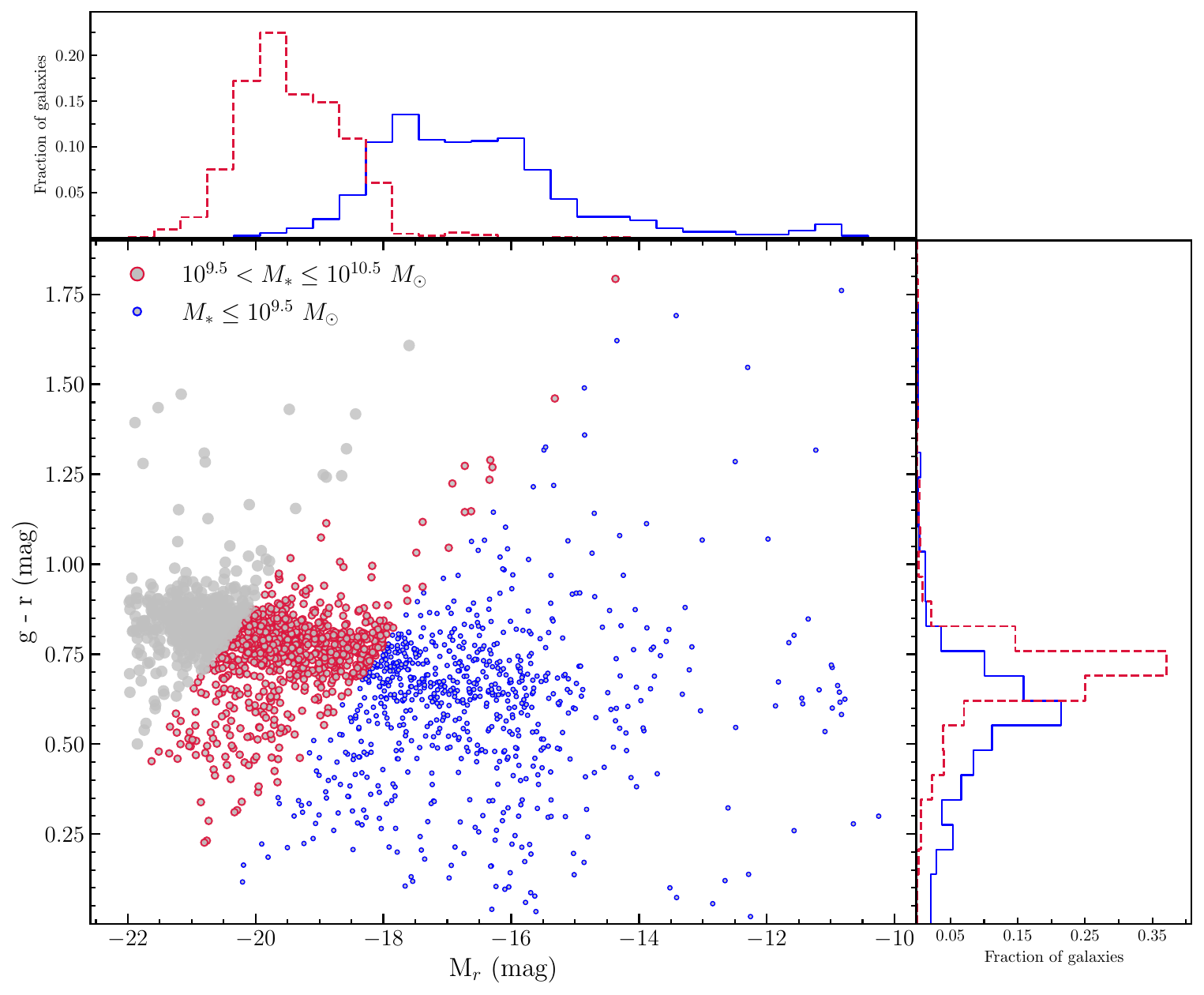}
	\caption{\small Distribution of cluster galaxies in the colour-magnitude plane, with magnitudes corrected for Galactic extinction. The full galaxy population is shown as \textit{filled grey circles}, with symbol sizes increasing from dwarf to giant galaxies. Dwarf galaxies ($\text{M}_*\leq 10^{9.5} \text{ M}_{\odot}$) are outlined in \textit{blue} and intermediate-mass galaxies ($10^{9.5}-10^{10.5} \text{M}_{\odot}$) in \textit{red}. The marginal histograms show the relative distributions of absolute $r$-band magnitude (top) and $g-r$ colour (right) for the dwarf (\textit{solid blue}) and intermediate-mass (\textit{dashed red}) populations. 
     Stellar masses of galaxies are derived using the relation $\log \Upsilon_{r} = -0.840 + (1.654 \times \textit{(g-r)})$ from \citet{zibetti2009}, where $\Upsilon_{r}$ is the stellar mass-to-light ratio of the galaxy in the $r$-band.}
    \label{fig:cmd}
\end{figure*}

To construct the member galaxy sample for each cluster, a search was conducted in the NASA Extragalactic Database (NED) for galaxies within $\pm3\sigma$ of the central velocity ($v_c$) of each galaxy cluster (where $\sigma$ is the rest-frame velocity dispersion\footnote{The cluster central velocities and velocity dispersions were adopted from the literature, with the corresponding references listed in Table \ref{tab:clustersample}. Where required, the observed velocity dispersions were corrected to the cluster rest frame using $\sigma=\sigma_{obs}/(1+z)$, where $z$ is the cluster redshift.}), selecting only those with spectroscopic redshifts available. The dataset was further restricted to galaxies projected within $2.5R_{500}$ of the cluster centre, where $R_{500}$ is the projected radius from the adopted cluster centre at which the mass density is 500-times the critical density of the Universe at the cluster's redshift. Galaxies were further filtered based on the availability of both $Chandra$ ACIS X-ray data and archival optical coverage from either SDSS, Pan-STARRS, DESI Legacy Imaging, or SkyMapper surveys\footnote{SDSS: Sloan Digital Sky Survey;  Pan-STARRS: Panoramic Survey Telescope and Rapid Response System; DESI: Dark Energy Spectroscopic Instrument}.
Additionally, only galaxies with at least one $Chandra$ ACIS observation within 8 arcminutes of the aimpoint were selected. The adopted values of $v_c$, $\sigma_{obs}$, the cluster centre, and $R_{500}$, along with the final number of galaxies in each cluster that satisfy the selection criteria, are listed in Table \ref{tab:clustersample}. In total, the 18 clusters have a combined exposure time of $\sim$22 Msec and include 735 dwarf and 823 intermediate-mass galaxies. The colour-magnitude diagram of the selected galaxies in the cluster sample is shown in Figure \ref{fig:cmd}.
\begin{table*}
\centering
\small
\setlength{\tabcolsep}{4pt}
\caption{Properties of galaxy clusters used in this work.}
\label{tab:clustersample}
\begin{tabular}{llllllllllll}
\hline

             Cluster$^{\rm a}$&$kT$& RA$^{\rm b}$ & Dec$^{\rm b}$ & ${v_\text{c}}$$^{\rm c}$& $\sigma_{obs}$ & $R_{500}$ & $D_L$& scale &Exp$^{\rm d}$&\# of galaxies$^{\rm e}$ \\  
            &(keV)&(deg)&(deg)&(km s$^{-1})$&(km s$^{-1}$)&($'$)&(Mpc)&(kpc/$\arcsec$)&(Msec)&
\\
			\hline
            Centaurus* (D)&3.7&192.2110&-41.3042&3480&840&59.0&36.8&0.237&1.04&39, 2, (44)\\
            A3627 (SM)&5.6&243.5938&-60.8686&4871&925&45.5&70.4&0.331&0.39&14,  20, (41)\\
            AWM7* (P)&3.7&43.6149&41.5791&5186&727&40.7&75.1&0.352&0.18&4, 14, (21)\\
            Perseus* (S)&6.4 & 49.9507 & 41.5117&5258&1040&60.9&76.0&0.356&2.81&11, 91, (149)\\
            A1367 (S)&3.6&176.1854&19.7322&6750&726&33.8&98.0&0.455&0.57&20, 24, (59)\\
            Coma (S)& 8.1&194.9389&27.9515& 7020&873&45.2&102.0&0.472&2.59&318, 111, (479)\\ 
            MKW8 (S)&3.3&220.1756&3.4716&8130&366&21.8&118.5&0.544&0.17&8, 11, (25)\\
           Ophiuchus* (P)&10.0&258.1155&-23.3686&8880&954&45.0&129.6&0.593&0.42&2, 11, (23)\\  
            A2199* (S)&4.3 &247.1592&39.5506&9300&705&25.6&135.9&0.620&0.22&54, 30, (97)\\
            A2052* (S)&3.1&229.1853&7.0218& 10640&736&20.2& 156.2&0.706&0.65&2, 19, (34)\\
            Hydra A* (D)&3.8&139.5250&-12.0934&16313&795&14.6&242.2&1.056&0.44&24, 34, (78)\\
            A754 (D)&9.0&137.3258&-9.6878&16374&995&20.8&243.4&1.061&0.22&6, 34, (73)\\ 
            A3667 (D)&6.3&303.1696&-56.8409&16558&1028&20.3& 246.3&1.073&0.75&39, 105, (201)\\
            A85* (S)&6.5&10.4602&-9.3032&16590&822&18.9&246.8&1.074&0.27&180, 138, (371)\\
            A2319 (P)&8.7&290.3000&43.9455&16710&1770&21.1&248.6&1.082&0.65&2, 29, (53)\\
            A133* (D)&4.0&15.6798&-21.8797&16760&813&13.6& 249.6&1.085&2.65&0, 10, (62)\\ 
            A1795* (S)&6.2 &207.2199&26.5929&18900&786&17.6&282.7&    1.213&5.85&6, 100, (168)\\
            Cygnus A* (P)&6.5&299.8682&40.7339&19008&1489&17.2&250.5&1.089&2.21&6, 40, (68)\\
\hline
\end{tabular}
	\vspace{0.5cm}
    \begin{tablenotes}
 \item[] \small{$^{\rm a}$ Cluster with * hosts a large cool core and is considered a cool-core cluster. Optical photometric data for galaxies were obtained from SDSS (abbreviated S in the table), the DESI Legacy Imaging Survey (D), Pan-STARRS survey (P), and SkyMapper (SM).
      $^{\rm b}$ RA and Dec in the J2000 epoch. 
      $^{\rm c}$ Central velocity $v_c=cz$. 
      $^{\rm d}$ The `Exp’ column lists the total Chandra ACIS exposure time summed over all observations of each cluster included in this work. The total exposure time summed across all clusters is $\sim$22 Msec.
      $^{\rm e}$ The number of cluster galaxies with $M_{*} \leq 10^{9.5} \text{ M}_{\odot}$, $M_{*} = 10^{9.5}-10^{10.5} \text{M}_{\odot}$, and in the complete sample (shown in brackets), with the brightest cluster galaxies (BCGs) included. The full sample spans a stellar mass range of $10^{6.0}-10^{11.8} \text{ M}_{\odot}$, with BCGs removed. In total, there are 735 dwarf and 823 intermediate-mass galaxies across the 18 clusters.      
      Galaxy cluster properties are taken from the literature -- \textbf{Centaurus}: \citet{angie2025}; \textbf{A3627}: \citet{woudt2008}, \citet{piffaretti2011} (P11); \textbf{AWM7}: \citet{awm7_2000}, P11; \textbf{Perseus}: \citet{giacintucci2019} (G19), \citet{aguerri2020}; \textbf{A1367}: \citet{zhang2011} (Z11), \citet{ge2019}, \citet{sohn2020} (S20); \textbf{Coma}: Z11, S20; \textbf{MKW8}: Z11, S20; \textbf{Ophiuchus}: \citet{durret2015}, G19; \textbf{A2199}: Z11, S20, \citet{yuan2023}; \textbf{A2052}: \citet{oegerle2001}, Z11; \textbf{Hydra A}: Z11, \citet{girardi2022}; \textbf{A754}: Z11, \citet{lauer2014} (L14); \textbf{A3667}: Z11, L14; \textbf{A85}: Z11, S20; \textbf{A2319}: \citet{struble_rood1999}, \citet{ghirardini2019}; \textbf{A133}: Z11, \citet{connor2018}; \textbf{A1795}: \citet{vikhlinin2006}, S20; \textbf{Cygnus A}: \citet{ledlow2005}, P11. Cygnus A has a bimodal velocity distribution; the full redshift range was used for galaxy selection.}
    \end{tablenotes}
\end{table*}\vspace{0.2cm}

\subsection{X-ray Data Analyses} 
\begin{figure*}
   \centering
\includegraphics[width=1.0\linewidth]{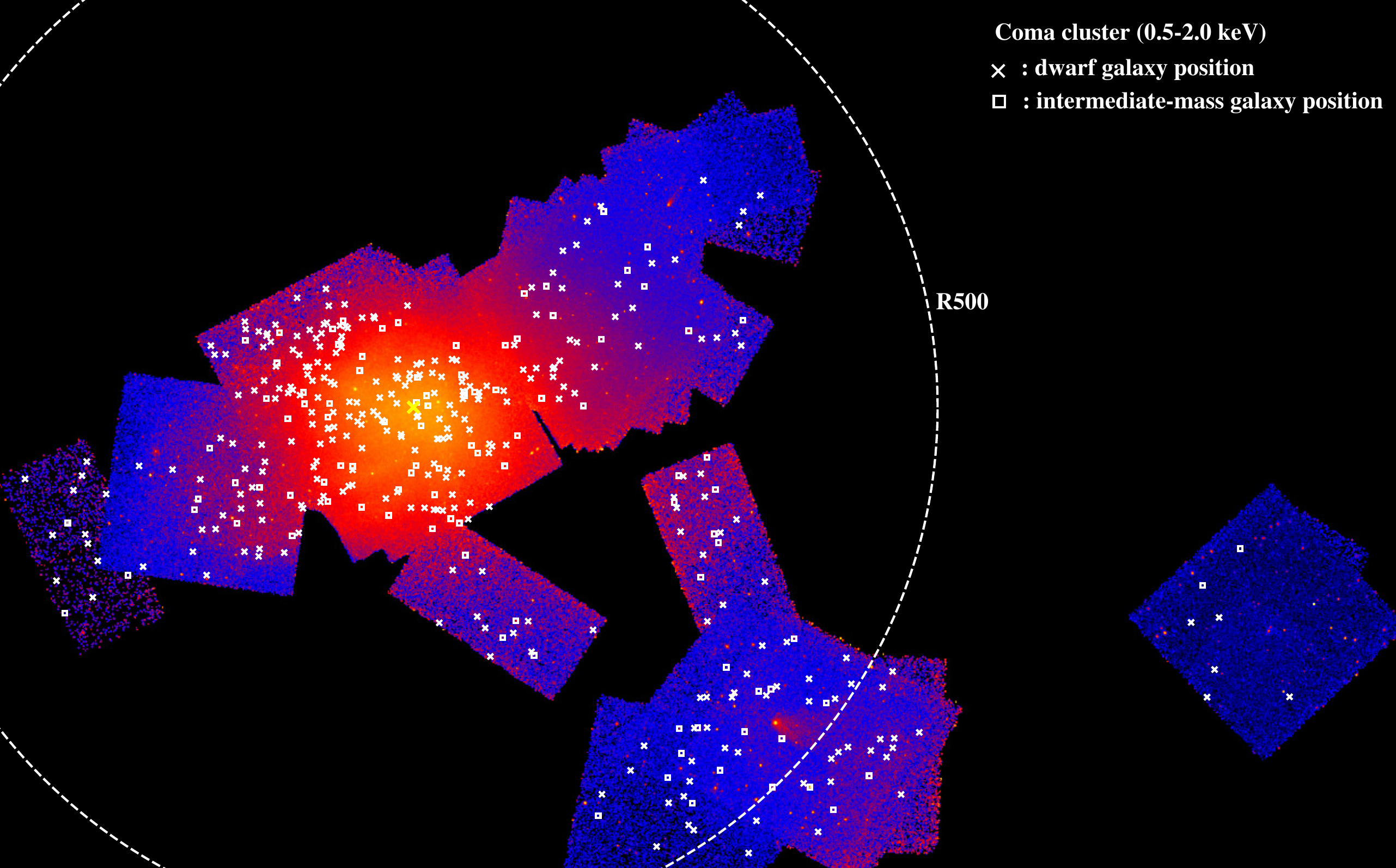}
  \caption{\small The \textit{Chandra} 0.5--2.0 keV flux image of the Coma galaxy cluster, with 2.56 Msec of cleaned data. \textit{White crosses} and \textit{white boxes} mark the positions of \textit{Chandra}-covered dwarf ($\text{M}_*\leq10^{9.5} \text{ M}_{\odot}$) and intermediate-mass  ($10^{9.5}< \text{M}_*\leq10^{10.5} \text{ M}_{\odot}$) galaxies in the cluster, respectively. The \textit{dashed white circle} denotes the radius $R_{500}$, centred at the cluster centre (marked by the \textit{yellow cross}). 
  Point sources are not masked in this image.
  \label{fig:coma_xray}}  
\end{figure*}

All \textit{Chandra} ACIS X-ray observations for the 18 galaxy clusters were retrieved from the Chandra Data Archive\footnote{\url{https://cda.harvard.edu/chaser/}} and analysed using the \textit{Chandra} Interactive Analysis of Observations (CIAO) software version 4.14. \textit{level=2} event files were generated using the $chandra\_repro$ script, and their light curves were filtered to exclude intervals affected by soft proton flares. We then used the CIAO script \textit{merge$\_$obs} to combine all \textit{Chandra} observations, generating X-ray mosaics of all the galaxy clusters used in this study. An example image resulting from this process is shown in Figure \ref{fig:coma_xray}, which presents the 0.5--2.0 keV $Chandra$ X-ray flux mosaic of the Coma cluster, one of the most deeply observed systems in our sample. Mosaics of the Perseus and A1795 clusters are also shown in Figure \ref{fig:perseus_xray} and Figure \ref{fig:a1795_xray}, respectively. These mosaics are useful for identifying direct signatures of X-ray thermal coronae and point sources at galaxy locations.

\begin{figure*}
\centering
  \hspace{-0.35cm}
\includegraphics[width=0.45\linewidth,height=6cm,keepaspectratio]{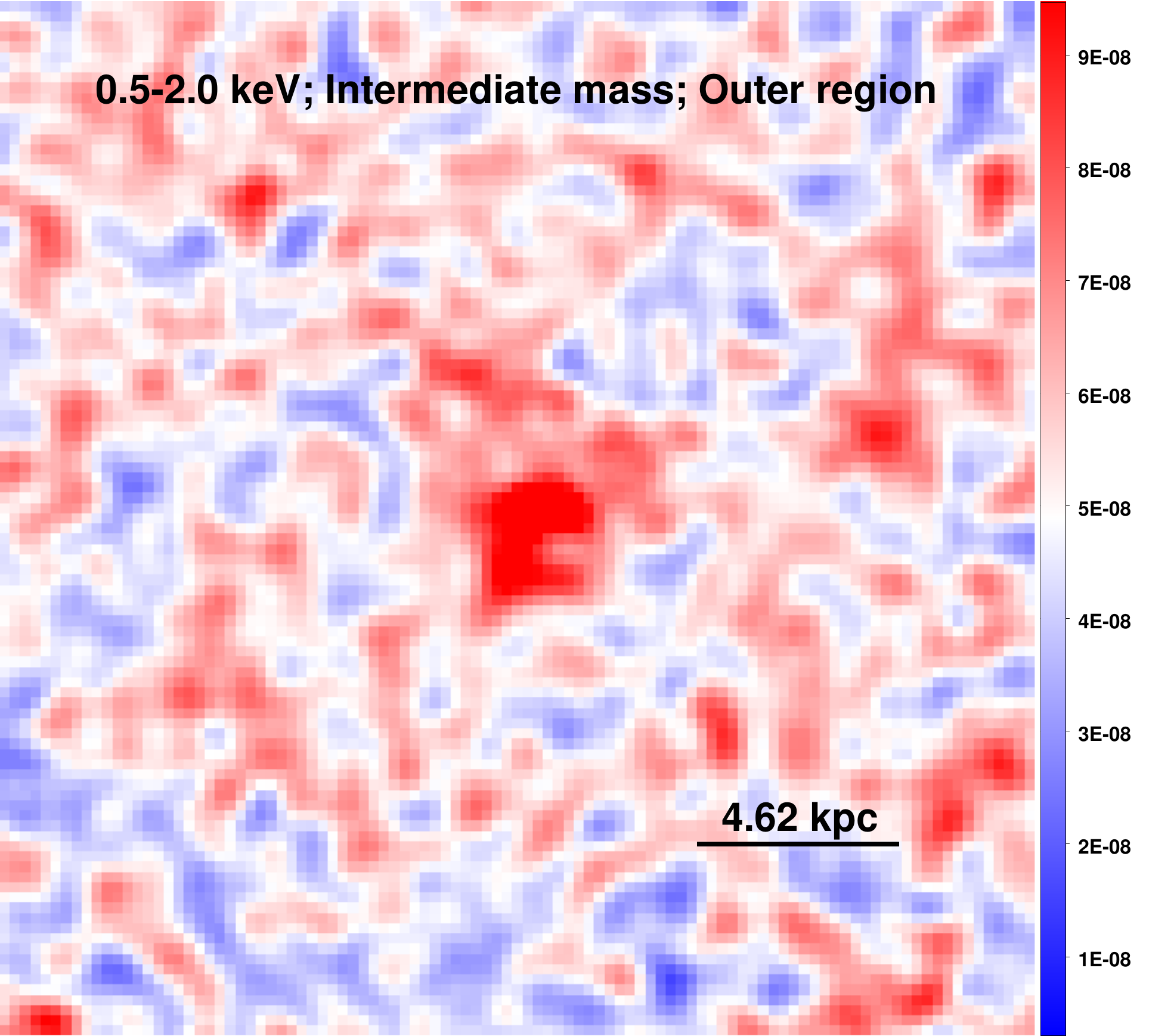}
\hspace{0.45cm} 
\includegraphics[width=0.45\linewidth,height=6cm,keepaspectratio]{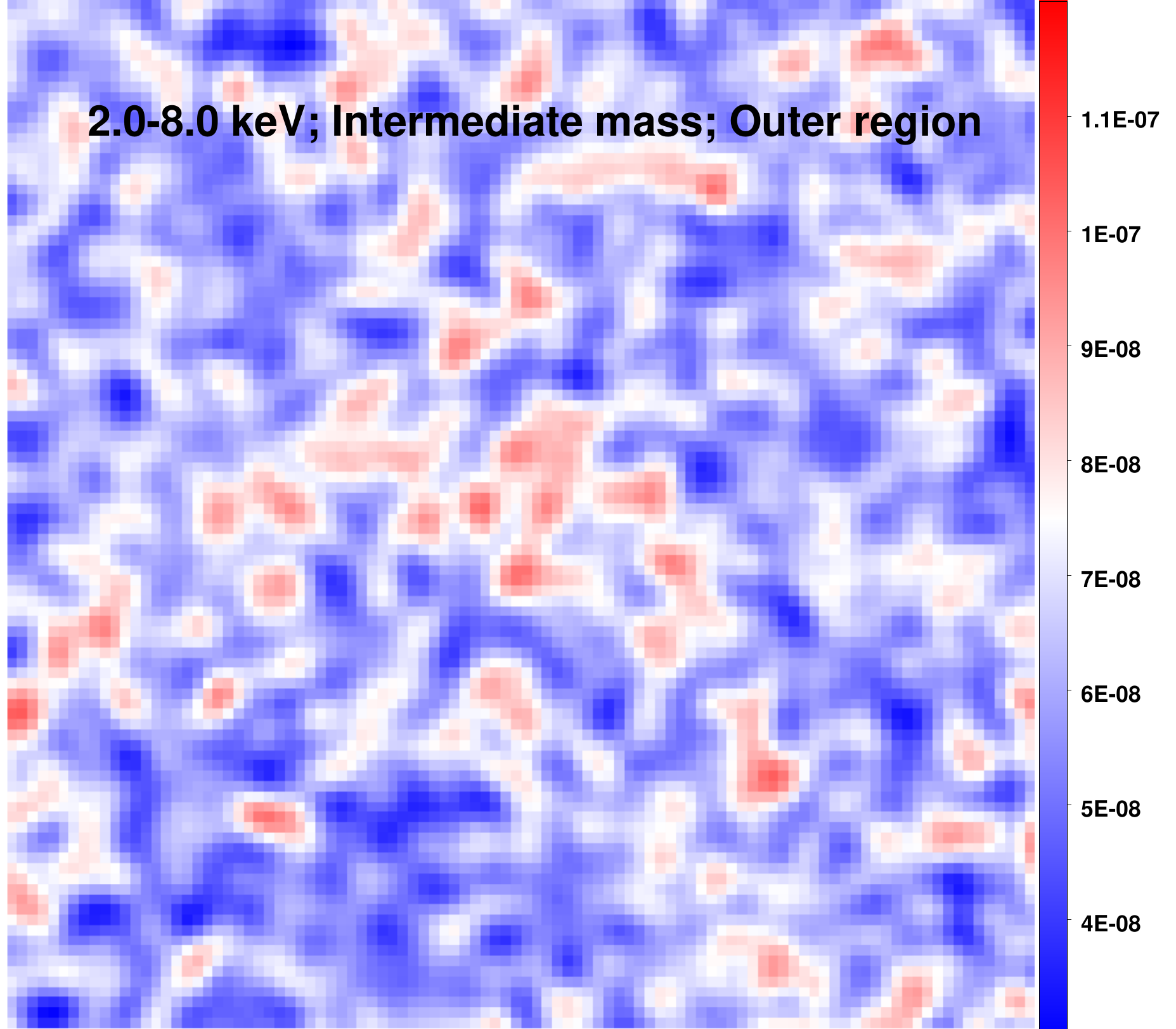}\\\vspace{0.3cm}
 \hspace{-0.2cm}
 \includegraphics[width=0.45\linewidth,height=6cm,keepaspectratio]{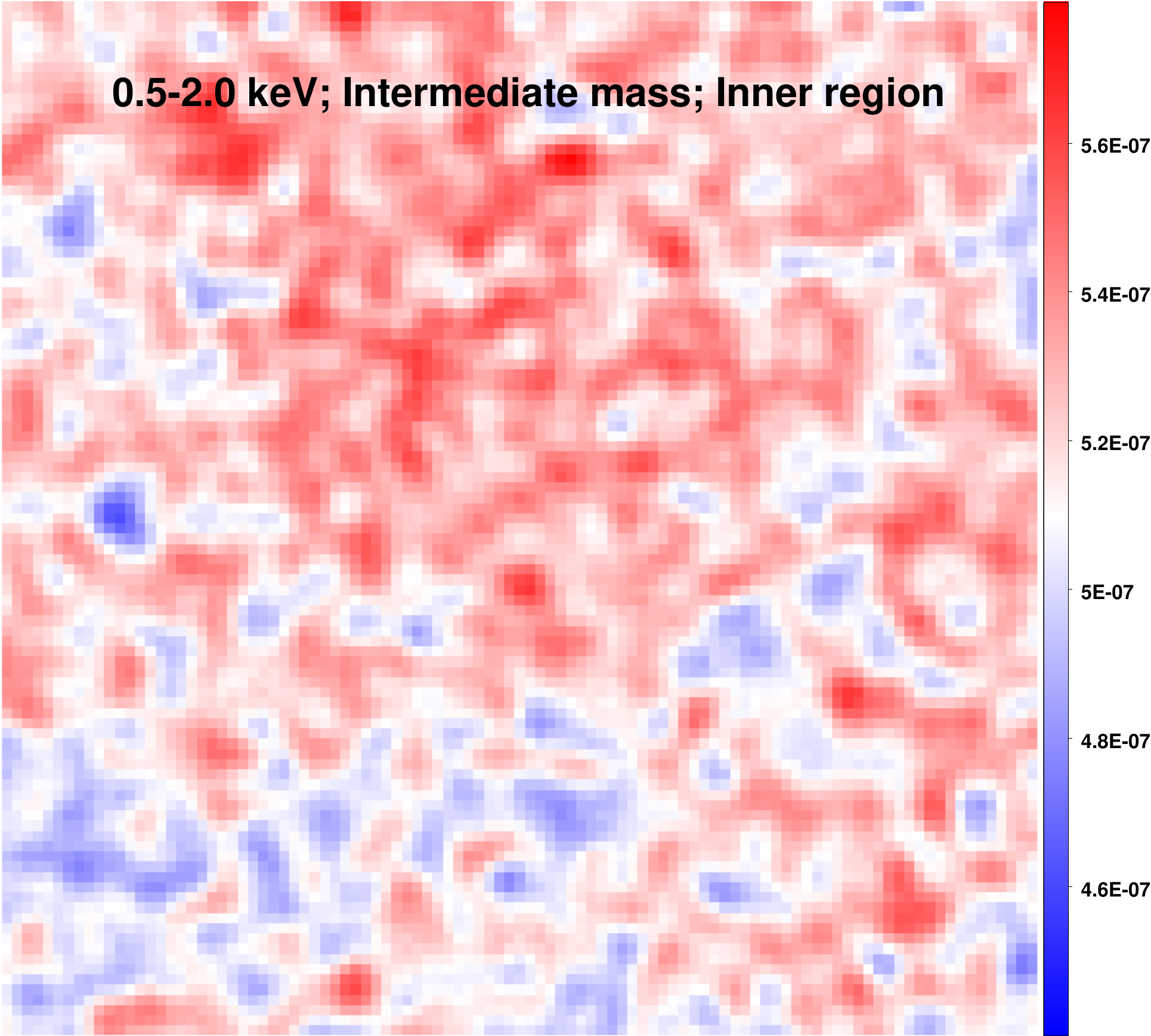}
\hspace{0.45cm} 
\includegraphics[width=0.45\linewidth,height=6cm,keepaspectratio]{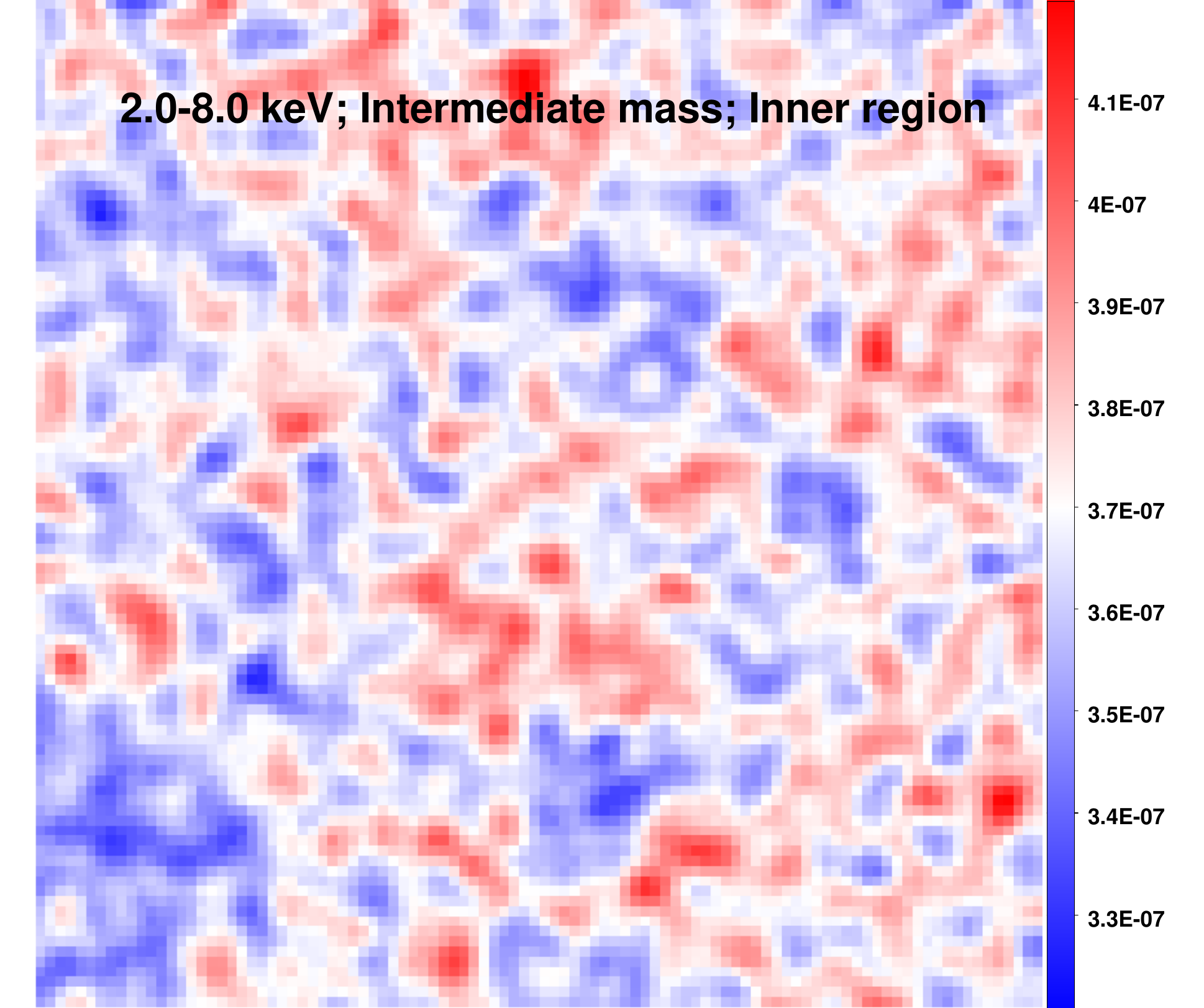}\\\vspace{0.3cm}
 \hspace{-0.08cm}
 \includegraphics[width=0.45\linewidth,height=6cm,keepaspectratio]{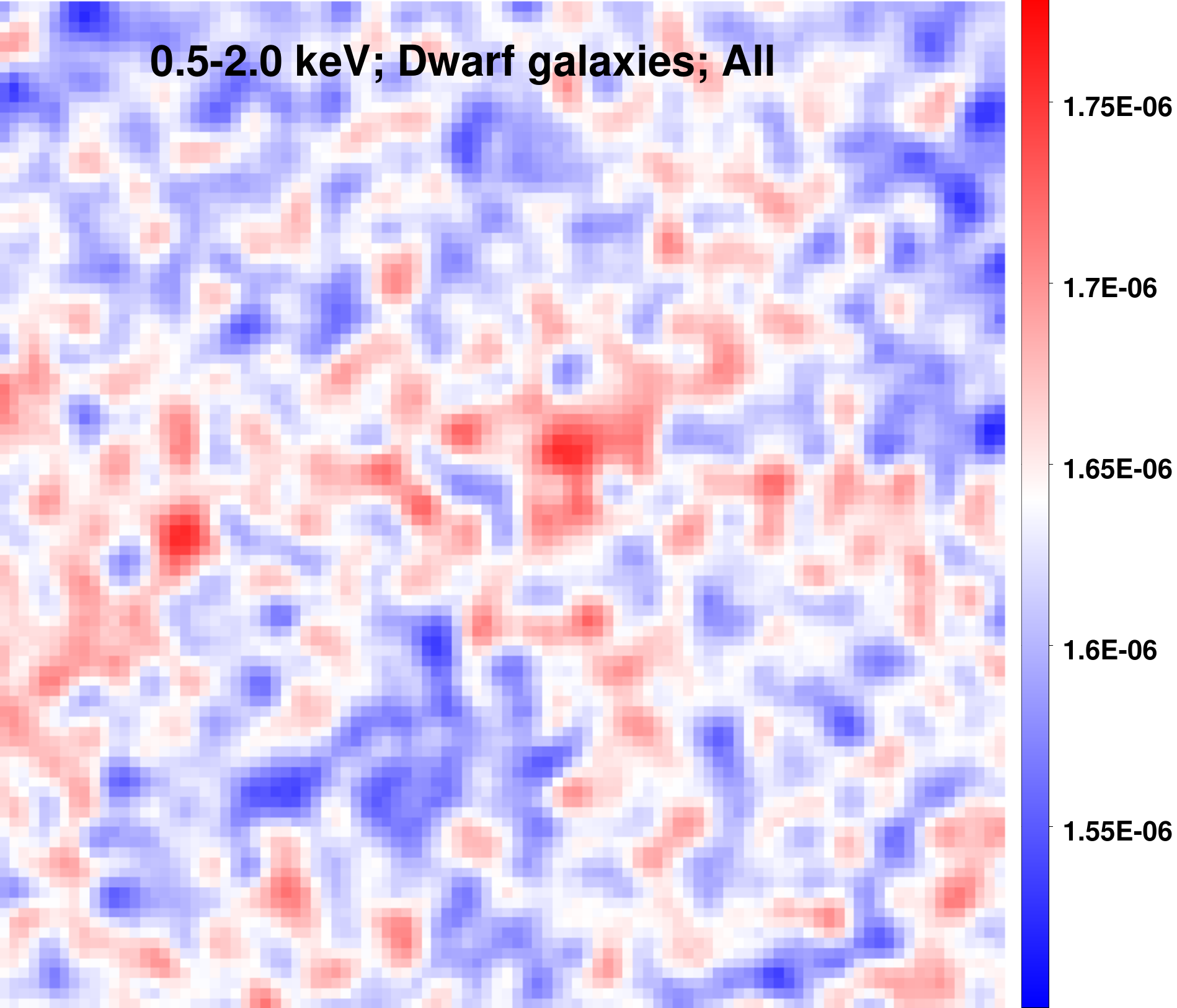}
 \hspace{0.18cm} 
 \includegraphics[width=0.45\linewidth,height=6cm,keepaspectratio]{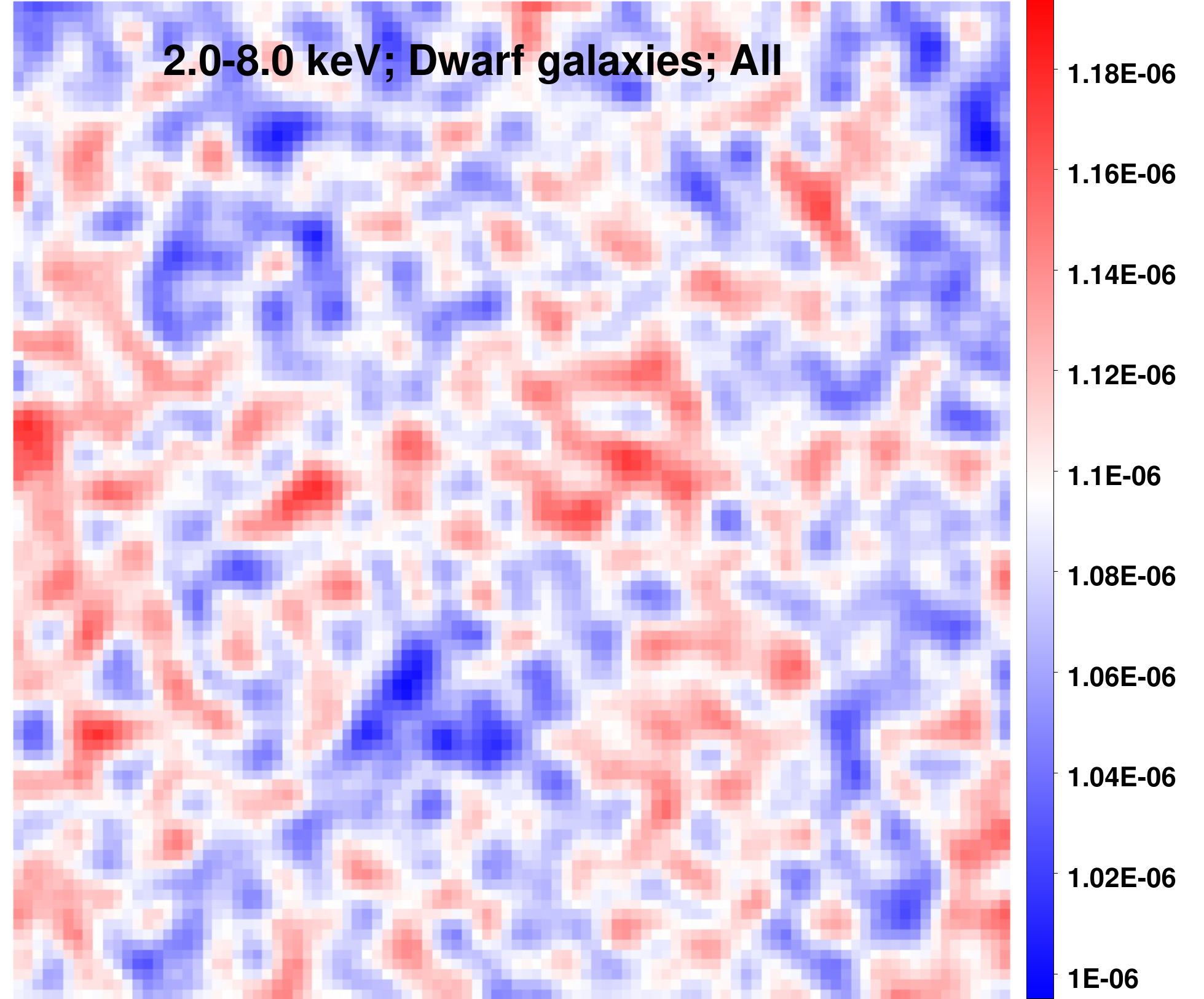}
  \caption{
  Stacked X-ray images of galaxies in different stellar mass and cluster-radial bins. Panels are ordered by rows, from left to right. The top row shows intermediate-mass galaxies beyond 0.5$R_{500}$ in the 0.5--2.0 keV (left) and 2.0--8.0 keV (right) bands, respectively. The middle row shows intermediate-mass galaxies within 0.5$R_{500}$ in the 0.5--2.0 keV (left) and 2.0--8.0 keV (right) bands, respectively. The bottom row shows all dwarf galaxies in the 0.5--2.0 keV (left) and 2--8 keV (right) bands; the dwarf subsamples in separate cluster-centric radial bins are not shown because they are also non-detections. Each image is 23.2 kpc on a side. Galaxies with individually confirmed coronae are excluded from the stack. Units are counts per second per pixel.}
  \label{fig:stackedimages}
\end{figure*}

\subsubsection{X-ray Image Stacking}\label{imagestacking}
Because individual X-ray observations are limited by exposure times and low photon counts, image stacking provides an effective means to enhance the signal-to-noise ratio (S/N), and thereby enable the detection of average emission from a source population where the individual source is too faint to be detected. To investigate stacked X-ray coronal emission from galaxies in our sample, we adopted the following procedure for each stellar mass range. 

For each X-ray observation used in this work, counts and exposure maps were generated using the CIAO task $fluximage$. For each galaxy, cutout images centred on the galaxy were created in the 0.5--2.0 keV and 2.0--8.0 keV energy bands. Prior to stacking, all detected X-ray point sources 
were masked in the cutout images to isolate diffuse thermal emission. The masked regions were defined using circles corresponding to the local 95\% encircled-energy radius of the Chandra PSF at 2 keV and were manually adjusted where necessary following visual inspection. Pixel values within the masked regions were set to zero. Corresponding exposure map cutouts were also generated for each observation. To construct a combined image for each individual galaxy, we summed the counts cutouts from all available observations using the CIAO tool $dmimgcalc$. A similar process was adopted for the exposure map cutouts. Finally, we divided the total stacked counts image by the total stacked exposure map to produce a combined flux image cutout for each galaxy.
The X-ray background in the studies of cluster galaxies is dominated by the X-ray emission of the hot ICM. Contributions from the non-X-ray background and the cosmic X-ray background are relatively small and accounted for using the local background used in all further analysis.
\smallskip\\
\textit{Stacking galaxies in a single cluster}: To generate stacked images for dwarf and intermediate-mass member galaxies within a given galaxy cluster, we extended the above methodology and applied it separately to each stellar mass range. Each galaxy’s counts and exposure map cutouts were created with uniform angular pixel size and identical image dimensions, ensuring a consistent spatial scale across all cutouts. Furthermore, each galaxy in the stack was positioned at the centre of its respective cutout image, ensuring spatial alignment of the galaxies when the cutouts from the selected member galaxies are combined, avoiding any positional mismatches. The stacked counts images and corresponding exposure maps were then summed across the selected member galaxies, producing a single combined counts image and exposure map for each cluster. Division of these images yielded the final stacked flux image, representing the average X-ray emission from the dwarf and intermediate-mass galaxy populations embedded within the cluster environment.
\smallskip\\
\textit{Stacking galaxies across galaxy clusters}: Given the variation in distance among galaxy clusters in our sample, direct comparisons of X-ray emission from galaxies require careful normalisation of physical scales. Although all source images share the same pixel sizes (in arcseconds), the physical size represented by each pixel (in kpc) differs due to the clusters’ varying redshifts. To address this, for each stellar mass range, we first generated stacked images of the corresponding galaxies within each individual cluster (as described above) using their native image scales. 
Subsequently, we reprojected each cluster-level stacked image to a uniform physical scale using the $reproject\_image\_grid$ tool. For the counts images, we applied \textit{method=sum} to maintain total photon counts, while the exposure maps were reprojected using \textit{method=average} to ensure correct flux normalisation after stacking. We selected a target pixel scale of 0.232 kpc/pixel, matching that of the Coma cluster. All reprojected images were mapped to a $100\times100$ pixel grid, which corresponds to 23.2 kpc on a side. This standardisation ensures that analyses -- such as measuring emission within a central region of a few kpc -- are physically consistent across all clusters. Finally, we stacked the reprojected, cluster-level counts and exposure images to produce a composite image that captures the combined X-ray emission from all galaxies in the stack and the ambient ICM background across the cluster sample. The stacking was performed separately for dwarf ($\text{M}_* \leq 10^{9.5}\,\text{M}_{\odot}$) and intermediate-mass ($10^{9.5} < \text{M}_*/\text{M}_{\odot} \leq 10^{10.5}$) galaxies, with each population further divided into inner ($R \leq 0.5R_{500}$) and outer ($0.5R_{500} < R \leq 2.5R_{500}$) regions based on projected cluster-centric radius. For stacking within individual clusters and across the cluster sample, we excluded galaxies with individually detected X-ray coronae and those located within $0.25R_{500}$ of the centres of cool-core clusters. Several of these stacked images are shown in Figure \ref{fig:stackedimages}.

\subsubsection{Spectral Analysis}\label{specextration}
The presence of thermal emission in galaxies, especially those with bright X-ray coronae, is often indicated by emission observed in their soft X-ray (0.5--2.0 keV or 0.5-1.5 keV) images, which benefit from $Chandra$’s excellent spatial resolution. While imaging is helpful for identifying potential thermal emission in less massive galaxies, it is not sufficient to confirm the presence of coronal emission. Spectral analysis is required and remains the most reliable method for robust assessment wherever the data are sufficiently deep. In this work, X-ray images are used primarily as a visual guide to identify potential coronal emission, while all conclusions are derived from spectral analysis.

Previous studies have shown that resolved galactic X-ray coronae embedded within clusters are typically compact, with characteristic median radii of $\sim$2--3 kpc (e.g., \citealt{sun2007,hou2021_virgo}). These results motivate the adoption of a 2.0 kpc physical radius as a representative aperture for capturing centrally concentrated coronal X-ray emission from our low-mass galaxies. However, the Chandra Point Spread Function (PSF) becomes broader at larger angles ($\theta$) from the optical axis, spreading source photons over larger regions. In addition, because 2.0 kpc corresponds to different angular sizes at different redshifts, this fixed physical aperture does not sample the PSF in a uniform way across the sample. For example, for nearby galaxies, the 2.0 kpc region is often larger than the PSF near the aim point, whereas for more distant galaxies it can become smaller than or comparable to the PSF even at similar off-axis angles. As a result, a fixed 2.0 kpc aperture can miss a varying fraction of the source flux depending on both redshift and detector position. To account for this, our spectral analysis uses a circular aperture of $R_{\rm corr}$ = 2.0 kpc + $R_{90}$, where $R_{90}$ is the local 90$\%$ encircled-energy radius at the galaxy's position, determined separately for each observation. Unless otherwise stated, all reported results in this work are based on the $R_{\rm corr}$ aperture. This includes galaxies with individually detected coronae. X-ray spectra and response files for each galaxy were extracted using the \texttt{specextract} task from circular regions of radius $R_{\rm corr}$ centred on each member galaxy. This approach improves flux recovery and ensures more uniform measurements across the sample. To estimate the local cluster background, an annular region located outside the chosen source aperture was used. Spectra of galaxies with multiple observations were co-added using $combine\_spectra$ to produce a combined spectrum along with exposure-weighted responses, and an area and exposure-weighted combined background.
The dependence of $R_{90}$ on off-axis angle relative to the aim point at a fixed energy of 1 keV for galaxies in our sample is shown in Figure \ref{fig:psf}.

\textit{X-ray Spectral Stacking}: Following the same galaxy subsample definitions adopted for image stacking (Section \ref{imagestacking}), we combined individual X-ray spectra and their corresponding response files, extracted within the circular source aperture for each galaxy in the sample, using $combine\_spectra$. To ensure the stacked spectra reflect the typical diffuse emission rather than being dominated by bright sources, we excluded galaxies with contaminating point sources in the central 2 kpc or those hosting individually detected coronae. Additionally, to avoid contamination from the bright cores of cool-core clusters, galaxies located within $0.25R_{500}$ of the centre in cool-core clusters were removed from the spectral stacking analysis, as done for the image stacking analysis (Section \ref{imagestacking}). The  backgrounds for all galaxies in each stellar mass range were similarly combined, with appropriate weighting by exposure time and source extraction area. The resulting stacked spectrum was then analysed to search for signatures of thermal coronal emission, thereby providing a well-constrained assessment of the average hot gas properties in the dwarf and intermediate-mass galaxy populations in clusters.
\smallskip\\
\textit{Spectral modelling}: Each X-ray spectrum was carefully inspected in XSPEC (\citealt{xspec}) for signatures of thermal coronal emission, in particular the presence of the iron L-line hump (Fe-L complex), a blend of emission lines arising from L-shell transitions of highly ionised iron that appears at $\sim$0.7--1.2 keV and is most sensitive to gas temperatures between 0.5 and 1.5 keV. In cases where such emission was seen, the ICM background-subtracted spectra were modelled in the 0.5--7.0 keV energy range using the model \textit{tbabs(apec + powerlaw +lnvmkl)}. We used Cash statistic ($cstat$) to fit the spectrum. The model component $tbabs$ (\citealt{tbabs}) accounts for Galactic absorption along the line-of-sight, the $apec$ component represents the soft thermal emission from the hot gas in the galaxy, the $powerlaw$ component models the contribution from unresolved low-mass X-ray binary (LMXB) population or any unresolved AGN. For individually detected coronae, it also accounts for emission from X-ray point sources within 2 kpc of the galaxy centre, which were retained in the extraction regions. The $lnvmkl$ accounts for emission from the unresolved cataclysmic variable (CV) and coronally active binary (AB) star population in galaxies (\citealt{ge2015}). For low-mass galaxies, the abundance in the $apec$ model was fixed at 0.3 $Z_{\odot}$, and for intermediate-mass galaxies at 0.4 $Z_{\odot}$, following the empirical trend of lower metallicity in less massive systems (e.g., \citealt{tremonti2004}). We adopted the $aspl$ (\citealt{aspl}) solar abundance table in the analysis. The photon index of the \textit{powerlaw} component was fixed at 1.7, typical for the integrated emission from LMXB and AGN populations. The normalisation of the $lnvmkl$ component was fixed to correspond to a $0.5-2.0$ keV emissivity of $6.0\times10^{27}$ erg s$^{-1}$ $\text M_{\odot}^{-1}$, representing the expected contribution from unresolved CVs and ABs (\citealt{ge2015}). The reported 0.5--2.0 keV X-ray luminosities of the hot-gas coronae are derived from the $apec$ component.

In faint X-ray sources with low metallicity -- such as the galactic X-ray halos in the mass range studied here -- a steep \textit{powerlaw} (with index $\gg1.7$) can sometimes mimic the soft diffuse emission in the 0.5--2.0 keV band. To test whether the soft emission truly originates from hot gas, we first fitted each spectrum with a single absorbed \textit{powerlaw} model, allowing the photon index to vary freely. Following \citet{sun2007}, only if the 1$\sigma$ lower bound of the best-fit photon index exceeded 2.4, we proceeded to fit the spectrum with the composite model described above.

\subsubsection{Upper Limits for Non-Detections}
In cases where no significant detection of coronal emission was found from the stacked galaxy population, we estimated $3\sigma$ upper limits on the 0.5--2.0 keV X-ray luminosity of thermal coronal emission, adopting the following procedure.

For each galaxy, we used the combined cutout counts image and the corresponding exposure map (in seconds) (Section 2.2.1) from all observations used in this work. The background noise level was estimated from the composite aperture $R_{\rm corr}$, and converted to a background count-rate uncertainty by dividing by the corresponding exposure time. To convert the noise in the count rate into a $1\sigma$ flux upper limit, we employed the $modelflux$ tool, using an absorbed $apec$ model along with the redistribution matrix file (RMF) and auxiliary response file (ARF) generated for the circular source aperture (Section 2.2.2). The adopted gas temperatures were 
0.6 keV for intermediate-mass galaxies, based on the average value of our individual detections, and 0.2 keV for dwarfs, chosen as a representative value for low-mass systems (e.g., \citealt{ngc7793}). Abundances were set to 0.4  $Z_{\odot}$ and 0.3 $Z_{\odot}$ for intermediate-mass and dwarf galaxies, respectively. This procedure was applied individually to each galaxy. The resulting $1\sigma$ flux upper limits were then converted to $1\sigma$ X-ray luminosity upper limits using the luminosity distance of each galaxy’s host cluster, allowing for direct comparison of intrinsic source luminosities.

To obtain a global constraint on the coronal soft X-ray luminosity of stacked galaxy population in clusters, we combined the individual upper limits using inverse-variance weighting. This approach assigns greater weight to measurements with smaller uncertainties. Mathematically, the uncertainty on the inverse-variance weighted mean luminosity, $\sigma_{L_X}$, is given by
\[
\sigma_{L_X} = \left( \sum_{i} \frac{1}{\sigma_{L_X, i}^2} \right)^{-\frac{1}{2}}
\]
where $\sigma_{L_X,i}$ is the $\boldsymbol{1\sigma}$ upper limit on the X-ray luminosity of the i-th galaxy. In the absence of a statistically significant detection, $\sigma_{L_X}$ is interpreted as a global $1\sigma$ upper limit on the coronal soft X-ray luminosity of the galaxy population. The corresponding $3\sigma$ upper limit is then given by 3$\sigma_{L_X}$. By employing this method, we derived a global constraint on the coronal soft X-ray emission, thus combining heterogeneous data with varying uncertainties.

\begin{table*}
\centering
\small
\setlength{\tabcolsep}{6pt}
\caption{Cluster members with individually detected hot-gas coronae and their properties; KUG 1259+280 is included for completeness.}
\label{tab:individual detections}
\begin{tabular}{lccclcccc}
\hline
Galaxy (Host cluster)&log (M$_*$/M$_{\odot}$)&$g-r$&$R/R_{500}$&$v^{(1)}$&$kT^{(2)}$&log $L_{X (0.5-2)}^{(3)}$ &PS$^{(4)}$& $L_{X (2-10)}^{(5)}$\\
&&&&(km s$^{-1})$&(keV)&(erg s$^{-1}$)&&($\times10^{39}$ erg s$^{-1}$)\\
\hline
NGC 3860B$^{a}$ (A1367)&9.42&0.33&0.08&
+1718&$0.68^{+0.15}_{-0.31}$&$39.47^{+0.13}_{-0.16}$&Y&$<2.34$\\
LEDA 1604354&9.93&0.69&0.21&
+1506&$0.57^{+0.36}_{-0.39}$&$39.25^{+0.20}_{-0.30}$&N&$<1.90$\\
CGCG 097-113&10.35&0.80&0.05&
-336&$0.96^{+0.58}_{-0.65}$&$38.90^{+0.24}_{-0.46}$&Y&$0.91^{+0.94}_{-0.86}$\\
CGCG 097-109&10.44&0.80&0.11&
-1&$0.90^{+0.23}_{-0.24}$&$39.35^{+0.12}_{-0.16}$&Y&$0.81^{+0.75}_{-0.73}$\\
\hline
LEDA 1827127$^{a}$ (Coma)&9.38&0.46&0.46&
+1183&$0.77^{+0.21}_{-0.35}$&$39.35^{+0.10}_{-0.12}$&N&$<1.18$\\
NGC 4858&9.63&0.38&0.30&
+2400&$0.68^{+0.11}_{-0.12}$&$39.95^{+0.05}_{-0.06}$&Y&$2.19^{+0.98}_{-0.97}$\\
D100&9.68&0.49&0.16&
-1892&$0.85^{+0.09}_{-0.10}$&$39.60^{+0.11}_{-0.13}$&Y&$<3.77$\\
KUG 1259+280$^*$&9.93&0.52&0.70&
+97&$0.26^{+0.03}_{-0.03}$&$40.8^{+0.06}_{-0.06}$&Y&$66.1^{+12.0}_{-12.2}$\\
LEDA 83727&10.23&0.78&0.34&
-12&$0.79^{+0.26}_{-0.27}$&$38.92^{+0.22}_{-0.32}$&Y&$0.83^{+0.46}_{-0.45}$\\
NGC 4864&10.24&0.93&0.16&
-265&$0.74^{+0.14}_{-0.16}$&$39.81^{+0.13}_{-0.18}$&Y&$2.13^{+1.64}_{-1.46}$\\
\hline
LEDA 86299 (A2199)&10.43&0.83&0.24&
-156&$0.94^{+0.22}_{-0.23}$&$39.87^{+0.11}_{-0.26}$&Y&$<7.50$\\
\hline
LEDA 93208 (A85)&10.44&0.70&0.51&
+93&$0.71^{+0.61}_{-0.48}$&$39.74^{+0.28}_{-0.60}$&N&$<8.80$\\
\hline
LEDA 1783873 (A1795)&10.27&0.46&0.36&
-1110&$0.62^{+0.20}_{-0.35}$&$40.1^{+0.12}_{-0.15}$&Y&$8.32^{+2.53}_{-2.41}$\\
\hline
W J201304.51-564856.3$^b$ (A3667)&9.90&0.57&0.18&
+1397&$0.58^{+0.26}_{-0.33}$&$39.79^{+0.17}_{-0.25}$&Y&$<7.8$\\
W J201252.55-565546.5$^b$&10.28&0.86&0.27&
+422&$0.28^{+0.44}_{-0.11}$&$39.53^{+0.27}_{-0.52}$&N&$2.23^{+1.82}_{-1.94}$\\
\hline
LEDA 822373 (A133)&10.35&0.80&0.09&
+125&$0.92^{+0.39}_{-0.22}$&$40.01^{+0.20}_{-0.32}$&Y&$15.8^{+9.6}_{-9.8}$\\
\hline
LEDA 197768 (AWM7)&10.24&0.76&0.05&
+1045&$0.63^{+0.17}_{-0.28}$&$39.6^{+0.13}_{-0.17}$&Y&$1.29^{+1.17}_{-1.16}$\\
NGC 1131&10.42&0.71&0.04&
+165&$0.58^{+0.16}_{-0.20}$&$39.83^{+0.10}_{-0.13}$&Y&$5.19^{+1.52}_{-1.50}$\\
\hline
2M J19210317+4353406 (A2319)&10.40&0.72&0.16&
-275&$0.58^{+0.68}_{-0.38}$&$39.85^{+0.24}_{-0.44}$&N&$<7.25$\\
\hline
WKK 6233 (A3627)&9.72&0.54&0.03&
+259&$0.87^{+0.27}_{-0.34}$&$39.48^{+0.14}_{-0.18}$&Y&$4.16^{+1.71}_{-1.66}$\\  
WKK 6229&9.77&0.51&0.04&
+433&$0.28^{+0.08}_{-0.04}$&$39.69^{+0.12}_{-0.16}$&Y&$8.33^{+0.93}_{-0.94}$\\
WKK 6242&9.97&0.49&0.04&
+355&$0.66^{+0.11}_{-0.13}$&$39.88^{+0.10}_{-0.12}$&Y&$4.10^{+1.58}_{-1.54}$\\ 
ESO 137-004&10.25&0.49&0.12&
-1403&$0.39^{+0.76}_{-0.19}$&$39.28^{+0.20}_{-0.32}$&Y&$1.01^{+0.85}_{-0.88}$\\  
\hline

\end{tabular}
	\vspace{0.5cm}
    \begin{tablenotes}
      \small{\item $^a$ Galaxies with stellar mass below $10^{9.5}$ M$_{\odot}$; $^b$W = WINGS
      \item $^*$ KUG 1259+280 is included for completeness; its soft X-ray excess originates from AGN related quasi-periodic eruptions (QPEs; \citealt{kug_qpe}).
      \item Column information: $^{(1)}$ lists the radial velocity of the galaxy relative to the cluster mean velocity; $^{(2)}$ and $^{(3)}$ list temperature and soft X-ray luminosity of the hot gas, respectively, from spectral analysis; $^{(4)}$ indicates if an X-ray point source is detected within 2 kpc of the galaxy centre; $^{(5)}$ lists the measured hard X-ray luminosity and 3-$\sigma$ upper limits of the powerlaw component from the composite model fit. All error bars represent 90$\%$ confidence intervals.}
    \end{tablenotes}
\end{table*}\vspace{0.2cm}

\begin{table*}
\small
\setlength{\tabcolsep}{6pt}
\caption{Number of galaxies and 3-$\sigma$ upper limits on X-ray luminosity (0.5--2.0 keV; units of erg s$^{-1}$) in stellar mass-selected stacks, further subdivided into inner and outer host environment.}
\label{tab:upperlimitsanddetections}
\begin{tabular}{lll|ll}
\hline
      &$M_{*}\leq10^{9.5} \text{ M}_\odot$ (603)&&$10^{9.5} \text{ M}_\odot<M_{*}\leq10^{10.5} \text{ M}_\odot$ (626)&\\
     \hline
     kT&--&&--&\\
     $L_X$&$<3.17\times10^{38}$&&$<1.93\times10^{38}$&\\
     \hline
     &$R\leq 0.5R_{500}$ (341)&$0.5R_{500} < R \leq 2.5R_{500}$ (262)&$R \leq 0.5R_{500}$ (292)&$0.5R_{500} < R \leq 2.5R_{500}$ (334)\\
     \hline
     kT&--&--&--&--\\
     $L_X$&$<4.16\times10^{38}$&$<4.91\times10^{38}$&$<2.32\times10^{38}$&$<3.52\times10^{38}$\\
\hline
\end{tabular}
	\vspace{0.5cm}
\end{table*}\vspace{0.2cm}

\begin{figure*}
  \centering
  \subfloat[]{\includegraphics[width=0.8\linewidth,height=0.55\linewidth]{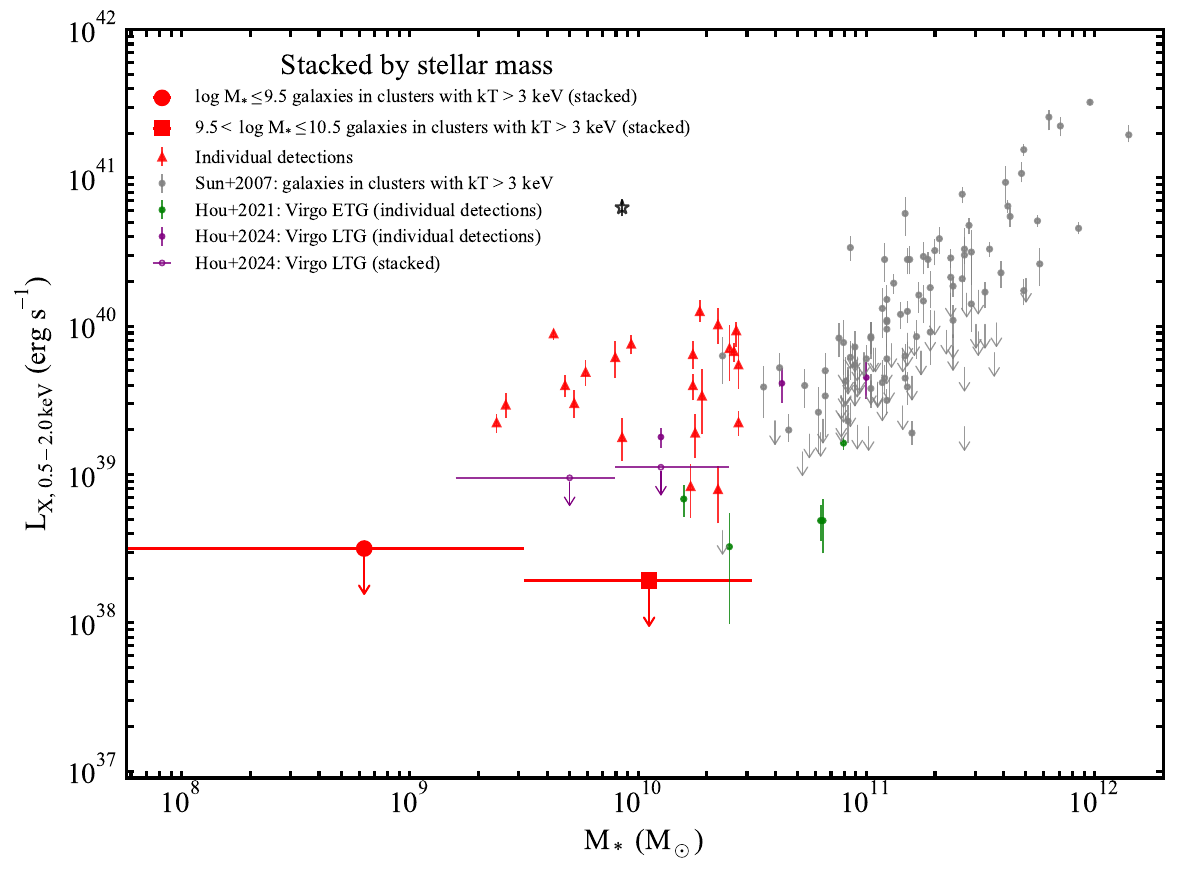}\label{}} \vspace{-0.5em}
  \subfloat[]{\includegraphics[width=0.8\linewidth,height=0.55\linewidth]{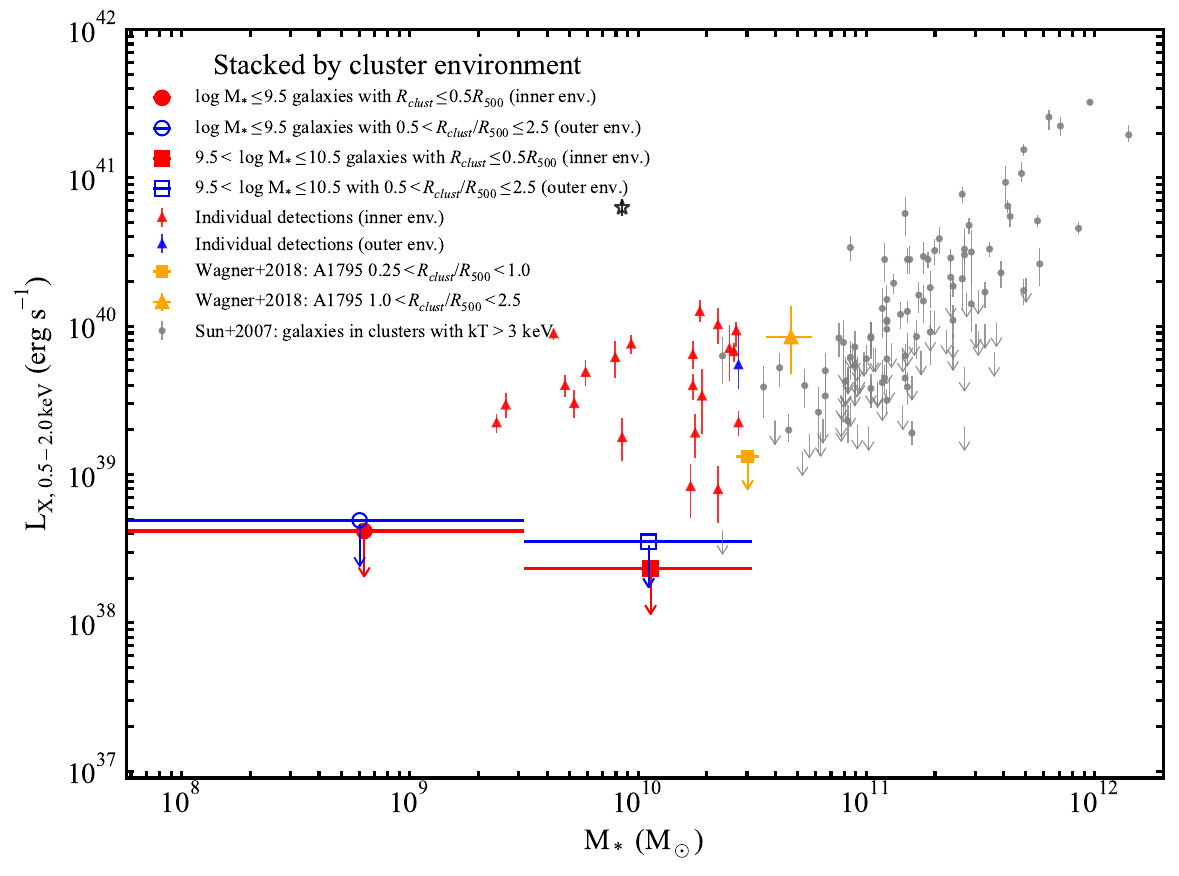}\label{}}\\
  \caption{0.5--2.0 keV X-ray gas luminosity versus galaxy stellar mass for individual detections and stacked bins. Panel (a) shows galaxies grouped by stellar mass: dwarf galaxies ($\text{M}_* \leq 10^{9.5} \text{ M}_\odot$; 603 galaxies) and intermediate-mass galaxies ($10^{9.5} \text{ M}_\odot < M_* \leq 10^{10.5} \text{ M}_\odot$; 626 galaxies). Galaxies individually detected with hot gas are plotted as \textit{solid red triangles}. The individual detection fraction of hot gaseous coronae in our low-mass galaxy sample is low, at $\sim$$1.4\%$. When split by stellar mass, the detection fractions are $\sim$$0.3\%$ for dwarf galaxies and $\sim$$2.4\%$ for intermediate-mass galaxies, respectively. These values are significantly lower than those reported by \citet{sun2007} for all types of massive galaxies ($\text{M}_* > 10^{10.5} \text{ M}_\odot$) excluding cD galaxies, with a coronal detection fraction of $\sim$$40\%$.
  Downward arrows indicate 3-$\sigma$ upper limits for non-detections. In panel (b), galaxies in each stellar mass bin are further segregated based on their host cluster environment -- galaxies within $0.5R_{500}$ are in the inner environment (shown in \textit{red}), while those with $0.5R_{500} < R_{clust} \leq 2.5R_{500}$ (\textit{blue}) are in the outer environment. Among dwarf galaxies, 341 (262) are in the inner (outer) environment. Among intermediate-mass galaxies, 292 (334) are in the inner (outer) environment. 
  The \textit{empty star} in both panels marks galaxy KUG 1259+280. Literature data from \citet{sun2007, wagner2018_a1795, hou2021_virgo, hou2024_virgo} are included for comparison. Error bars represent 1-$\sigma$ uncertainties. For plotting purposes, the lower bound of the dwarf galaxy stellar mass bin is clipped at $5.73\times10^7 \text{ M}_\odot$, with $90\%$ of dwarf galaxies having M$_*$ above this value.}
  \label{fig:LX-M*}
\end{figure*}

\begin{figure}
    \centering
    \includegraphics[width=\columnwidth]{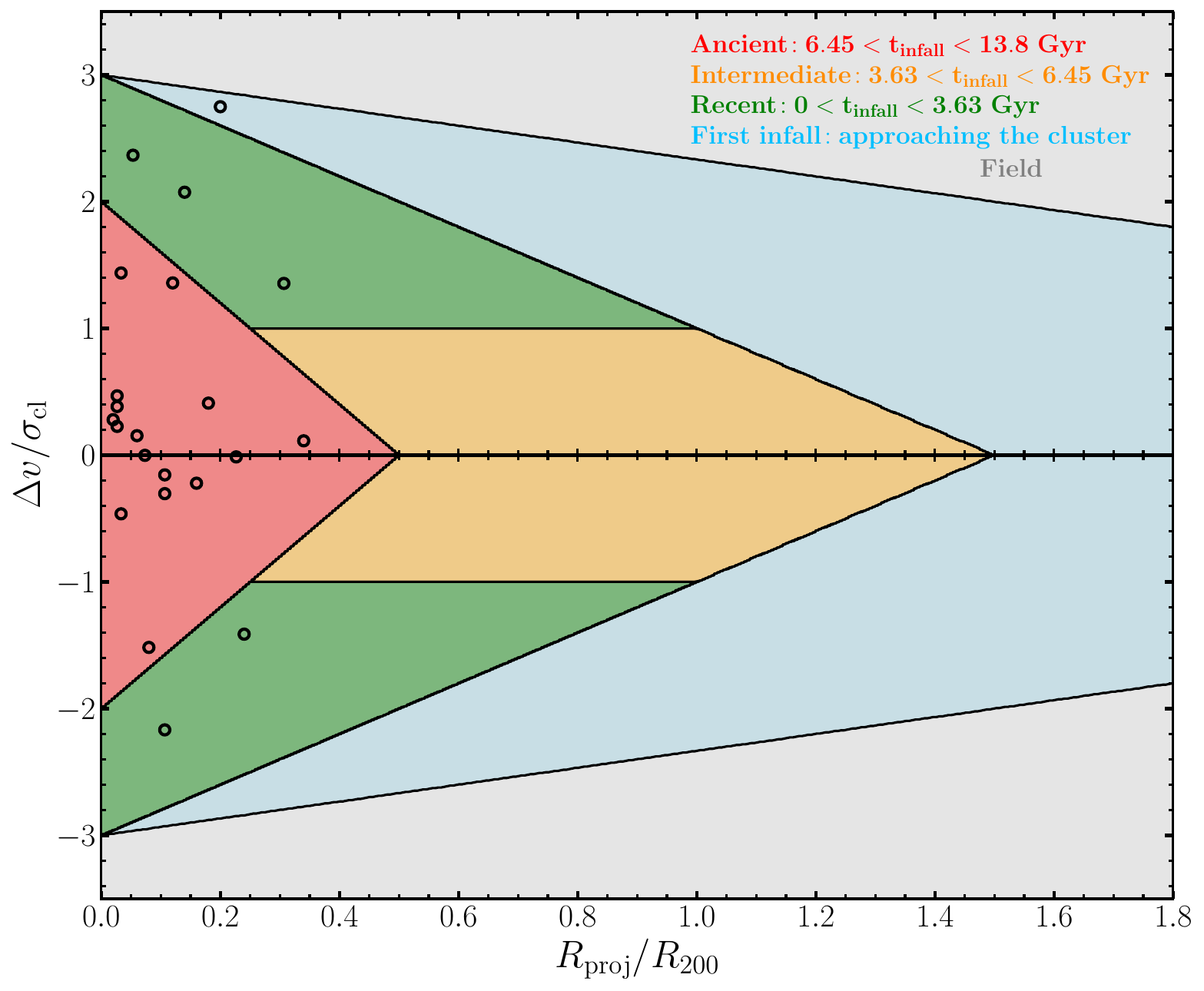}
    \caption{Phase-space diagram with coloured regions adopted from \citet{mun2021}, defined according to the time since infall ($t_{\rm infall}$) into the galaxy cluster, following the classification of \citet{rhee2017}. Circles mark galaxies with individual hot-gas detections in this work.}
    \label{fig:phase}
\end{figure}

\section{Results}\label{sec:results}
To investigate the presence of hot gaseous coronae in cluster galaxies, we examined X-ray thermal emission in intermediate-mass galaxies ($10^{9.5}< \text{M}_*\leq10^{10.5} \text{ M}_{\odot}$) and dwarf galaxies ($\text{M}_*\leq10^{9.5} \text{ M}_{\odot}$) across 18 nearby clusters using the methods described in Section 2.2.2.  A careful galaxy-by-galaxy inspection
resulted in 22 galaxies with X-ray coronae (hot phase ISM) confirmed via the presence of the iron-L line hump in their spectra. Their properties are listed in Table \ref{tab:individual detections} and spectra shown in Figure \ref{fig:individualdetectionsspectrum}. The detected individual coronae exhibit soft X-ray luminosities of a few times $10^{39}$ erg s$^{-1}$, temperatures ranging from 0.3 to 1.0 keV, with average electron gas densities of a few times $10^{-2}$ cm$^{-3}$.
  
The remaining galaxies were stacked to enhance the signal, as described in Section 2.2.2, and study the average coronal properties of the intermediate-mass and dwarf galaxy populations. We then divided the galaxy sample to explore potential dependencies on host environment and galaxy colour.
\subsection{Thermal Emission in Intermediate-mass Galaxies in Clusters}

First, we stacked all intermediate-mass galaxies without distinguishing by environment or colour. Using the $R_{\rm corr}$ aperture, which accounts for PSF smearing effects, we find no significant detection of soft thermal X-ray emission in the stacked spectrum, including no evidence for Fe-L complex emission. We therefore derive a 3$\sigma$ upper limit of $1.93\times10^{38} \text{ erg} \text{ s}^{-1}$ on the 0.5--2.0 keV X-ray luminosity (Table \ref{tab:upperlimitsanddetections}), which we adopt as our primary constraint on the average coronal emission from the intermediate-mass galaxy population. Next, to assess potential cluster-to-cluster variation, we inspected individual cluster stacks. No individual cluster stack shows a significant detection. In subsequent analysis, we examine subsamples split by environment and colour to investigate potential dependencies of the hot gas emission in intermediate-mass galaxies.

\subsubsection{Stacking by environment}
To investigate environmental effects, galaxies were classified based on their projected distance from the cluster centre, separating inner ($R< 0.5R_{500}$) and outer populations (between $0.5R_{500}$ and $2.5R_{500}$), which were then stacked separately for analysis.
No statistically significant X-ray emission is detected in either subsample, and we therefore derive 3$\sigma$ upper limits on the 0.5--2.0 keV X-ray luminosity for both the inner and outer populations. These values are reported in Table \ref{tab:upperlimitsanddetections}. 

We note that some enhancement in soft X-ray emission is visible in the central region of the PSF-convolved stacked image of the outer-environment subsample of intermediate-mass galaxies (Figure \ref{fig:stackedimages} top-left). However, tests using several different extraction apertures show no evidence for a thermal component in the stacked spectra.

\subsubsection{Stacking by colour}
To investigate possible colour dependence, we divided our intermediate-mass galaxy sample into blue ($g-r<0.6$) and red ($g-r\geq0.6$) populations. Each colour-segregated subset was stacked separately following the procedure described in Section 2.2.2. Qualitatively, neither stack shows a detectable thermal signature. Given the small fraction of blue galaxies in the sample ($\sim$7$\%$), we do not report quantitative results for the colour-based stacks.
\subsection{Coronal Emission in Dwarf Galaxies in Clusters}
Following the same procedure used for intermediate-mass galaxies (Section 3.1), we stacked galaxies in the dwarf regime ($\text{M}_* \leq 10^{9.5} \text{M}_{\odot}$). No significant emission is detected when dwarf galaxies are stacked by environment, by colour, or across the entire sample. The X-ray spectrum of the full dwarf galaxy stack is shown in Figure \ref{fig:dwarfstackspec}.
Even when examining dwarf galaxies within individual clusters, neither the stacked images nor the spectra show any unambiguous evidence of coronal X-ray emission. Therefore, as for the intermediate-mass galaxy population, we derive upper limits on the X-ray luminosity of soft thermal coronal emission in the dwarf galaxy population within clusters, which are listed in Table \ref{tab:upperlimitsanddetections}.

\section{Discussion}\label{sec:discussion}

In this work, we investigate the presence of hot X-ray gas content in intermediate-mass ($10^{9.5}< \text{M}_*\leq10^{10.5} \text{ M}_{\odot}$) and dwarf galaxies ($\text{M}_*\leq10^{9.5} \text{ M}_{\odot}$) across 18 massive galaxy clusters ($kT>3$ keV) in the nearby Universe ($0.011 < z < 0.063$) -- a population that has not been systematically studied in the literature, as previous studies have focused primarily on massive galaxies (e.g., \citealt{sun2007}) or less rich environments (e.g., \citealt{hou2021_virgo,hou2024_virgo}). Using archival $Chandra$ ACIS observations spanning over two decades, we search for this gas in a sample of 823 intermediate-mass and 735 dwarf galaxies. 

\subsection{Individual detections}

We detect X-ray coronae in a total of 22 galaxies in our dwarf and intermediate-mass galaxy sample. The fraction of galaxies hosting individual X-ray coronae is $\sim$2.4$\%$ for intermediate-mass galaxies and $\sim$0.3$\%$ for dwarfs, highlighting the rarity of these systems in the low-stellar mass regime. For comparison, among massive galaxies with $\text{M}_*\gtrsim10^{10.5} \text{ M}_{\odot}$, \citet{sun2007} reported a coronal detection fraction of $\sim$$40\%$ across all morphological types, excluding cD galaxies. This fraction increases to $\sim$$60\%$ for the more massive sample of elliptical galaxies with $\text{M}_*\gtrsim10^{11.0} \text{ M}_{\odot}$. The difference between these fractions reflects the higher stellar-mass threshold and the restriction to elliptical galaxies in the latter sample. Figure \ref{fig:LX-M*}a presents the soft X-ray luminosities of our detected galaxies as a function of their stellar mass, together with literature data, showing the general trend of higher luminosities at larger stellar masses. These properties suggest that even low-mass galaxies can form and maintain small yet resilient X-ray cool cores capable of surviving the harsh cluster environment, albeit with low detection fractions, highlighting that under favourable conditions, the retention of dense, X-ray cool cores can occur in low-mass systems and is not restricted to massive early-type galaxies. Among the 22 galaxies, central stellar velocity dispersions are available for 9 systems, spanning 21--197 km s$^{-1}$. For these systems, the $\beta$ parameter, defined in Section \ref{sec:intro} as the ratio of the specific stellar kinetic energy to the specific thermal energy of the gas, ranges from 0.02 to 0.5, well below unity. In an idealised scenario where the gas temperature is determined primarily by the depth of the gravitational potential, $\beta$ is expected to be close to unity. The low $\beta$ values found here indicate that the hot gas possesses substantially more thermal energy than expected from the stellar velocity dispersion alone. Similar departures from $\beta\sim1$ have been reported in massive early-type galaxies (e.g., \citealt{forman1985,davis1996,mathhews2003,sun2007}), and are generally interpreted as evidence for additional heating associated with stellar feedback and, possibly, AGN activity. Given the shallow gravitational potentials of the low-mass systems studied here, the very low $\beta$ ($<0.2$ in some cases) values imply that non-gravitational heating processes contribute significantly to the thermal energy of the hot gas.

Interestingly, 16 out of 22 galaxies with individually detected hot gas occupy the ancient infall region of the cluster phase-space diagram shown in Figure \ref{fig:phase}, prompting the question of how these galaxies have retained their compact gas reservoirs over long timescales ($t_{\rm infall}>6.45 \text{ Gyr}$). Nearly all of these galaxies are early-type systems, based on their optical colours and morphologies. In general, these ancient infallers formed the bulk of their stars early, with light-weighted ages of $\sim$8-9 Gyr or more (e.g., \citealt{gallazi2021}), at a time when galaxies of similar mass were smaller and hence more compact (e.g., \citealt{vanderwel2014,shibuya2015}). This early formation likely produced dense central stellar concentrations (e.g., \citealt{vandokkum2014}), creating deep gravitational potentials in their cores. Such deep potentials likely play a key role in resisting ICM ram-pressure stripping, enabling these galaxies to retain mini-cool cores over billions of years. Figure \ref{fig:phase} also suggests two groups in $|\Delta v/\sigma_{cl}|$, one below $\sim$0.5 and the other above $\sim$1.3. The low $|\Delta v/\sigma_{cl}|$ group may represent older galaxies that have resisted stripping, whereas galaxies in the higher $|\Delta v/\sigma_{cl}|$ group may have fallen into the cluster more recently and have not yet lost their gas. However, the small sample does not allow us to confirm that these are two separate populations. At the same time, being an ancient infaller is not a sufficient condition for hot gas survival. When restricting the full low-mass sample to the ancient infall region (758 galaxies), we find a retention fraction of $\sim$2.1$\%$, demonstrating that hot gas survival remains rare even within this subset. This indicates that the survival of hot gas in cluster galaxies is a multi-factor phenomenon, with dense central potentials being only one contributing factor. Other factors may include pressure confinement by the ICM, suppression of heat conduction and evaporation (e.g., \citealt{lu2011,rv2017a,rv2017b}), and the orbital paths of galaxies within the cluster, which determine their exposure to RPS (e.g., \citealt{mccarthy2008,wetzel2011,lotz2019}).

Optical spectra from SDSS/DESI are available for nine galaxies with individual hot-gas detections, and indicate that the sample is not spectroscopically homogeneous. Two galaxies -- LEDA 1604354 and LEDA 1827127 -- are star-forming, while three galaxies -- NGC 3860B, NGC 4858, and D100 -- are classified as starburst galaxies.  Notably, D100 and NGC 4858 are well-studied ram-pressure stripped galaxies (see e.g., \citealt{yagi2007, yagi2010, jachym2017, cramer2019, chen2020, sun2021, boselli2022, souchereau2025}). Four galaxies -- CGCG 097-113, LEDA 83727, NGC 4864, and LEDA 86299 -- exhibit absorption-line spectra, consistent with quiescent systems. KUG 1259+280 is not included in either the star-forming or quiescent classes. The presence of star-forming and starburst galaxies indicates that, in at least a subset of the individually detected sample, stellar feedback associated with ongoing or recent star formation may contribute to the observed thermal X-ray emission, with the surrounding ICM likely aiding in confining the hot gas and supporting its survival. The mixed spectroscopic properties of the galaxies with individual coronal detections therefore point toward a combination of internal and environmental processes governing their hot gas reservoirs. Given the small number of individual detections in our sample, we cannot confidently attribute hot gas retention to any single cause for individual galaxies.

\subsection{Pressure confinement by the ICM}
Evidence for pressure confinement in surviving galaxy coronae of early-type galaxies is apparent from a comparison with those in the Virgo cluster (Figure \ref{fig:LX-M*}a), which hosts a substantially lower-pressure ICM than the massive clusters studied here. At fixed stellar mass, early-type Virgo galaxies typically host less luminous X-ray-emitting gas, or lack detectable hot coronae altogether. For example, VCC 1630 ($10^{10.2}\text{ M}_{\odot}$) and VCC 944 ($10^{10.4} \text{ M}_{\odot}$) in Virgo (\citealt{hou2021_virgo}), located at projected cluster-centric distances of $\sim$0.45--0.5$R_{500}$, are cooler and fainter in X-rays than galaxies of similar stellar mass in our sample at equal or smaller cluster-centric distances, which host hotter and more luminous compact coronae (Table \ref{tab:individual detections}, Figure \ref{fig:LX-M*}a). This suggests stronger pressure confinement by the ICM in massive clusters. While building deep central potentials may be necessary for long-term gas retention in low-mass galaxies, the ambient ICM pressure also plays a key role in regulating the thermodynamic state and the X-ray visibility of the surviving hot gas. Harsh as it may seem, under favourable conditions the host ICM can gently sustain the survival of compact coronae, even in the lower stellar mass regime. 

By comparison, most present-day galaxies of similar stellar mass in less-rich environments (e.g., the field, filaments, or low-density groups) formed later, as suggested by their younger luminosity-weighted ages (e.g., \citealt{clemens2006,tiwari2020}), and are likely less compact on average based on observed trends in galaxy size-mass evolution (e.g., \citealt{vanderwel2014,shibuya2015}), which could contribute to shallower central potentials. This, combined with the absence of a dense ICM providing external pressure confinement, and the contribution of stellar feedback driven outflows, may make it more difficult for these galaxies to retain long-lived hot gas. This is consistent with the observed rarity of detectable X-ray gas in nearby low-mass non-cluster early-type galaxies (e.g., comparison field sample from \citealt{hou2021_virgo}; NGC 147, NGC 185, and NGC 205 in the Local Group from \citealt{ge2015}), which show only stellar emission. Some nearby low-mass irregular and disk galaxies (e.g., \citealt{bogdan2011,ngc7793}), by contrast, do retain hot X-ray-emitting gas. In disk systems such as NGC 7793 (\citealt{ngc7793}), this gas is more spatially extended ($\sim$6 kpc, nearly 30$\%$ beyond the stellar disk), has lower characteristic electron densities ($\sim$0.002 cm$^{-3}$), and does not resemble a compact, pressure-confined mini–cool core, but appears influenced by disk geometry and ongoing or recent star formation.

\subsection{Constraints on X-ray Coronal Emission Across Galaxy Mass and Environments}

Although soft X-ray coronal emission is individually detected in a small number of galaxies, no significant stacked emission is detected from the remaining intermediate-mass and dwarf galaxies across the cluster sample. We therefore derive the first population-level upper limits on the typical coronal luminosity for both galaxy populations (Figure \ref{fig:LX-M*}a; Table \ref{tab:upperlimitsanddetections}). These limits provide constraints on the average hot gas emission of dwarf and intermediate-mass cluster galaxies, independent of individual detections. For comparison, \citet{hou2021_virgo} detected stacked X-ray emission in low-mass ($10^{8.7} \leq M_*/M_{\odot} < 10^{9.9}$; 45 galaxies) and intermediate-mass ($10^{9.9} \leq M_*/M_{\odot} \leq 10^{10.9}$; 25 galaxies) Virgo early-type galaxies, but found it consistent with unresolved stellar populations (LMXBs, CVs, and ABs), without requiring an additional diffuse hot-gas component. Similarly, \citet{hou2024_virgo} found no significant stacked coronal emission in nearly edge-on Virgo late-type galaxies in either their low-mass ($10^{9.2} < M_*/M_{\odot} < 10^{9.9}$; 9 galaxies) or high-mass ($10^{9.9} < M_*/M_{\odot} < 10^{10.4}$; 7 galaxies) subsamples. The results from these studies are consistent with our non-detections of coronal emission in stacked populations of lower-mass galaxies. We further subdivide the sample by projected cluster-centric radius and repeat the analysis for inner and outer regions for both galaxy populations. In both environmental regimes, we find no significant stacked detection of soft X-ray coronal emission for either dwarf or intermediate-mass galaxies, and instead obtain consistent population-level upper limits (Figure \ref{fig:LX-M*}b; Table \ref{tab:upperlimitsanddetections}).   
Within the sensitivity of the current data, the average stacked coronal X-ray emission from the individually undetected dwarf and intermediate-mass galaxies remains below the derived upper limits, with no evidence for a detectable signal in either population across the cluster environments considered. The absence of significant stacked coronal emission, together with the low fraction of individual detections, suggests that most lower-mass galaxies are susceptible to environmental processes in massive clusters, although certain favourable conditions may aid the survival of compact coronae in a small subset of these systems, as discussed in Section 4.2.
\vspace{4mm}\\
In summary, the retention of hot gas in cluster galaxies depends on a combination of factors. Stellar mass plays an important role, as evident from the low detection fraction and lack of significant stacked emission among low-mass galaxies. 
The concentration of individual detections among ancient infallers suggests that early formation and deep central gravitational potentials may help galaxies retain their gas. Comparisons with Virgo galaxies indicate that ICM pressure confinement may also aid the survival of compact coronae. The low-$\beta$ values of individually detected coronae suggest that internal galaxy processes such as stellar feedback may contribute to the heating of the hot gas. Other factors, including individual galaxy orbits, ICM microphysics and AGN feedback, may also be relevant but are not studied in this work.

\subsection{X-ray-Luminous AGN in Less Massive Galaxy Hosts}
We examine all dwarf and intermediate-mass galaxies in our cluster sample associated with X-ray point sources and identify six cluster members hosting X-ray AGN with the rest-frame 2–10 keV luminosities exceeding $10^{41}$ erg s$^{-1}$. Optical spectra are available for all six galaxies, and the resulting spectroscopic redshifts confirm their cluster membership. In all six galaxies, the X-ray source is spatially coincident with the optical nucleus, with an X-ray-optical positional separation of less than $0.5^{\prime\prime}$ (Table \ref{tab:agndetections}), supporting a nuclear origin for the X-ray emission. Optical and X-ray images of these galaxies, together with their X-ray spectra, are presented in Figure \ref{fig:agn}. Four of the six AGN show evidence of intrinsic absorption. The six detections among the 1558 dwarf and intermediate-mass cluster members correspond to an observed luminous X-ray AGN fraction of $\sim$$0.4\%$ for $L_{X_{(2-10\text{ keV})}} > 10^{41} \text{erg s}^{-1}$.

Three of the six galaxies -- WISEA J134746.87+262611.0, WISEA J134834.95+263109.8, and WISEA J135034.34+265656.0 -- are independently classified as AGN based on their SDSS optical spectra. WISEA J134746.87+262611.0 is notable, as it hosts a luminous X-ray AGN despite its particularly low stellar mass of $10^{9.2}\text{ M}_{\odot}$. It represents a rare example of a low-mass galaxy hosting a luminous X-ray AGN in a nearby cluster. WISEA J134834.95+263109.8 and WISEA J135034.34+265656.0 have the SDSS spectral classification \texttt{QSO STARBURST BROADLINE} and are also the most X-ray-luminous objects in the sample, with luminosities approaching $10^{43}$ erg s$^{-1}$. The SDSS spectral classifications identify CGCG 097-125 and WISEA J135043.48+265402.5 as star-forming galaxies.

Previous studies provide useful context for these detections. The AGN Multiwavelength Survey of Early-type Galaxies in Virgo (AMUSE-Virgo; \citealt{gallo2008,gallo2010}) included galaxies with stellar masses down to $10^{8.7}\text{ M}_{\odot}$, but did not identify comparably luminous X-ray AGN among its low-mass members. Similarly, \citet{seydasen2023} identified X-ray-emitting dwarf candidates with stellar masses down to $10^{8.4}\text{ M}_{\odot}$
across 12 nearby clusters, while the eROSITA census of Virgo and Fornax detected nuclear X-ray sources in galaxies with masses as low as $10^{7.7}\text{ M}_{\odot}$ \citep{hou2024}. However, neither study found comparably luminous X-ray AGN in the dwarf-galaxy population. The present sample therefore adds six spectroscopically confirmed cluster members with nuclear X-ray emission to the limited population of luminous X-ray AGN known in less massive cluster galaxies.

All six AGN hosts are classified as first or recent infallers based on their projected phase-space locations. Previous studies have likewise found AGN to be more common among infalling galaxies and at larger clustercentric radii (e.g., \citealt{hwang2012} for  $0.02 < z < 0.14$; \citealt{haines2012} for $0.15 < z < 0.30$). These galaxies may therefore still retain enough gas to fuel their central black holes during the early stages of infall. Further environmental processing, including RPS, may later reduce the available gas supply and suppress black hole accretion, particularly in low-mass galaxies.

Taken together, these sources represent rare examples of luminous X-ray AGN hosted by dwarf and intermediate-mass galaxies in nearby clusters. KUG 1259+280 (Table \ref{tab:individual detections}) provides an additional example of AGN-related soft nuclear X-ray emission in a less massive cluster galaxy. However, its rest-frame 2--10 keV luminosity falls below the $10^{41}\text{ erg s}^{-1}$ threshold adopted here, and it is therefore not included in the luminous-AGN sample.

\subsection{Considerations for Spectral Stacking and Model Assumptions}
Given the low counts of the individual galaxy spectra, the spectral stacking is performed in the observed frame without shifting each galaxy spectrum to its rest frame energy. The expected coronal emission is dominated by the broad Fe-L complex, which spans roughly 0.7--1.1 keV in the rest frame. The galaxies in the full stack cover a redshift range of $z = 0.0116-0.0634$, corresponding to the bulk ($>90\%$) of the cluster galaxy population. This translates to observed-frame energy shifts of $\sim$30--55 eV across the redshift range. Accounting for the full velocity dispersion of the clusters, the effective redshift range extends from 0.0036 to 0.077, yielding maximum energy shifts of $\sim$30--80 eV. These shifts are smaller than the average ACIS CCD spectral resolution, which is $\sim$100--120 eV (Full Width at Half Maximum) at 1 keV (see Chandra Proposer’s Observatory Guide), with an additional uncertainty of $\sim$30 eV below 1 keV (based on Chandra Calibration Status Report 2017), and are also well below the intrinsic width of the Fe-L feature ($\sim$400 eV). 

When deriving upper limits on the X-ray luminosity of the coronae, we adopt temperatures of 0.2 and 0.6 keV and abundances of 0.3 and 0.4 $Z_{\odot}$ for dwarf and intermediate-mass galaxies, respectively. Variations of $\pm0.2$ keV in $kT$ at fixed abundance change the derived upper limits by less than $17\%$, while varying the abundance between 0.2 and 0.6 $Z_{\odot}$ at fixed $kT$ changes the limits by less than $2\%$ over the full temperature range considered ($kT=0.2-0.8$ keV).

\begin{table*}
\centering
\small
\setlength{\tabcolsep}{4pt}
\caption{Properties of dwarf and intermediate-mass cluster members hosting X-ray AGN with rest-frame 2--10 keV luminosities $>10^{41}$ erg s$^{-1}$. Here, $\Gamma$ is the photon index of the best-fitting powerlaw model, and $N_{H_{intrinsic}}$ is the intrinsic equivalent hydrogen column density.}
\label{tab:agndetections}
\begin{tabular}{lcccccccc}
\hline
Galaxy (Host cluster)&log (M$_*$/M$_{\odot}$)&$g-r$&$R/R_{500}$&$v$&X-ray--optical&$\Gamma$&$L_{X (2-10\text{ keV})}$&$N_{H_{intrinsic}}$\\[-0.05cm]
&&&&&separation&&&\\
&&&&(km s$^{-1})$&(arcsec, pc)&&(10$^{42}$ erg s$^{-1}$)&($10^{23}$ cm$^{-2}$)\\
\hline
WISEA J134746.87+262611.0 (A1795) & 9.20&0.41&0.99&-1471&0.06, 73&$1.6^{+0.6}_{-0.6}$&$2.2^{+0.4}_{-0.3}$&1.6\\
WISEA J125710.75+272417.6 (Coma)&10.22&0.69&1.05&-821&0.12, 56&$1.6^{+0.4}_{-0.4}$&$0.39\pm{0.03}$&0.17\\
WISEA J134834.95+263109.8 (A1795) &10.29&0.52&0.34&-1233&0.12, 146&$1.9^{+0.1}_{-0.1}$&$9.7\pm{0.4}$&--\\
CGCG 097-125 (A1367)&10.37&0.66&0.11&+1469&0.0, 0.0&$1.7^{+0.1}_{-0.1}$&$0.34\pm{0.03}$&2.6\\
WISEA J135034.34+265656.0 (A1795) &10.38&0.77&1.77&+1024&0.06, 73&$1.3^{+0.2}_{-0.2}$&$11\pm{1}$&0.28\\
WISEA J135043.48+265402.5 (A1795)&10.42&0.56&1.75&+1590&0.30, 364&$1.6^{+0.3}_{-0.3}$&$0.24\pm{0.04}$&--\\
\hline
\end{tabular}
\end{table*}

\section{Conclusions and Future Prospects}
In this work, we present the first systematic X-ray study of 735 dwarf ($\text{M}_*\leq10^{9.5} \text{ M}_{\odot}$) and 823 intermediate-mass ($10^{9.5}< \text{M}_*\leq10^{10.5} \text{ M}_{\odot}$) galaxies in 18 nearby massive galaxy clusters ($0.011 < z < 0.063$) with average ICM temperatures in the range $3-10$ keV, using $\sim$22 Msec of $Chandra$ archival observations spanning over two decades. We investigate the presence of diffuse soft X-ray emitting gas in these low-mass member galaxies through both individual and stacked spectral analyses across stellar mass and environment. Additionally, we identify luminous X-ray AGN with rest-frame 2--10 keV luminosities exceeding $10^{41}$ erg s$^{-1}$ among the dwarf and intermediate-mass galaxies. Our main findings are as follows:
\begin{itemize}
    \item We identify 22 galaxies with individual detections of hot X-ray coronae, confirmed through the presence of the characteristic Fe–L feature in their spectra. These detections correspond to $\sim$2.4$\%$ of intermediate-mass galaxies and $\sim$0.3$\%$ of dwarf galaxies, highlighting the rarity of such systems at low stellar masses.
    \item The individually detected coronae are compact and centrally concentrated, with temperatures of $kT \sim 0.3-1.0$ keV and soft X-ray luminosities of a few times $10^{39}$ erg s$^{-1}$, consistent with small, surviving cool cores embedded within the cluster environment.
    \item Among the individual detections, nine galaxies have available stellar velocity dispersions, yielding $\beta$ values in the range 0.02--0.5, well below unity. These low values suggest that additional heating processes, likely associated with stellar feedback and possibly nuclear activity, contribute significantly to the thermal energy of the hot gas in these low-mass systems.
    \item Most individual detections are associated with ancient infallers, yet the low retention fraction among this population indicates that hot gas survival is a multi-factor phenomenon. The presence of both quiescent and star-forming/starburst galaxies among the detections suggests that a combination of internal and environmental processes (e.g., pressure confinement by the surrounding ICM) governs the survival of hot gas.
    \item Across the full sample, we find no significant population-wide detection of soft X-ray emitting gas in stacked spectra of either dwarf or intermediate-mass galaxies in clusters. We place stringent constraints on hot gas X-ray emission from low-mass cluster members, constraining the soft (0.5–2.0 keV) X-ray luminosity to $< 3.17 \times 10^{38}$ erg s$^{-1}$ and $< 1.93 \times 10^{38}$ erg s$^{-1}$, for dwarfs and intermediate-mass populations, respectively. This implies that any hot-gas component in low-mass cluster galaxies is generally absent or below current detection thresholds.
    \item We find no clear dependence of embedded thermal coronae on host cluster environment in the stacked analysis.
    \item In addition to the thermal coronae analysis, we identify a rare population of luminous nuclear X-ray AGN in one dwarf and five intermediate-mass cluster members, with $L_{X_{(2-10 \text{ keV)}}}>10^{41}\text{ erg s}^{-1}$, corresponding to an observed AGN fraction of $\sim$$0.4\%$. Two intermediate-mass galaxies host very luminous AGN with $L_X \sim 10^{43}$ erg s$^{-1}$, while the dwarf host is particularly notable for its low stellar mass of $10^{9.2}$ M$_{\odot}$ and its high X-ray luminosity of $2.2^{+0.4}_{-0.3}\times10^{42}$ erg s$^{-1}$. All six hosts are classified as first or recent infallers. 
     \item Overall, our results reinforce the picture that low-mass galaxies in clusters are highly susceptible to environmental effects. Galaxy hot gas retention is largely determined by its stellar mass, with additional factors -- such as early central potential formation, ICM pressure confinement, and galaxy orbital history -- influencing deviations from general trends.
    \end{itemize}
The general lack of detected hot X-ray coronae in low-mass cluster galaxies suggests that their hot CGM is also likely stripped or strongly suppressed in a outside-in manner following infall into cluster environments. This leaves these galaxies with little to no hot gas to sustain or replenish their gas reservoirs, which can significantly affect their future evolution in dense environments, likely reducing the fuel available for future star formation and AGN activity, and influencing their long-term evolution.

Future deeper observations will be essential to probe below current detection limits and to determine whether hot coronae that remain undetected in the present data exist in low mass cluster galaxies. With deeper data, similar analyses may also be extended to study the CGM, which is expected to be significantly fainter and therefore more challenging to detect, in galaxy populations across nearby
clusters. Such studies may provide further insight into the environmental dependence of galaxy coronae in low-mass galaxies, particularly interesting in light of the environmental trends reported for more massive galaxies (e.g., in A1795; \citealt{wagner2018_a1795}). Additionally, recent simulation studies also suggest that satellite galaxies may contribute $\sim$$10\%$ of the soft X-ray emission to the ICM in outer regions (e.g., \citealt{rohr2024}), highlighting the potential cumulative impact of such surviving hot-gas halos in galaxy clusters.

\section*{Acknowledgements}
We thank Annalisa Pillepich and Yi Zhang for discussions. We thank Chong Ge for providing the code used to model the \textit{lnvmkl} component. We thank the referee for their helpful suggestions.
Support for this work was provided by the National Aeronautics and Space Administration through {\em Chandra} Award Number AR2-23013X, GO2-23114X and GO3-24115X issued by the {\em Chandra} X-ray Center, which is operated by the Smithsonian Astrophysical Observatory for and on behalf of the National Aeronautics and Space Administration under contract NAS8-03060. 
JT and MS also acknowledge support from the NSF grant 2407821.
This article has made use of archival data from X-ray observations with the Advanced CCD Imaging Spectrometer (ACIS) onboard the \textit{Chandra} Observatory. We have used photometric data from data release 18 of the Sloan Digital Sky Survey-V, data release 2 of the Pan-STARRS1 Survey, data release 4 of the SkyMapper Southern Survey \citep{skymapperdr4}, and data release 10 of DESI Legacy Imaging Surveys. Funding for the Sloan Digital Sky Survey-V has been provided by the Alfred P. Sloan Foundation, the Heising-Simons Foundation, the National Science Foundation, and the Participating Institutions. SDSS acknowledges support and resources from the centre for High-Performance Computing at the University of Utah. SDSS telescopes are located at Apache Point Observatory, funded by the Astrophysical Research Consortium and operated by New Mexico State University, and at Las Campanas Observatory, operated by the Carnegie Institution for Science. The SDSS web site is \url{www.sdss.org}. The Pan-STARRS1 Surveys (PS1) and the PS1 public science archive have been made possible through contributions by the Institute for Astronomy, the University of Hawaii, the Pan-STARRS Project Office, the Max-Planck Society and its participating institutes, the Max Planck Institute for Astronomy, Heidelberg and the Max Planck Institute for Extraterrestrial Physics, Garching, The Johns Hopkins University, Durham University, the University of Edinburgh, the Queen's University Belfast, the Harvard-Smithsonian Center for Astrophysics, the Las Cumbres Observatory Global Telescope Network Incorporated, the National Central University of Taiwan, the Space Telescope Science Institute, the National Aeronautics and Space Administration under Grant No. NNX08AR22G issued through the Planetary Science Division of the NASA Science Mission Directorate, the National Science Foundation Grant No. AST–1238877, the University of Maryland, Eotvos Lorand University (ELTE), the Los Alamos National Laboratory, and the Gordon and Betty Moore Foundation. The national facility capability for SkyMapper has been funded through ARC LIEF grant LE130100104 from the Australian Research Council, awarded to the University of Sydney, the Australian National University, Swinburne University of Technology, the University of Queensland, the University of Western Australia, the University of Melbourne, Curtin University of Technology, Monash University and the Australian Astronomical Observatory. SkyMapper is owned and operated by The Australian National University's Research School of Astronomy and Astrophysics. The survey data were processed and provided by the SkyMapper Team at ANU. The SkyMapper node of the All-Sky Virtual Observatory (ASVO) is hosted at the National Computational Infrastructure (NCI). Development and support of the SkyMapper node of the ASVO has been funded in part by Astronomy Australia Limited (AAL) and the Australian Government through the Commonwealth's Education Investment Fund (EIF) and National Collaborative Research Infrastructure Strategy (NCRIS), particularly the National eResearch Collaboration Tools and Resources (NeCTAR) and the Australian National Data Service Projects (ANDS). The DESI Legacy Imaging Surveys consist of three individual and complementary projects: the Dark Energy Camera Legacy Survey (DECaLS), the Beijing-Arizona Sky Survey (BASS), and the Mayall z-band Legacy Survey (MzLS). DECaLS, BASS and MzLS together include data obtained, respectively, at the Blanco telescope, Cerro Tololo Inter-American Observatory, NSF’s NOIRLab; the Bok telescope, Steward Observatory, University of Arizona; and the Mayall telescope, Kitt Peak National Observatory, NOIRLab. NOIRLab is operated by the Association of Universities for Research in Astronomy (AURA) under a cooperative agreement with the National Science Foundation. Pipeline processing and analyses of the data were supported by NOIRLab and the Lawrence Berkeley National Laboratory (LBNL). Legacy Surveys also uses data products from the Near-Earth Object Wide-field Infrared Survey Explorer (NEOWISE), a project of the Jet Propulsion Laboratory/California Institute of Technology, funded by the National Aeronautics and Space Administration. Legacy Surveys was supported by: the Director, Office of Science, Office of High Energy Physics of the U.S. Department of Energy; the National Energy Research Scientific Computing Center, a DOE Office of Science User Facility; the U.S. National Science Foundation, Division of Astronomical Sciences; the National Astronomical Observatories of China, the Chinese Academy of Sciences and the Chinese National Natural Science Foundation. LBNL is managed by the Regents of the University of California under contract to the U.S. Department of Energy. This research has made use of SAOImageDS9, developed by Smithsonian Astrophysical Observatory, the HEASoft FTOOLS (\url{http://heasarc.gsfc.nasa.gov/ftools}), and the NASA/IPAC Extragalactic Database (NED), which is funded by the National Aeronautics and Space Administration and operated by the California Institute of Technology. The use of TOPCAT (\url{https://www.star.bristol.ac.uk/mbt/topcat/}) has been very useful for this study. OpenAI ChatGPT (GPT-5.6 Sol) was used for language polishing of the manuscript.

\section*{Data Availability}
All X-ray data used in this study are publicly accessible via the Chandra Data Archive: \url{ https://cda.harvard.edu/chaser/}. The Chandra ObsIDs used in this work are listed in Table \ref{tab:obsidlist}. The photometric data from Sloan Digital Sky Survey-V, Pan-STARRS1 Survey, data release 4 of the SkyMapper Southern Survey, and data release 10 of DESI Legacy Imaging Surveys are also publicly available.

\bibliography{xraycoronae}

\begin{thebibliography}{}
\expandafter\ifx\csname natexlab\endcsname\relax\def\natexlab#1{#1}\fi

\bibitem[{{Aguerri} {et~al.}(2020){Aguerri}, {Girardi}, {Agulli}, {Negri},
  {Dalla Vecchia}, \& {Dom{\'\i}nguez Palmero}}]{aguerri2020}
{Aguerri}, J.~A.~L., {Girardi}, M., {Agulli}, I., {et~al.} 2020, \mnras, 494,
  1681

\bibitem[{{Arnaud}(1996)}]{xspec}
{Arnaud}, K.~A. 1996, in Astronomical Society of the Pacific Conference Series,
  Vol. 101, Astronomical Data Analysis Software and Systems V, ed. G.~H.
  {Jacoby} \& J.~{Barnes}, 17

\bibitem[{{Arnold} {et~al.}(2009){Arnold}, {Martini}, {Mulchaey}, {Berti}, \&
  {Jeltema}}]{arnold2009}
{Arnold}, T.~J., {Martini}, P., {Mulchaey}, J.~S., {Berti}, A., \& {Jeltema},
  T.~E. 2009, \apj, 707, 1691

\bibitem[{{Arthur} {et~al.}(2019){Arthur}, {Pearce}, {Gray}, {Knebe}, {Cui},
  {Elahi}, {Power}, {Yepes}, {Arth}, {De Petris}, {Dolag}, {Garratt-Smithson},
  {Old}, {Rasia}, \& {Stevens}}]{arthur2019}
{Arthur}, J., {Pearce}, F.~R., {Gray}, M.~E., {et~al.} 2019, \mnras, 484, 3968

\bibitem[{{Asplund} {et~al.}(2009){Asplund}, {Grevesse}, {Sauval}, \&
  {Scott}}]{aspl}
{Asplund}, M., {Grevesse}, N., {Sauval}, A.~J., \& {Scott}, P. 2009, \araa, 47,
  481

\bibitem[{{Bah{\'e}} {et~al.}(2013){Bah{\'e}}, {McCarthy}, {Balogh}, \&
  {Font}}]{bahe2013}
{Bah{\'e}}, Y.~M., {McCarthy}, I.~G., {Balogh}, M.~L., \& {Font}, A.~S. 2013,
  \mnras, 430, 3017

\bibitem[{{Bogd{\'a}n} \& {Gilfanov}(2011)}]{bogdan2011}
{Bogd{\'a}n}, {\'A}., \& {Gilfanov}, M. 2011, \mnras, 418, 1901

\bibitem[{{Boselli} {et~al.}(2022){Boselli}, {Fossati}, \& {Sun}}]{boselli2022}
{Boselli}, A., {Fossati}, M., \& {Sun}, M. 2022, \aapr, 30, 3

\bibitem[{{Chen} {et~al.}(2020){Chen}, {Sun}, {Yagi}, {Bravo-Alfaro}, {Brinks},
  {Kenney}, {Combes}, {Sivanandam}, {Jachym}, {Fossati}, {Gavazzi}, {Boselli},
  {Nulsen}, {Sarazin}, {Ge}, {Yoshida}, \& {Roediger}}]{chen2020}
{Chen}, H., {Sun}, M., {Yagi}, M., {et~al.} 2020, \mnras, 496, 4654

\bibitem[{{Clemens} {et~al.}(2006){Clemens}, {Bressan}, {Nikolic}, {Alexander},
  {Annibali}, \& {Rampazzo}}]{clemens2006}
{Clemens}, M.~S., {Bressan}, A., {Nikolic}, B., {et~al.} 2006, \mnras, 370, 702

\bibitem[{{Cole} {et~al.}(2000){Cole}, {Lacey}, {Baugh}, \& {Frenk}}]{cole2000}
{Cole}, S., {Lacey}, C.~G., {Baugh}, C.~M., \& {Frenk}, C.~S. 2000, \mnras,
  319, 168

\bibitem[{{Connor} {et~al.}(2018){Connor}, {Kelson}, {Mulchaey}, {Vikhlinin},
  {Patel}, {Balogh}, {Joshi}, {Kraft}, {Nagai}, \& {Starikova}}]{connor2018}
{Connor}, T., {Kelson}, D.~D., {Mulchaey}, J., {et~al.} 2018, \apj, 867, 25

\bibitem[{{Cramer} {et~al.}(2019){Cramer}, {Kenney}, {Sun}, {Crowl}, {Yagi},
  {J{\'a}chym}, {Roediger}, \& {Waldron}}]{cramer2019}
{Cramer}, W.~J., {Kenney}, J.~D.~P., {Sun}, M., {et~al.} 2019, \apj, 870, 63

\bibitem[{{Davis} \& {White}(1996)}]{davis1996}
{Davis}, D.~S., \& {White}, III, R.~E. 1996, \apjl, 470, L35

\bibitem[{{Dewangan} {et~al.}(2000){Dewangan}, {Singh}, {Mayya}, \&
  {Anupama}}]{dewangan2000}
{Dewangan}, G.~C., {Singh}, K.~P., {Mayya}, Y.~D., \& {Anupama}, G.~C. 2000,
  \mnras, 318, 309

\bibitem[{{Durret} {et~al.}(2015){Durret}, {Wakamatsu}, {Nagayama}, {Adami}, \&
  {Biviano}}]{durret2015}
{Durret}, F., {Wakamatsu}, K., {Nagayama}, T., {Adami}, C., \& {Biviano}, A.
  2015, \aap, 583, A124

\bibitem[{{Font} {et~al.}(2008){Font}, {Bower}, {McCarthy}, {Benson}, {Frenk},
  {Helly}, {Lacey}, {Baugh}, \& {Cole}}]{font2008}
{Font}, A.~S., {Bower}, R.~G., {McCarthy}, I.~G., {et~al.} 2008, \mnras, 389,
  1619

\bibitem[{{Forman} {et~al.}(1985){Forman}, {Jones}, \& {Tucker}}]{forman1985}
{Forman}, W., {Jones}, C., \& {Tucker}, W. 1985, \apj, 293, 102

\bibitem[{{Gallazzi} {et~al.}(2021){Gallazzi}, {Pasquali}, {Zibetti}, \&
  {Barbera}}]{gallazi2021}
{Gallazzi}, A.~R., {Pasquali}, A., {Zibetti}, S., \& {Barbera}, F.~L. 2021,
  \mnras, 502, 4457

\bibitem[{{Gallo} {et~al.}(2008){Gallo}, {Treu}, {Jacob}, {Woo}, {Marshall}, \&
  {Antonucci}}]{gallo2008}
{Gallo}, E., {Treu}, T., {Jacob}, J., {et~al.} 2008, \apj, 680, 154

\bibitem[{{Gallo} {et~al.}(2010){Gallo}, {Treu}, {Marshall}, {Woo}, {Leipski},
  \& {Antonucci}}]{gallo2010}
{Gallo}, E., {Treu}, T., {Marshall}, P.~J., {et~al.} 2010, \apj, 714, 25

\bibitem[{{Ge} {et~al.}(2015){Ge}, {Li}, {Xu}, {Gu}, {Wang}, {Roberts},
  {Kraft}, {Jones}, \& {Forman}}]{ge2015}
{Ge}, C., {Li}, Z., {Xu}, X., {et~al.} 2015, \apj, 812, 130

\bibitem[{{Ge} {et~al.}(2019){Ge}, {Sun}, {Liu}, {Rudnick}, {Sarazin},
  {Forman}, {Jones}, {Chen}, {Liu}, {Yagi}, {Boselli}, {Fossati}, \&
  {Gavazzi}}]{ge2019}
{Ge}, C., {Sun}, M., {Liu}, R.-Y., {et~al.} 2019, \mnras, 486, L36

\bibitem[{{Ghirardini} {et~al.}(2019){Ghirardini}, {Eckert}, {Ettori},
  {Pointecouteau}, {Molendi}, {Gaspari}, {Rossetti}, {De Grandi}, {Roncarelli},
  {Bourdin}, {Mazzotta}, {Rasia}, \& {Vazza}}]{ghirardini2019}
{Ghirardini}, V., {Eckert}, D., {Ettori}, S., {et~al.} 2019, \aap, 621, A41

\bibitem[{{Giacintucci} {et~al.}(2019){Giacintucci}, {Markevitch}, {Cassano},
  {Venturi}, {Clarke}, {Kale}, \& {Cuciti}}]{giacintucci2019}
{Giacintucci}, S., {Markevitch}, M., {Cassano}, R., {et~al.} 2019, \apj, 880,
  70

\bibitem[{{Girardi} {et~al.}(2022){Girardi}, {Boschin}, {Nonino}, {Innocentin},
  \& {De Grandi}}]{girardi2022}
{Girardi}, M., {Boschin}, W., {Nonino}, M., {Innocentin}, C., \& {De Grandi},
  S. 2022, \aap, 658, A159

\bibitem[{{Giustini} {et~al.}(2020){Giustini}, {Miniutti}, \&
  {Saxton}}]{kug_qpe}
{Giustini}, M., {Miniutti}, G., \& {Saxton}, R.~D. 2020, \aap, 636, L2

\bibitem[{{Haines} {et~al.}(2012){Haines}, {Pereira}, {Sanderson}, {Smith},
  {Egami}, {Babul}, {Edge}, {Finoguenov}, {Moran}, \& {Okabe}}]{haines2012}
{Haines}, C.~P., {Pereira}, M.~J., {Sanderson}, A.~J.~R., {et~al.} 2012, \apj,
  754, 97

\bibitem[{{He} {et~al.}(2025){He}, {Li}, {Hou}, {Du}, {Fang}, \&
  {Cui}}]{ngc7793}
{He}, L., {Li}, Z., {Hou}, M., {et~al.} 2025, \apj, 983, 149

\bibitem[{{Hou} {et~al.}(2024{\natexlab{a}}){Hou}, {Hu}, \& {Li}}]{hou2024}
{Hou}, M., {Hu}, Z., \& {Li}, Z. 2024{\natexlab{a}}, \apjl, 965, L24

\bibitem[{{Hou} {et~al.}(2021){Hou}, {Li}, {Jones}, {Forman}, \&
  {Su}}]{hou2021_virgo}
{Hou}, M., {Li}, Z., {Jones}, C., {Forman}, W., \& {Su}, Y. 2021, \apj, 919,
  141

\bibitem[{{Hou} {et~al.}(2024{\natexlab{b}}){Hou}, {He}, {Hu}, {Li}, {Jones},
  {Forman}, {Su}, {Wang}, \& {Ho}}]{hou2024_virgo}
{Hou}, M., {He}, L., {Hu}, Z., {et~al.} 2024{\natexlab{b}}, \apj, 961, 249

\bibitem[{{Hu} {et~al.}(2023){Hu}, {Su}, {Li}, {Hess}, {Kraft}, {Forman},
  {Nulsen}, {Sridhar}, {Stroe}, {Baek}, {Chung}, {Grupe}, {Chen}, {Irwin},
  {Jones}, {Randall}, \& {Roediger}}]{antlia2023}
{Hu}, Z., {Su}, Y., {Li}, Z., {et~al.} 2023, \apj, 956, 104

\bibitem[{{Hwang} {et~al.}(2012){Hwang}, {Park}, {Elbaz}, \&
  {Choi}}]{hwang2012}
{Hwang}, H.~S., {Park}, C., {Elbaz}, D., \& {Choi}, Y.-Y. 2012, \aap, 538, A15

\bibitem[{{J{\'a}chym} {et~al.}(2017){J{\'a}chym}, {Sun}, {Kenney}, {Cortese},
  {Combes}, {Yagi}, {Yoshida}, {Palou{\v{s}}}, \& {Roediger}}]{jachym2017}
{J{\'a}chym}, P., {Sun}, M., {Kenney}, J. D.~P., {et~al.} 2017, \apj, 839, 114

\bibitem[{{Jeltema} {et~al.}(2008){Jeltema}, {Binder}, \&
  {Mulchaey}}]{jeltema2008}
{Jeltema}, T.~E., {Binder}, B., \& {Mulchaey}, J.~S. 2008, \apj, 679, 1162

\bibitem[{{Jung} {et~al.}(2018){Jung}, {Choi}, {Wong}, {Kimm}, {Chung}, \&
  {Yi}}]{jung2018}
{Jung}, S.~L., {Choi}, H., {Wong}, O.~I., {et~al.} 2018, \apj, 865, 156

\bibitem[{{Kauffmann} {et~al.}(1993){Kauffmann}, {White}, \&
  {Guiderdoni}}]{kauffmann1993}
{Kauffmann}, G., {White}, S.~D.~M., \& {Guiderdoni}, B. 1993, \mnras, 264, 201

\bibitem[{{Koranyi} \& {Geller}(2000)}]{awm7_2000}
{Koranyi}, D.~M., \& {Geller}, M.~J. 2000, \aj, 119, 44

\bibitem[{{Lauer} {et~al.}(2014){Lauer}, {Postman}, {Strauss}, {Graves}, \&
  {Chisari}}]{lauer2014}
{Lauer}, T.~R., {Postman}, M., {Strauss}, M.~A., {Graves}, G.~J., \& {Chisari},
  N.~E. 2014, \apj, 797, 82

\bibitem[{{Ledlow} {et~al.}(2005){Ledlow}, {Owen}, \& {Miller}}]{ledlow2005}
{Ledlow}, M.~J., {Owen}, F.~N., \& {Miller}, N.~A. 2005, \aj, 130, 47

\bibitem[{{Lotz} {et~al.}(2019){Lotz}, {Remus}, {Dolag}, {Biviano}, \&
  {Burkert}}]{lotz2019}
{Lotz}, M., {Remus}, R.-S., {Dolag}, K., {Biviano}, A., \& {Burkert}, A. 2019,
  \mnras, 488, 5370

\bibitem[{{Lu} \& {Wang}(2011)}]{lu2011}
{Lu}, Z., \& {Wang}, Q.~D. 2011, \mnras, 413, 347

\bibitem[{{Martini} {et~al.}(2007){Martini}, {Mulchaey}, \&
  {Kelson}}]{martini2007}
{Martini}, P., {Mulchaey}, J.~S., \& {Kelson}, D.~D. 2007, \apj, 664, 761

\bibitem[{{Mathews} \& {Brighenti}(2003)}]{mathhews2003}
{Mathews}, W.~G., \& {Brighenti}, F. 2003, \araa, 41, 191

\bibitem[{{McCarthy} {et~al.}(2008){McCarthy}, {Frenk}, {Font}, {Lacey},
  {Bower}, {Mitchell}, {Balogh}, \& {Theuns}}]{mccarthy2008}
{McCarthy}, I.~G., {Frenk}, C.~S., {Font}, A.~S., {et~al.} 2008, \mnras, 383,
  593

\bibitem[{{Mostoghiu} {et~al.}(2021){Mostoghiu}, {Arthur}, {Pearce}, {Gray},
  {Knebe}, {Cui}, {Welker}, {Cora}, {Murante}, {Dolag}, \&
  {Yepes}}]{mostoghiu2021}
{Mostoghiu}, R., {Arthur}, J., {Pearce}, F.~R., {et~al.} 2021, \mnras, 501,
  5029

\bibitem[{{Mun} {et~al.}(2021){Mun}, {Hwang}, {Lee}, {Chung}, {Yoon}, \&
  {Lee}}]{mun2021}
{Mun}, J.~Y., {Hwang}, H.~S., {Lee}, M.~G., {et~al.} 2021, Journal of Korean
  Astronomical Society, 54, 17

\bibitem[{{Oegerle} \& {Hill}(2001)}]{oegerle2001}
{Oegerle}, W.~R., \& {Hill}, J.~M. 2001, \aj, 122, 2858

\bibitem[{{Oman} {et~al.}(2013){Oman}, {Hudson}, \& {Behroozi}}]{oman2013}
{Oman}, K.~A., {Hudson}, M.~J., \& {Behroozi}, P.~S. 2013, \mnras, 431, 2307

\bibitem[{{Onken} {et~al.}(2024){Onken}, {Wolf}, {Bessell}, {Chang}, {Luvaul},
  {Tonry}, {White}, \& {Da Costa}}]{skymapperdr4}
{Onken}, C.~A., {Wolf}, C., {Bessell}, M.~S., {et~al.} 2024, \pasa, 41, e061

\bibitem[{{Piffaretti} {et~al.}(2011){Piffaretti}, {Arnaud}, {Pratt},
  {Pointecouteau}, \& {Melin}}]{piffaretti2011}
{Piffaretti}, R., {Arnaud}, M., {Pratt}, G.~W., {Pointecouteau}, E., \&
  {Melin}, J.-B. 2011, \aap, 534, A109

\bibitem[{{Rhee} {et~al.}(2017){Rhee}, {Smith}, {Choi}, {Yi}, {Jaff{\'e}},
  {Candlish}, \& {S{\'a}nchez-J{\'a}nssen}}]{rhee2017}
{Rhee}, J., {Smith}, R., {Choi}, H., {et~al.} 2017, \apj, 843, 128

\bibitem[{{Rohr} {et~al.}(2024){Rohr}, {Pillepich}, {Nelson}, {Ayromlou}, \&
  {Zinger}}]{rohr2024}
{Rohr}, E., {Pillepich}, A., {Nelson}, D., {Ayromlou}, M., \& {Zinger}, E.
  2024, \aap, 686, A86

\bibitem[{{Rohr} {et~al.}(2023){Rohr}, {Pillepich}, {Nelson}, {Zinger},
  {Joshi}, \& {Ayromlou}}]{rohr2023}
{Rohr}, E., {Pillepich}, A., {Nelson}, D., {et~al.} 2023, \mnras, 524, 3502

\bibitem[{{\c{S}}en {et~al.}(2023){\c{S}}en, G{\"o}{\u{g}}{\"u}{\c{s}},
  Peletier, Choque-Challapa, \& Amiri}]{seydasen2023}
{\c{S}}en, {\c{S}}., G{\"o}{\u{g}}{\"u}{\c{s}}, E., Peletier, R.~F.,
  Choque-Challapa, N., \& Amiri, A. 2023, \mnras, 525, 1182

\bibitem[{{Shibuya} {et~al.}(2015){Shibuya}, {Ouchi}, \&
  {Harikane}}]{shibuya2015}
{Shibuya}, T., {Ouchi}, M., \& {Harikane}, Y. 2015, \apjs, 219, 15

\bibitem[{{Sivakoff} {et~al.}(2008){Sivakoff}, {Martini}, {Zabludoff},
  {Kelson}, \& {Mulchaey}}]{sivakoff2008}
{Sivakoff}, G.~R., {Martini}, P., {Zabludoff}, A.~I., {Kelson}, D.~D., \&
  {Mulchaey}, J.~S. 2008, \apj, 682, 803

\bibitem[{{Sohn} {et~al.}(2020){Sohn}, {Geller}, {Diaferio}, \&
  {Rines}}]{sohn2020}
{Sohn}, J., {Geller}, M.~J., {Diaferio}, A., \& {Rines}, K.~J. 2020, \apj, 891,
  129

\bibitem[{{Souchereau} {et~al.}(2025){Souchereau}, {Kenney}, {J{\'a}chym},
  {Sun}, {Cramer}, {Yagi}, {Boselli}, {Brinks}, {Combes}, {Cortese}, {Deshev},
  {Fossati}, {Grossov{\'a}}, {Luo}, {Palou{\v{s}}}, \&
  {Scott}}]{souchereau2025}
{Souchereau}, H.~J., {Kenney}, J. D.~P., {J{\'a}chym}, P., {et~al.} 2025, \apj,
  988, 72

\bibitem[{{Steinhauser} {et~al.}(2016){Steinhauser}, {Schindler}, \&
  {Springel}}]{steinhauser2016}
{Steinhauser}, D., {Schindler}, S., \& {Springel}, V. 2016, \aap, 591, A51

\bibitem[{{Struble} \& {Rood}(1999)}]{struble_rood1999}
{Struble}, M.~F., \& {Rood}, H.~J. 1999, \apjs, 125, 35

\bibitem[{{Sun}(2009)}]{sun2009}
{Sun}, M. 2009, \apj, 704, 1586

\bibitem[{{Sun} {et~al.}(2007){Sun}, {Jones}, {Forman}, {Vikhlinin}, {Donahue},
  \& {Voit}}]{sun2007}
{Sun}, M., {Jones}, C., {Forman}, W., {et~al.} 2007, \apj, 657, 197

\bibitem[{{Sun} {et~al.}(2005){Sun}, {Vikhlinin}, {Forman}, {Jones}, \&
  {Murray}}]{sun2005}
{Sun}, M., {Vikhlinin}, A., {Forman}, W., {Jones}, C., \& {Murray}, S.~S. 2005,
  \apj, 619, 169

\bibitem[{{Sun} {et~al.}(2021){Sun}, {Ge}, {Luo}, {Yagi}, {J{\'a}chym},
  {Boselli}, {Fossati}, {Nulsen}, {Yoshida}, \& {Gavazzi}}]{sun2021}
{Sun}, M., {Ge}, C., {Luo}, R., {et~al.} 2021, Nature Astronomy, 6, 270

\bibitem[{{Tiwari} {et~al.}(2020){Tiwari}, {Mahajan}, \& {Singh}}]{tiwari2020}
{Tiwari}, J., {Mahajan}, S., \& {Singh}, K.~P. 2020, \na, 81, 101417

\bibitem[{{Tremonti} {et~al.}(2004){Tremonti}, {Heckman}, {Kauffmann},
  {Brinchmann}, {Charlot}, {White}, {Seibert}, {Peng}, {Schlegel}, {Uomoto},
  {Fukugita}, \& {Brinkmann}}]{tremonti2004}
{Tremonti}, C.~A., {Heckman}, T.~M., {Kauffmann}, G., {et~al.} 2004, \apj, 613,
  898

\bibitem[{{van der Wel} {et~al.}(2014){van der Wel}, {Franx}, {van Dokkum},
  {Skelton}, {Momcheva}, {Whitaker}, {Brammer}, {Bell}, {Rix}, {Wuyts},
  {Ferguson}, {Holden}, {Barro}, {Koekemoer}, {Chang}, {McGrath},
  {H{\"a}ussler}, {Dekel}, {Behroozi}, {Fumagalli}, {Leja}, {Lundgren},
  {Maseda}, {Nelson}, {Wake}, {Patel}, {Labb{\'e}}, {Faber}, {Grogin}, \&
  {Kocevski}}]{vanderwel2014}
{van der Wel}, A., {Franx}, M., {van Dokkum}, P.~G., {et~al.} 2014, \apj, 788,
  28

\bibitem[{{van Dokkum} {et~al.}(2014){van Dokkum}, {Bezanson}, {van der Wel},
  {Nelson}, {Momcheva}, {Skelton}, {Whitaker}, {Brammer}, {Conroy},
  {F{\"o}rster Schreiber}, {Fumagalli}, {Kriek}, {Labb{\'e}}, {Leja},
  {Marchesini}, {Muzzin}, {Oesch}, \& {Wuyts}}]{vandokkum2014}
{van Dokkum}, P.~G., {Bezanson}, R., {van der Wel}, A., {et~al.} 2014, \apj,
  791, 45

\bibitem[{{Veronica} {et~al.}(2025){Veronica}, {Reiprich}, {Pacaud}, {Sanders},
  {Gattuzz}, {Yeung}, {Bulbul}, {Ghirardini}, {Liu}, {Mannes}, {Morelli}, \&
  {Ota}}]{angie2025}
{Veronica}, A., {Reiprich}, T.~H., {Pacaud}, F., {et~al.} 2025, \aap, 694, A168

\bibitem[{{Vijayaraghavan} \& {Ricker}(2015)}]{rv2015}
{Vijayaraghavan}, R., \& {Ricker}, P.~M. 2015, \mnras, 449, 2312

\bibitem[{{Vijayaraghavan} \& {Sarazin}(2017{\natexlab{a}})}]{rv2017b}
{Vijayaraghavan}, R., \& {Sarazin}, C. 2017{\natexlab{a}}, \apj, 848, 63

\bibitem[{{Vijayaraghavan} \& {Sarazin}(2017{\natexlab{b}})}]{rv2017a}
---. 2017{\natexlab{b}}, \apj, 841, 22

\bibitem[{{Vikhlinin} {et~al.}(2006){Vikhlinin}, {Kravtsov}, {Forman}, {Jones},
  {Markevitch}, {Murray}, \& {Van Speybroeck}}]{vikhlinin2006}
{Vikhlinin}, A., {Kravtsov}, A., {Forman}, W., {et~al.} 2006, \apj, 640, 691

\bibitem[{{Vikhlinin} {et~al.}(2001){Vikhlinin}, {Markevitch}, {Forman}, \&
  {Jones}}]{vikhlinin2001}
{Vikhlinin}, A., {Markevitch}, M., {Forman}, W., \& {Jones}, C. 2001, \apjl,
  555, L87

\bibitem[{{Wagner} {et~al.}(2018){Wagner}, {McDonald}, \&
  {Courteau}}]{wagner2018_a1795}
{Wagner}, C.~R., {McDonald}, M., \& {Courteau}, S. 2018, \apj, 867, 14

\bibitem[{{Wetzel}(2011)}]{wetzel2011}
{Wetzel}, A.~R. 2011, \mnras, 412, 49

\bibitem[{{Wilms} {et~al.}(2000){Wilms}, {Allen}, \& {McCray}}]{tbabs}
{Wilms}, J., {Allen}, A., \& {McCray}, R. 2000, \apj, 542, 914

\bibitem[{{Woudt} {et~al.}(2008){Woudt}, {Kraan-Korteweg}, {Lucey}, {Fairall},
  \& {Moore}}]{woudt2008}
{Woudt}, P.~A., {Kraan-Korteweg}, R.~C., {Lucey}, J., {Fairall}, A.~P., \&
  {Moore}, S.~A.~W. 2008, \mnras, 383, 445

\bibitem[{{Xie} {et~al.}(2025){Xie}, {De Lucia}, {Fossati}, {Fontanot}, \&
  {Hirschmann}}]{xie2025}
{Xie}, L., {De Lucia}, G., {Fossati}, M., {Fontanot}, F., \& {Hirschmann}, M.
  2025, \aap, 698, A73

\bibitem[{{Yagi} {et~al.}(2007){Yagi}, {Komiyama}, {Yoshida}, {Furusawa},
  {Kashikawa}, {Koyama}, \& {Okamura}}]{yagi2007}
{Yagi}, M., {Komiyama}, Y., {Yoshida}, M., {et~al.} 2007, \apj, 660, 1209

\bibitem[{{Yagi} {et~al.}(2010){Yagi}, {Yoshida}, {Komiyama}, {Kashikawa},
  {Furusawa}, {Okamura}, {Graham}, {Miller}, {Carter}, {Mobasher}, \&
  {Jogee}}]{yagi2010}
{Yagi}, M., {Yoshida}, M., {Komiyama}, Y., {et~al.} 2010, \aj, 140, 1814

\bibitem[{{Yuan} {et~al.}(2023){Yuan}, {Han}, {B{\"o}hringer}, {Wen}, \&
  {Chon}}]{yuan2023}
{Yuan}, Z.~S., {Han}, J.~L., {B{\"o}hringer}, H., {Wen}, Z.~L., \& {Chon}, G.
  2023, \mnras, 523, 1364

\bibitem[{{Zhang} {et~al.}(2011){Zhang}, {Andernach}, {Caretta}, {Reiprich},
  {B{\"o}hringer}, {Puchwein}, {Sijacki}, \& {Girardi}}]{zhang2011}
{Zhang}, Y.~Y., {Andernach}, H., {Caretta}, C.~A., {et~al.} 2011, \aap, 526,
  A105

\bibitem[{{Zibetti} {et~al.}(2009){Zibetti}, {Charlot}, \& {Rix}}]{zibetti2009}
{Zibetti}, S., {Charlot}, S., \& {Rix}, H.-W. 2009, \mnras, 400, 1181

\end{thebibliography}

\appendix
\section*{Appendix A: Additional Figures and Tables}
This appendix presents the following additional figures and tables:
\begin{itemize}
    \item The 0.5--2.0 keV X-ray flux mosaics of Perseus and A1795 (Figure \ref{fig:perseus_xray} and Figure \ref{fig:a1795_xray}).
    \item The dependence of $R_{90}$ on off-axis angle at 1 keV for galaxies in our sample (Figure \ref{fig:psf}).
   \item The X-ray spectra of the individually detected galactic coronae, with emphasis on distinguishing their thermal emission from the soft X-ray excess observed in KUG 1259+280 (Figure \ref{fig:individualdetectionsspectrum} and Figure \ref{fig:extended}).
   \item X-ray spectrum of the full dwarf galaxy stack (Figure \ref{fig:dwarfstackspec}).
   \item Optical and X-ray images of low-mass galaxies hosting X-ray AGN with rest-frame 2--10 keV luminosities exceeding $10^{41}$ erg s$^{-1}$, and their corresponding X-ray spectra (Figure \ref{fig:agn}).
   \item List of Chandra observations used in this work, including the total exposure time and total cleaned exposure time for each galaxy cluster (Table \ref{tab:obsidlist}).
\end{itemize}
\renewcommand{\thefigure}{A\arabic{figure}}
\setcounter{figure}{0}

\begin{figure*}
   \centering
\includegraphics[width=1.0\linewidth]{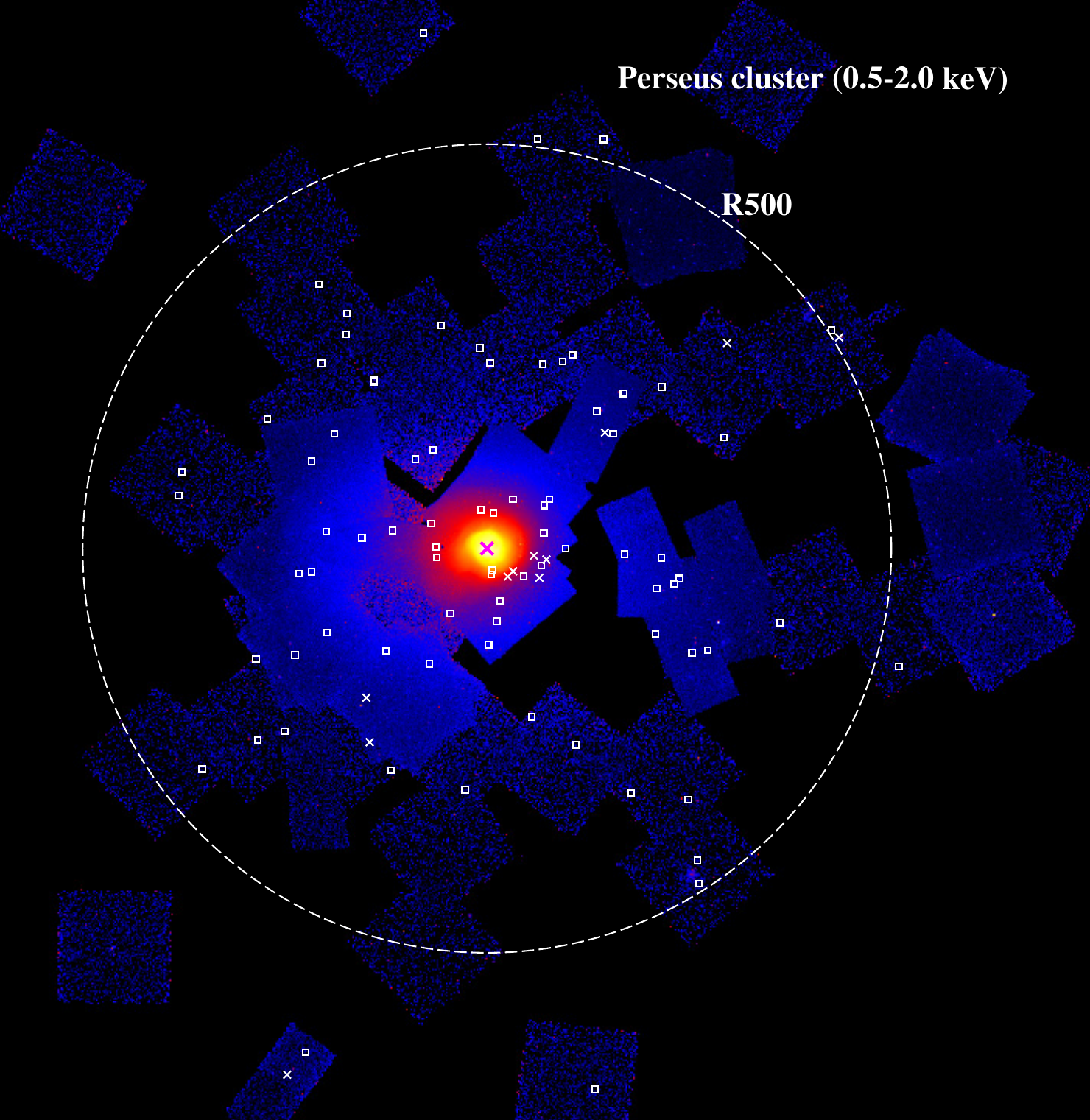}
  \caption{\small The \textit{Chandra} X-ray flux image of the Perseus galaxy cluster in the energy range 0.5--2.0 keV, combining 2.74 Msec of cleaned exposure. Symbols are as in Figure \ref{fig:coma_xray}. 
  \label{fig:perseus_xray}}  
\end{figure*}

\begin{figure*}
   \centering
\includegraphics[width=1.0\linewidth]{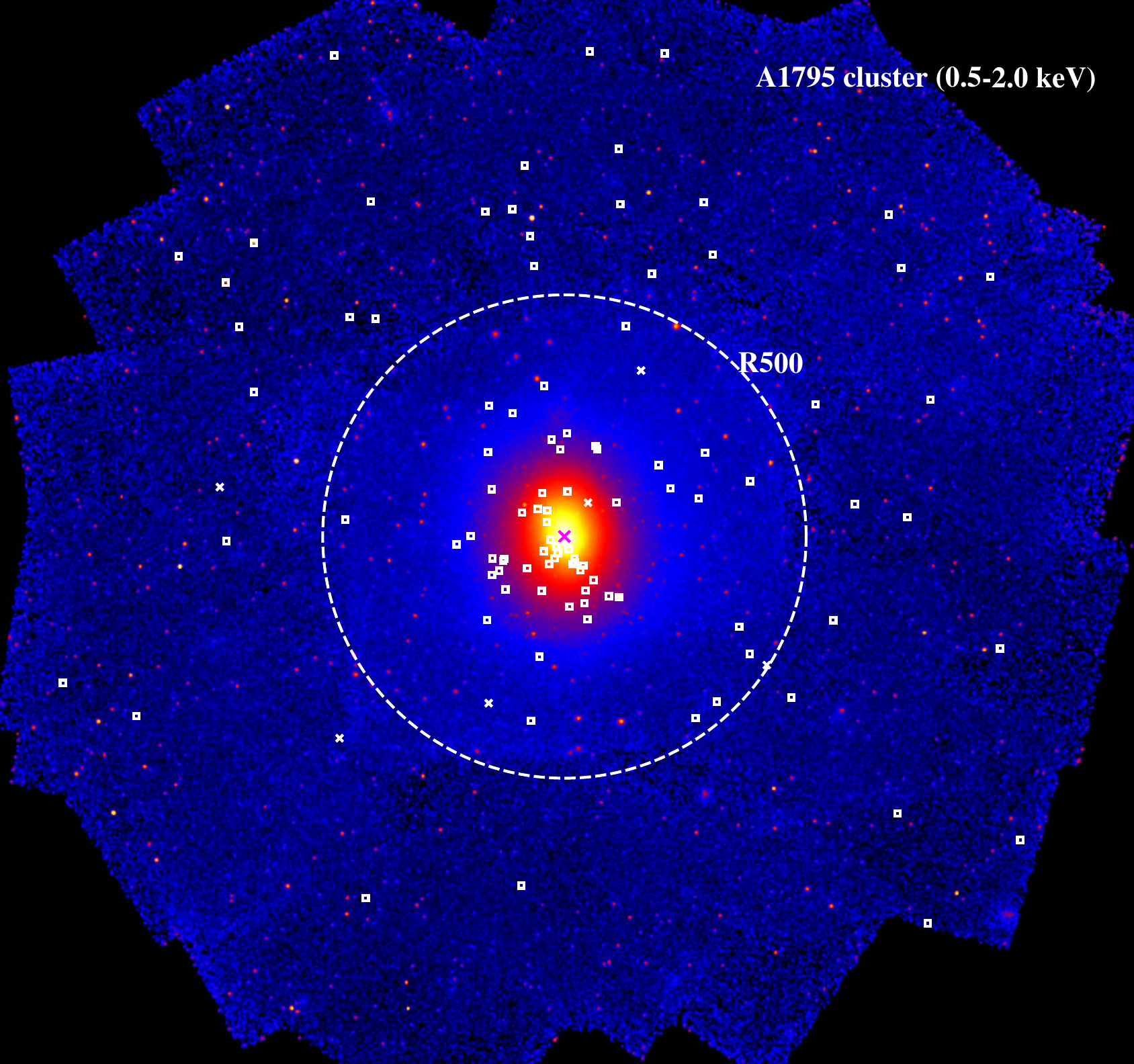}
  \caption{\small The \textit{Chandra} X-ray flux image of A1795 in the energy range 0.5--2.0 keV, combining $\sim$5.8 Msec of cleaned exposure. Symbols are as in Figure \ref{fig:coma_xray}. 
  \label{fig:a1795_xray}}  
\end{figure*}

\begin{figure*}
   \centering
\includegraphics[width=0.7\linewidth]{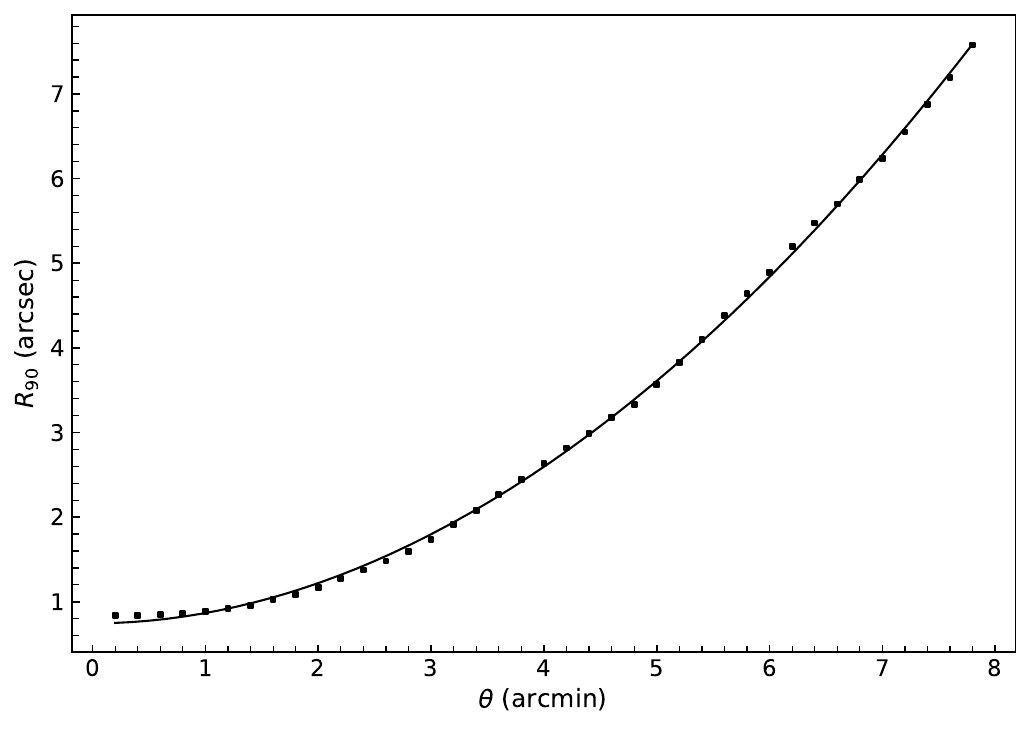}
  \caption{The dependence of $R_{90}$ on off-axis angle ($\theta$) at 1 keV for galaxies in our sample. Each black point represents the median $R_{90}$ in bins of off-axis angle. The dependence is well described by the relation: $R_{90}(\theta)=0.74 + 0.12\ \theta^{1.96}$.
  \label{fig:psf}}  
\end{figure*}
\begin{figure*}
  \centering
  \subfloat[]{\includegraphics[width=0.23\linewidth,angle=270]{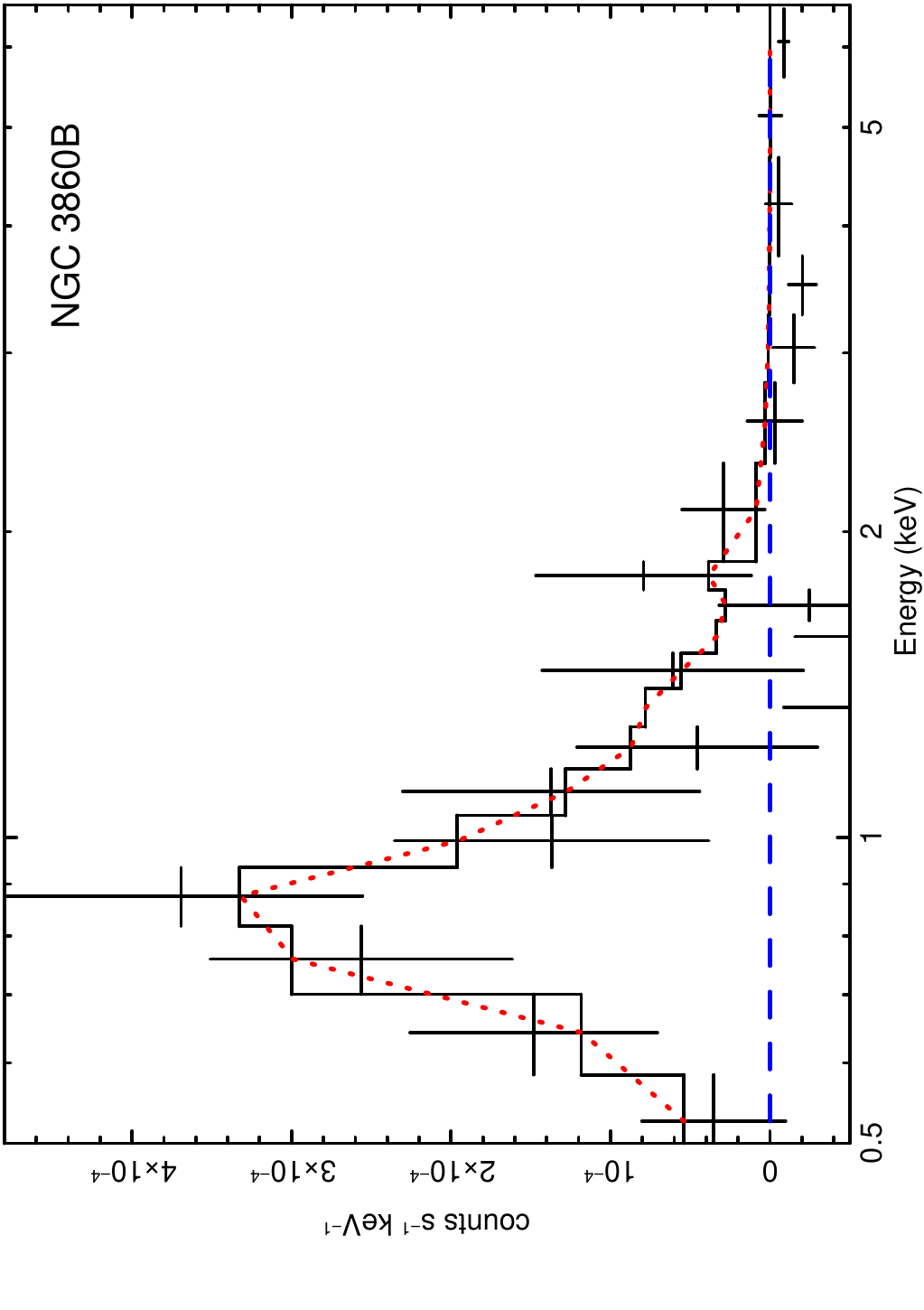}\label{}}\hfill
  \subfloat[]{
  \raisebox{0.9mm}{
    \includegraphics[width=0.236\linewidth,angle=270]{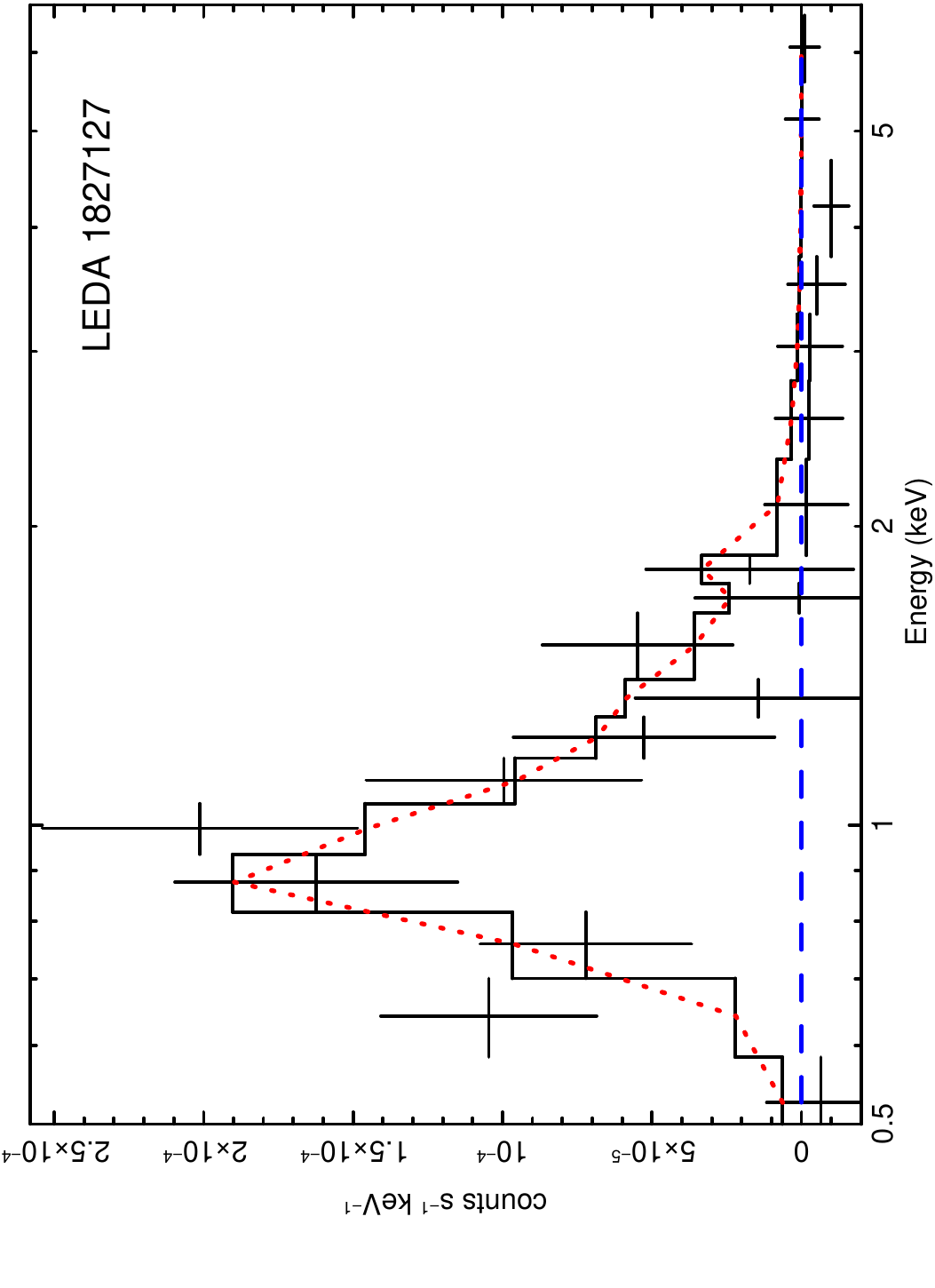}
  }}\hfill
  \subfloat[]{\raisebox{-0.2mm}{\includegraphics[width=0.23\linewidth, angle=270]{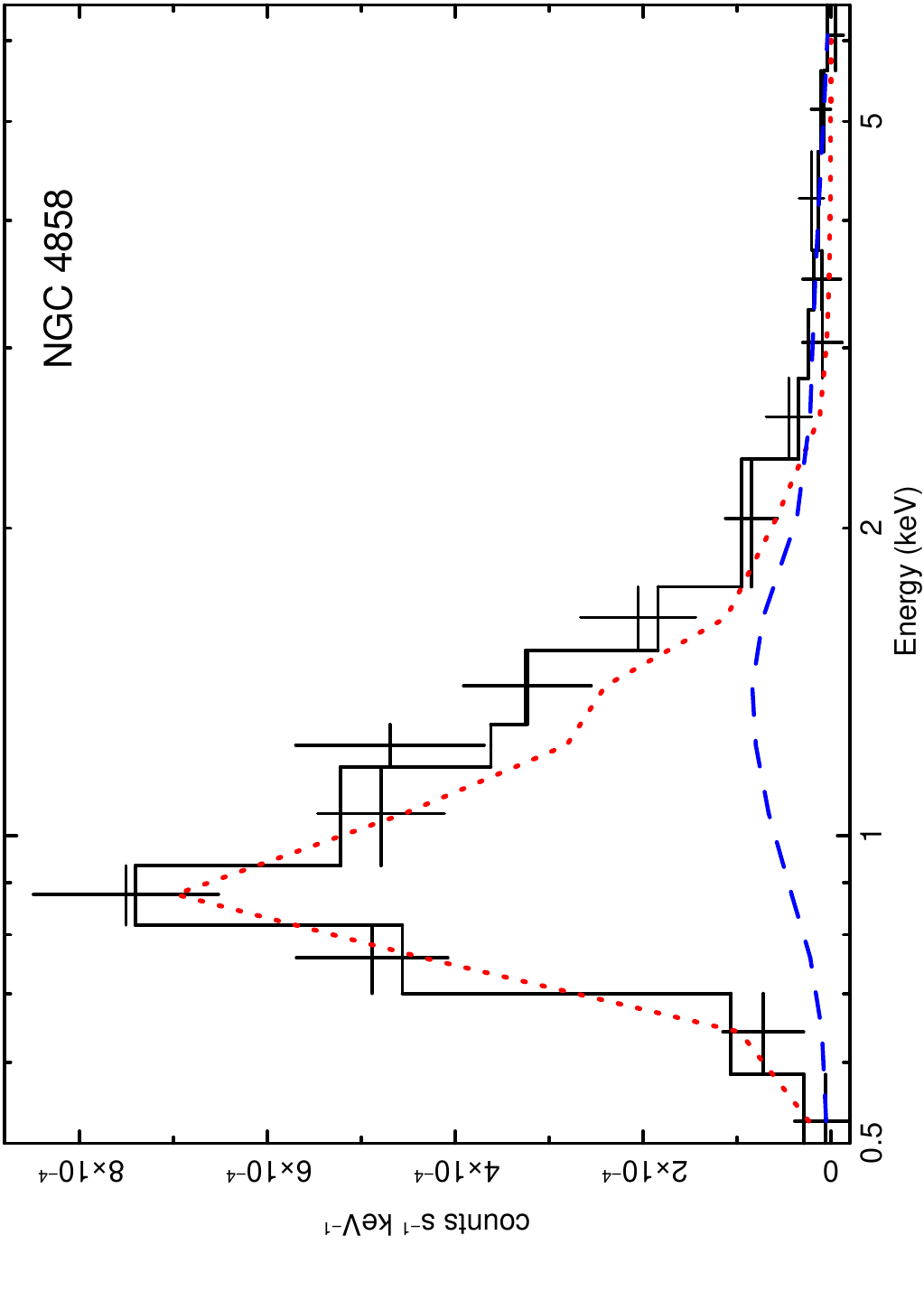}}}\\

  \subfloat[]{\includegraphics[width=0.23\linewidth, angle=270]{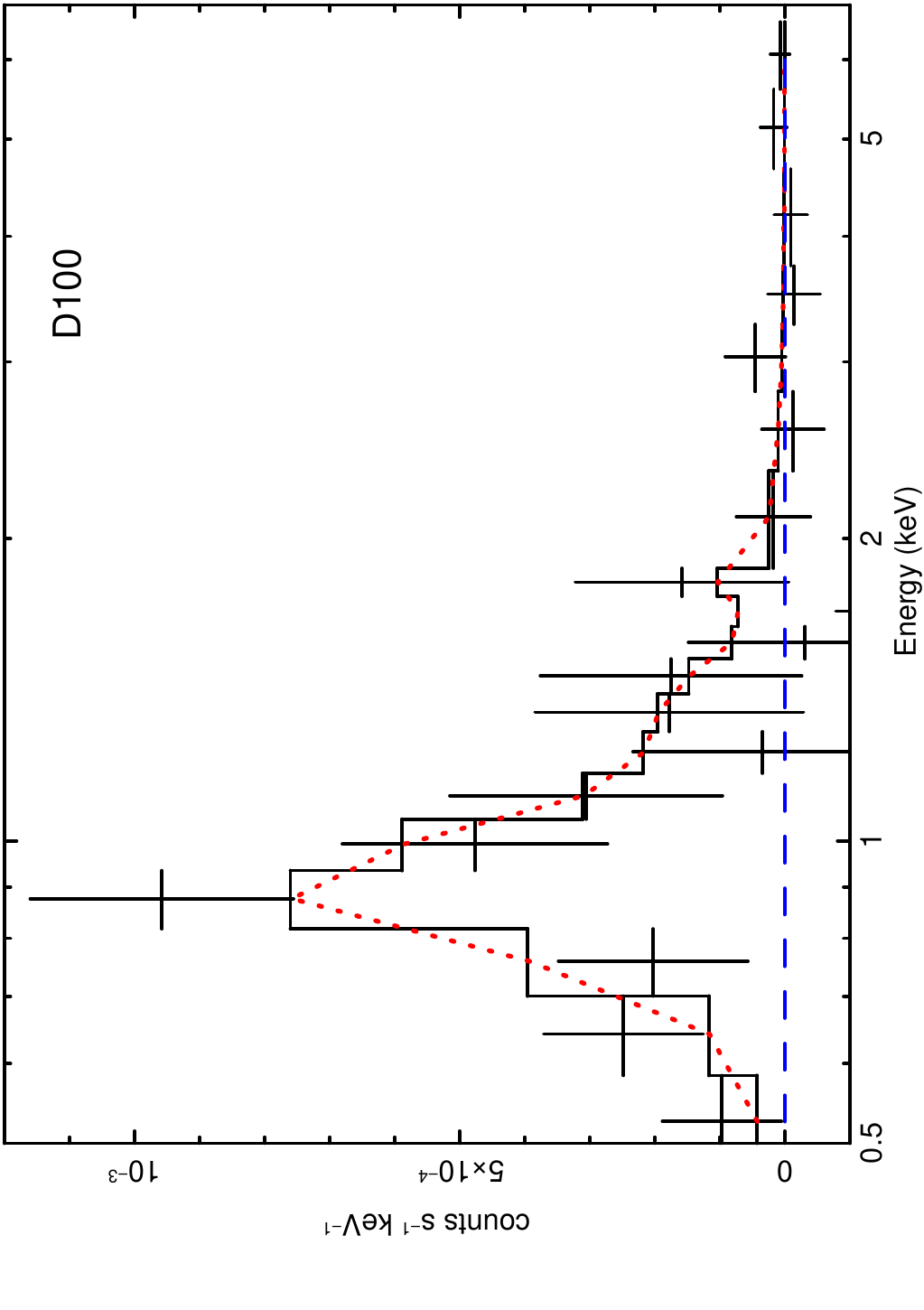}\label{}}\hfill
  \subfloat[]{\includegraphics[width=0.23\linewidth, angle=270]{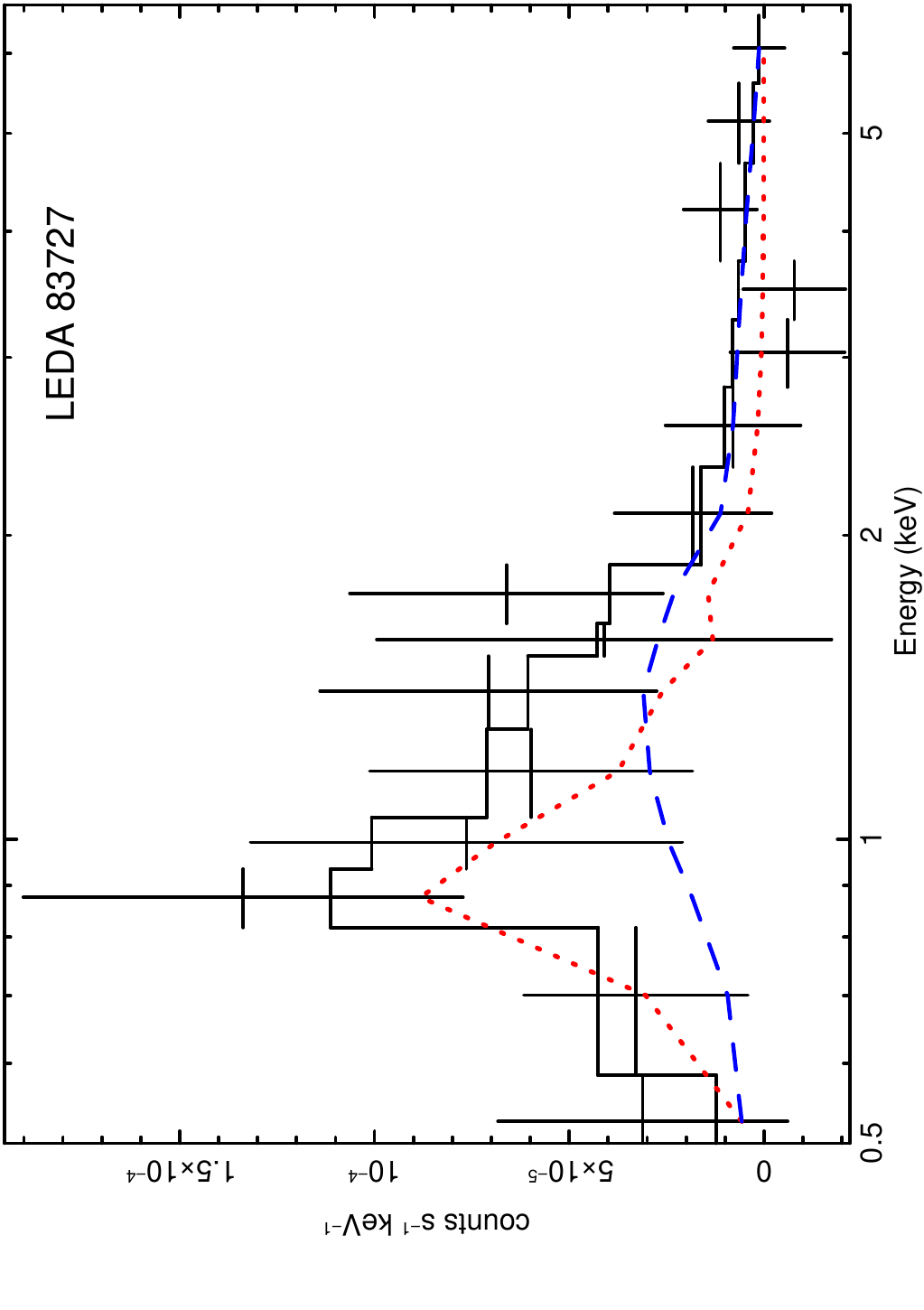}\label{}}\hfill
  \subfloat[]{\includegraphics[width=0.23\linewidth, angle=270]{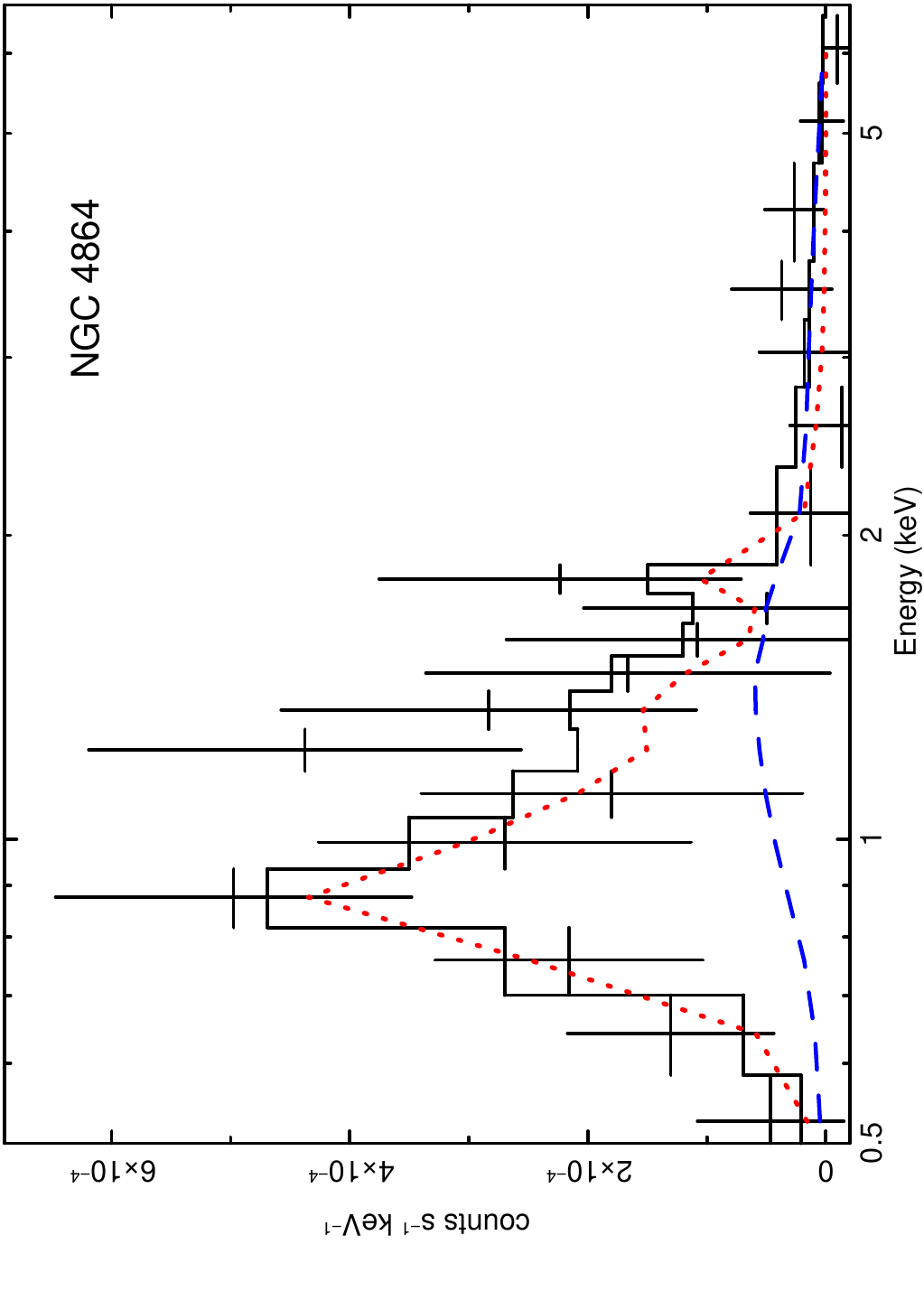}\label{}}\\
  
  \subfloat[]{\includegraphics[width=0.23\linewidth, angle=270]{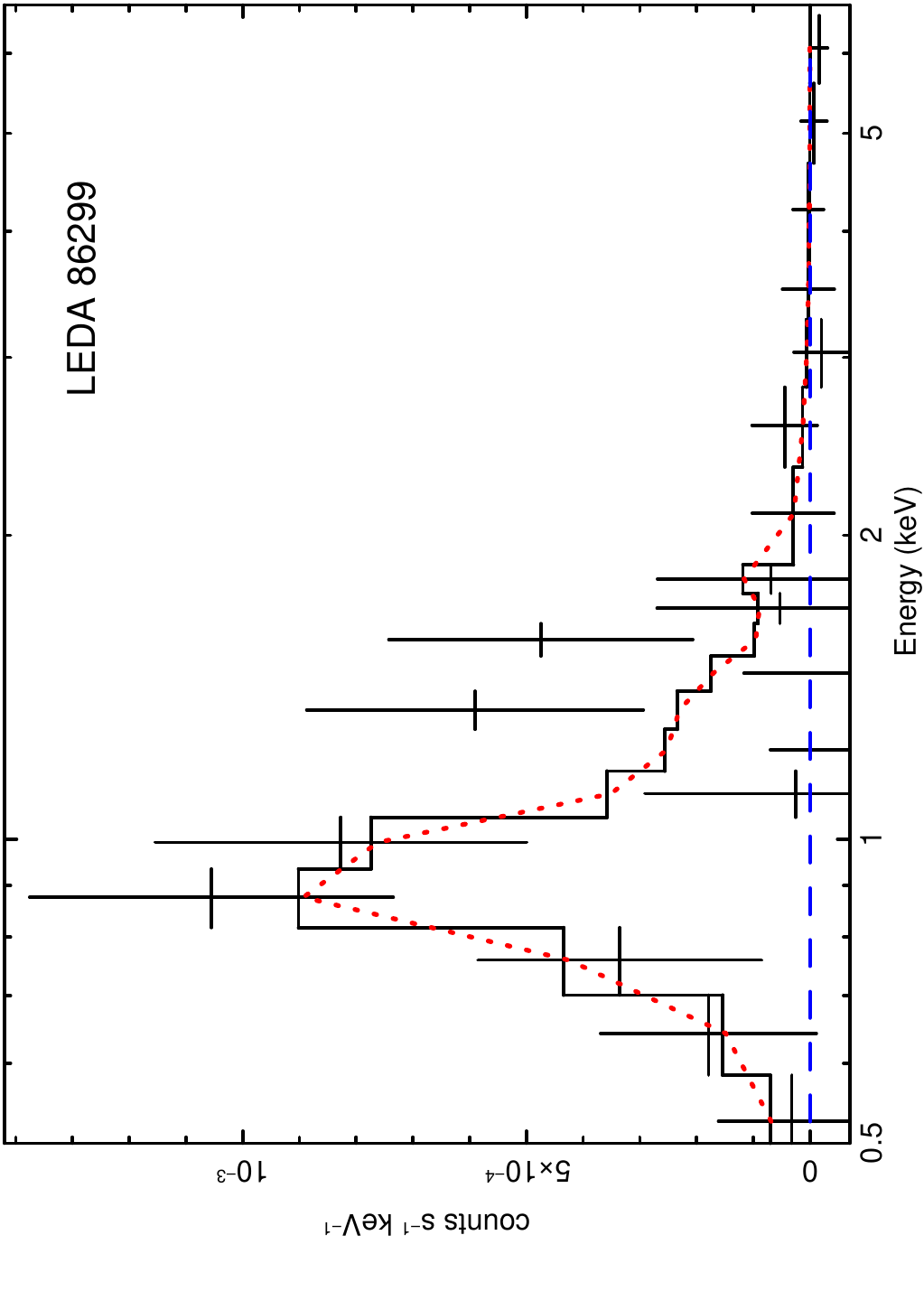}\label{}}\hfill
  \subfloat[]{\includegraphics[width=0.23\linewidth, angle=270]{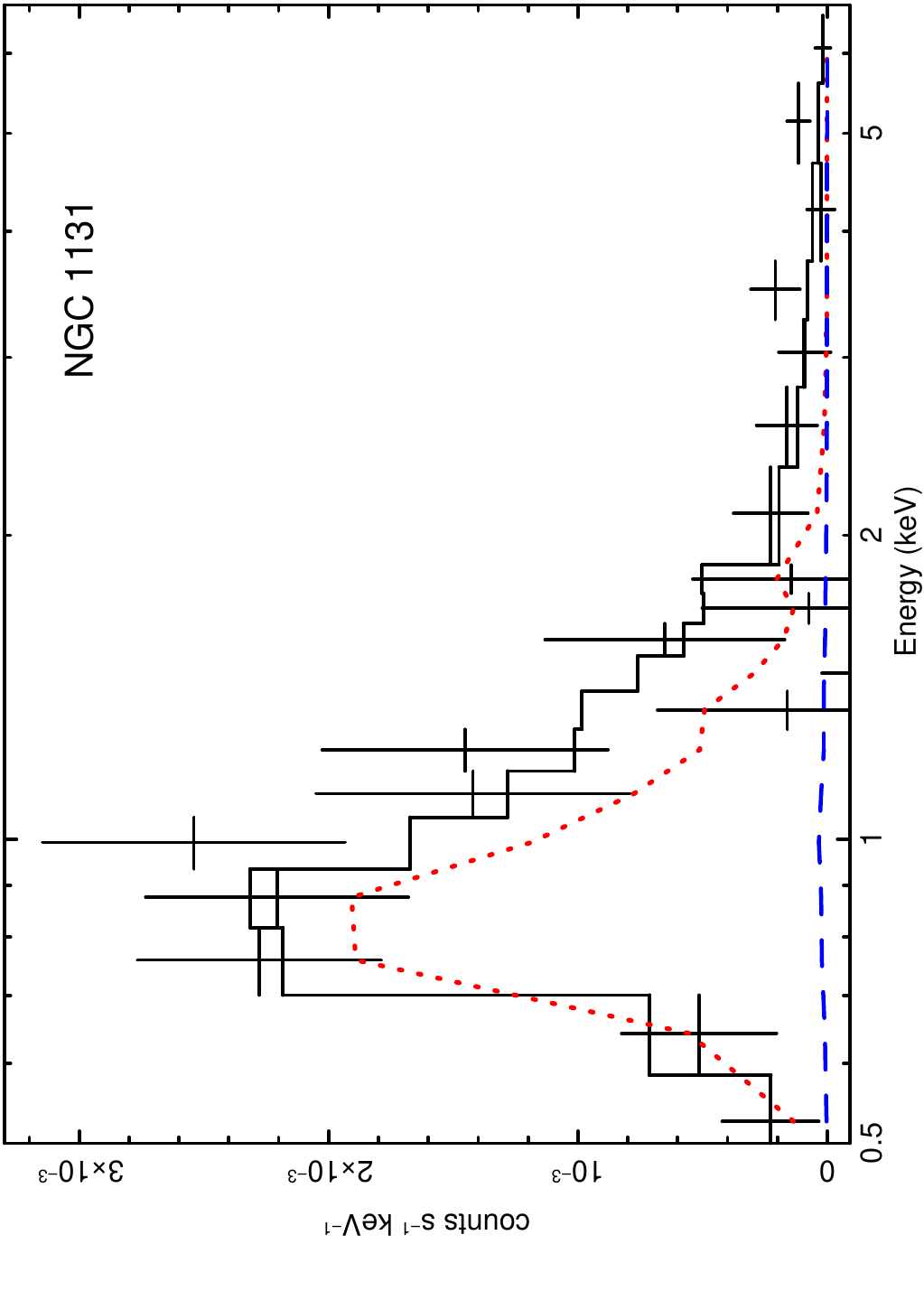}\label{}}\hfill
  \subfloat[]{\includegraphics[width=0.23\linewidth, angle=270]{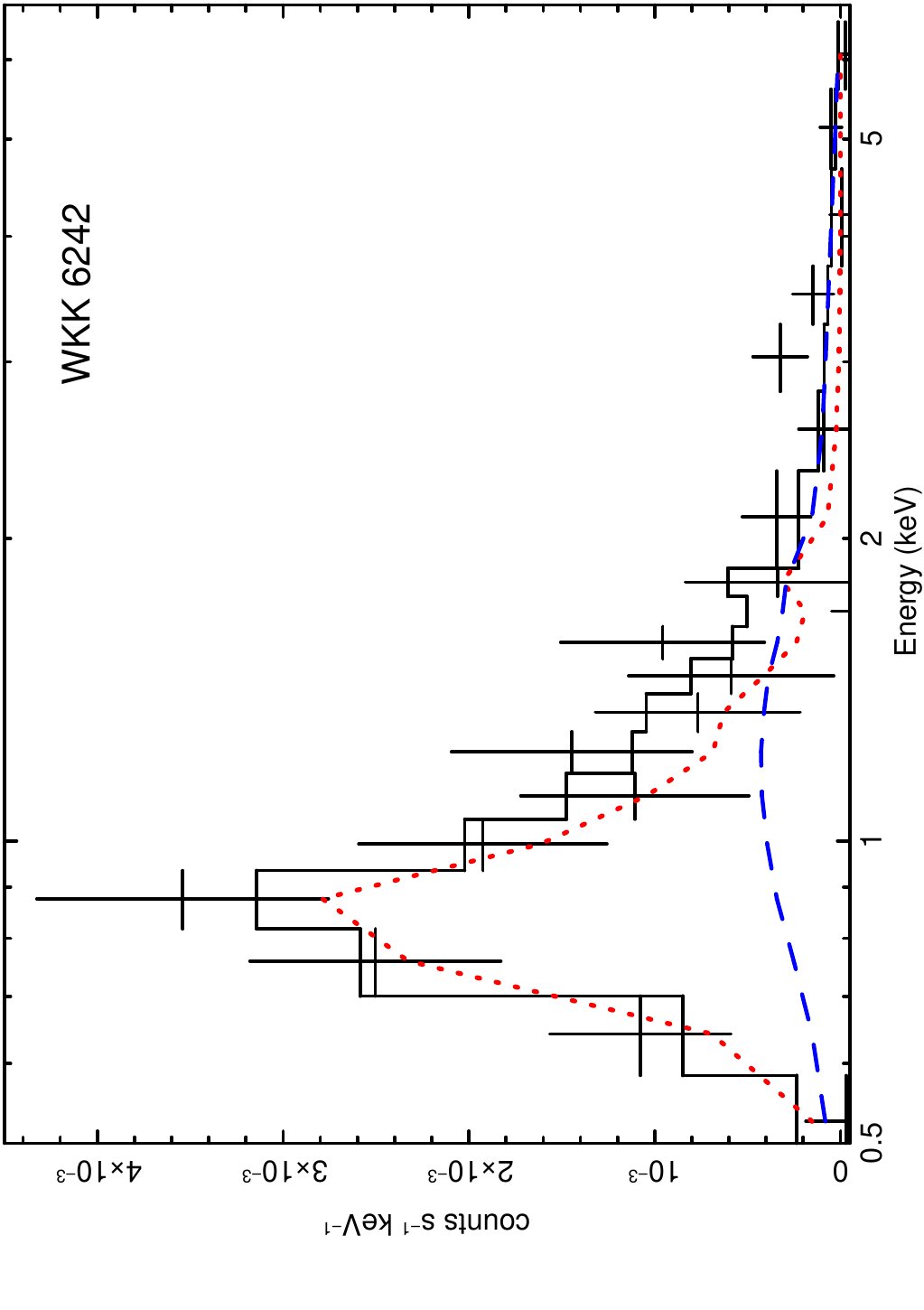}\label{}}\\
  
  \subfloat[]{\includegraphics[width=0.23\linewidth, angle=270]{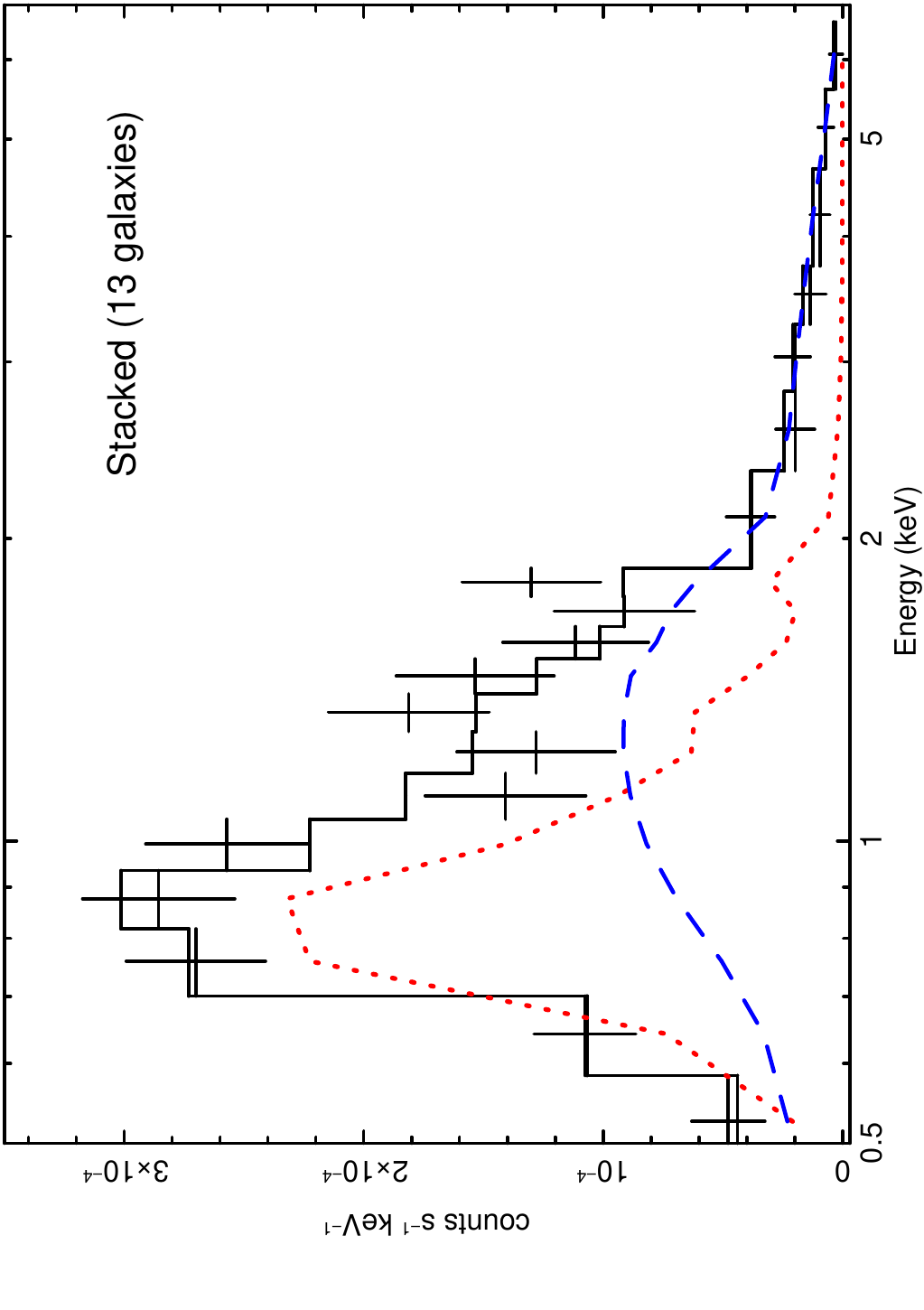}\label{}}
  \hspace{0.08\linewidth}
  \subfloat[]{  
  \setlength{\fboxsep}{1pt}
  \setlength{\fboxrule}{1pt}
  \fcolorbox{red!55!black}{white}{\includegraphics[width=0.23\linewidth,
      angle=270]{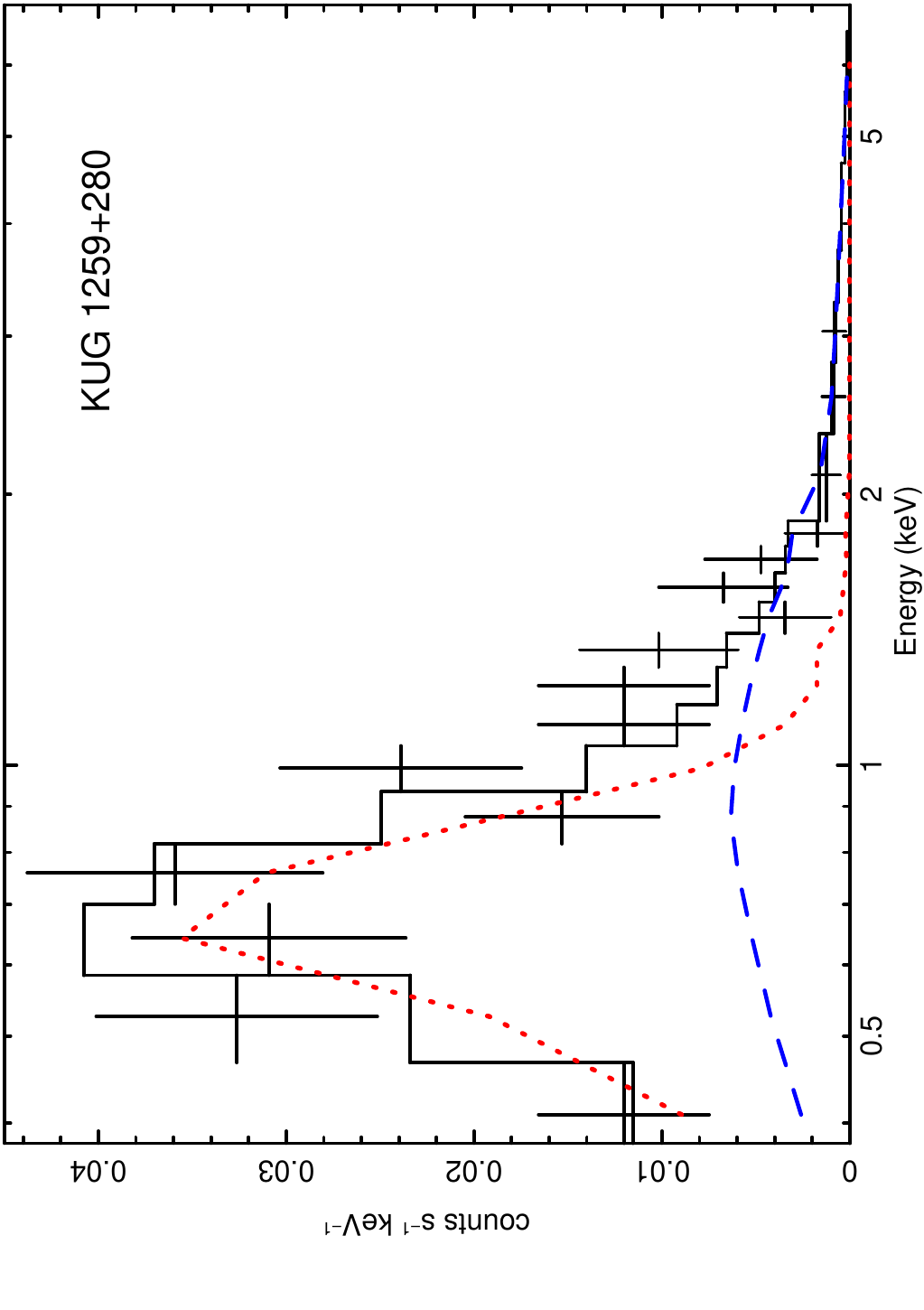}}}
  \caption{
(a)--(i) 0.5--7.0 keV X-ray spectra of nine galaxies with individual hot-gas detections in this work. The iron L-shell hump is clearly visible, centred at $\sim$0.8--1.0 keV.
The \textit{red dotted} line shows the hot-gas component modelled with \textit{apec} and the \textit{blue dashed} line shows the \textit{powerlaw} component representing the contribution from unresolved XRBs and AGN. The total best-fit model is shown in \textit{solid black}. The contribution from the $lnvmkl$ component representing the contribution from unresolved CVs and ABs, is negligible and therefore not shown. (j) Stacked spectrum of the remaining 13 galaxies with individual hot-gas detections, also showing the presence of the iron L-shell hump. (k) X-ray spectrum of KUG~1259+280, highlighted to distinguish it from the coronal sources. The source is an AGN with a pronounced soft X-ray excess (\citealt{dewangan2000}) associated with QPEs (\citealt{kug_qpe}).
It is 5--80 times more luminous than the individual coronae and also softer.
  }
  \label{fig:individualdetectionsspectrum}
\end{figure*}

\begin{figure*}
    \centering
    \begin{overpic}[width=\linewidth]{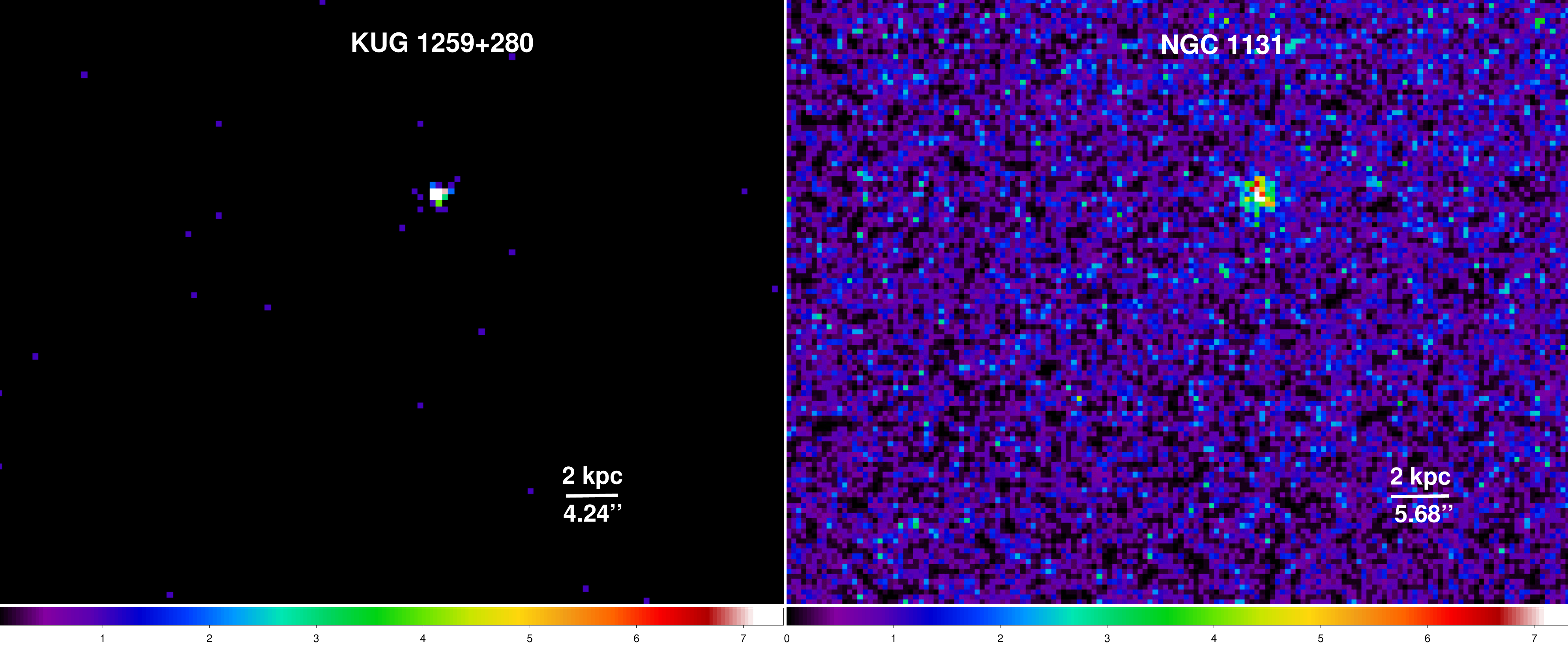}
    \put(-0.2,2.8){
    \includegraphics[width=0.21\linewidth]{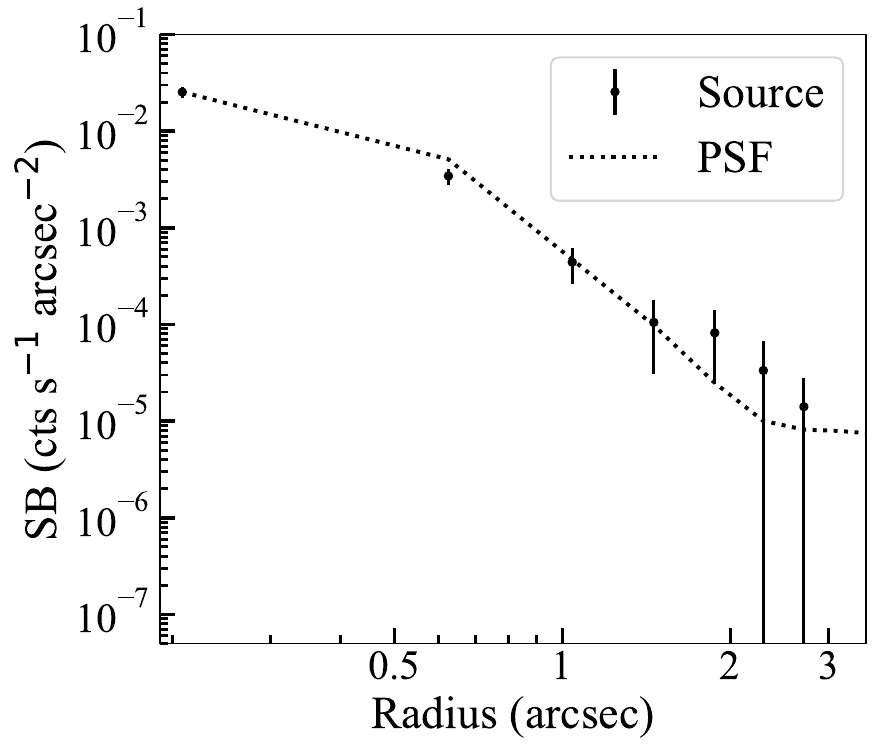}
     }
     \put(50,2.8){
    \includegraphics[width=0.21\linewidth]{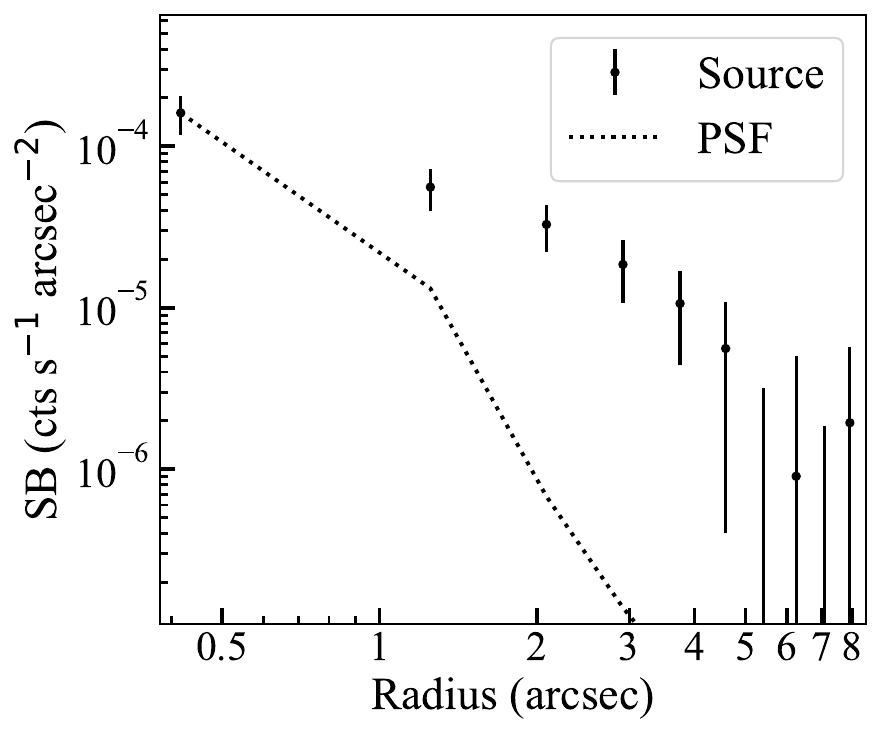}
     }
    \end{overpic}
    \caption{The 0.5--2.0 keV count images of KUG~1259+280 in Coma (left, 5 ksec) and NGC~1131 in AWM7 (right, 133.7 ksec), are shown as an example of unresolved versus extended soft X-ray emission.
    Insets in the lower left of each panel show the local background-subtracted surface brightness profiles of the sources together with the 1 keV PSF (\textit{dotted curve}) at the corresponding detector locations, normalised to match the source surface brightness in the innermost radial bin. The radial profile of KUG~1259+280 is consistent with the local PSF, indicating that the source is unresolved.
    Indeed, KUG~1259+280 hosts an X-ray AGN with a pronounced soft X-ray excess (\citealt{dewangan2000}).
    In contrast, NGC 1131 shows emission more extended than the PSF.
    }
    \label{fig:extended}
\end{figure*}

\begin{figure*}
   \centering
\includegraphics[width=0.5\linewidth, angle=270]{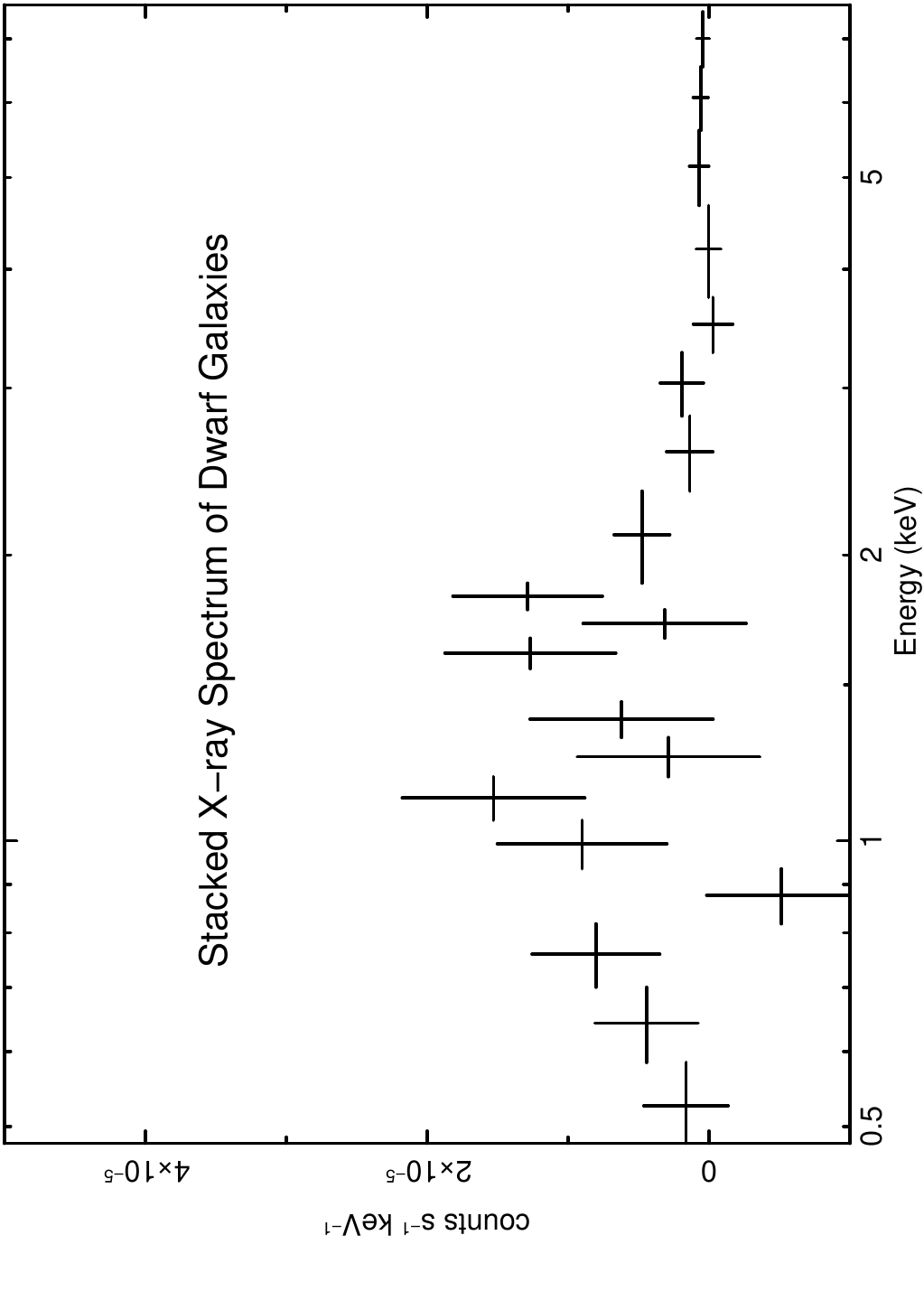}
  \caption{X-ray spectrum of the full dwarf galaxy stack, showing no clear Fe-L emission hump.}
  \label{fig:dwarfstackspec}
\end{figure*}

\begin{figure*}
\centering

\begin{minipage}{0.3\linewidth}
\centering
\includegraphics[width=\linewidth]{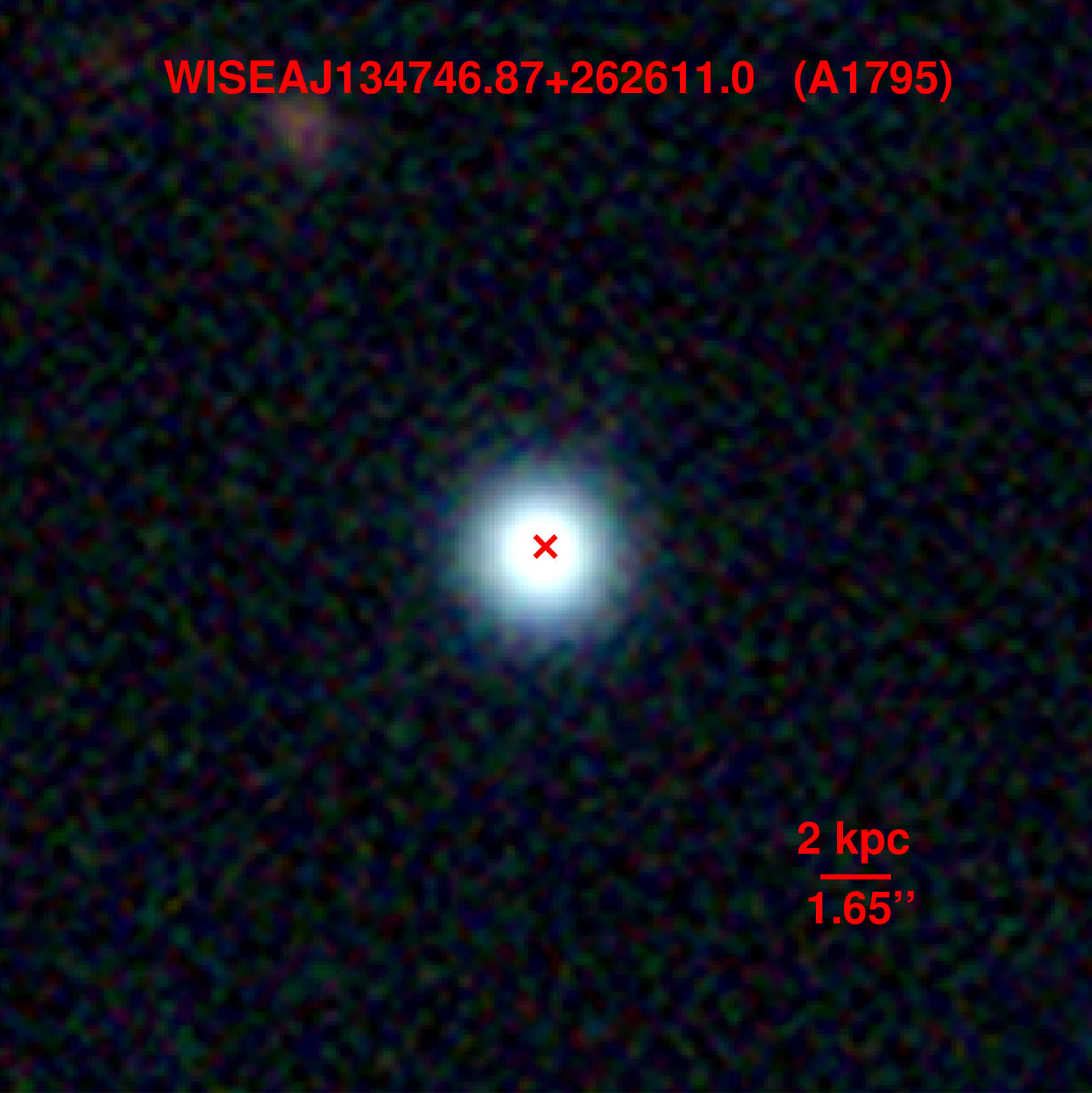}
\end{minipage}
\hspace{0.01\linewidth}
\begin{minipage}{0.3\linewidth}
\centering
\includegraphics[width=\linewidth]{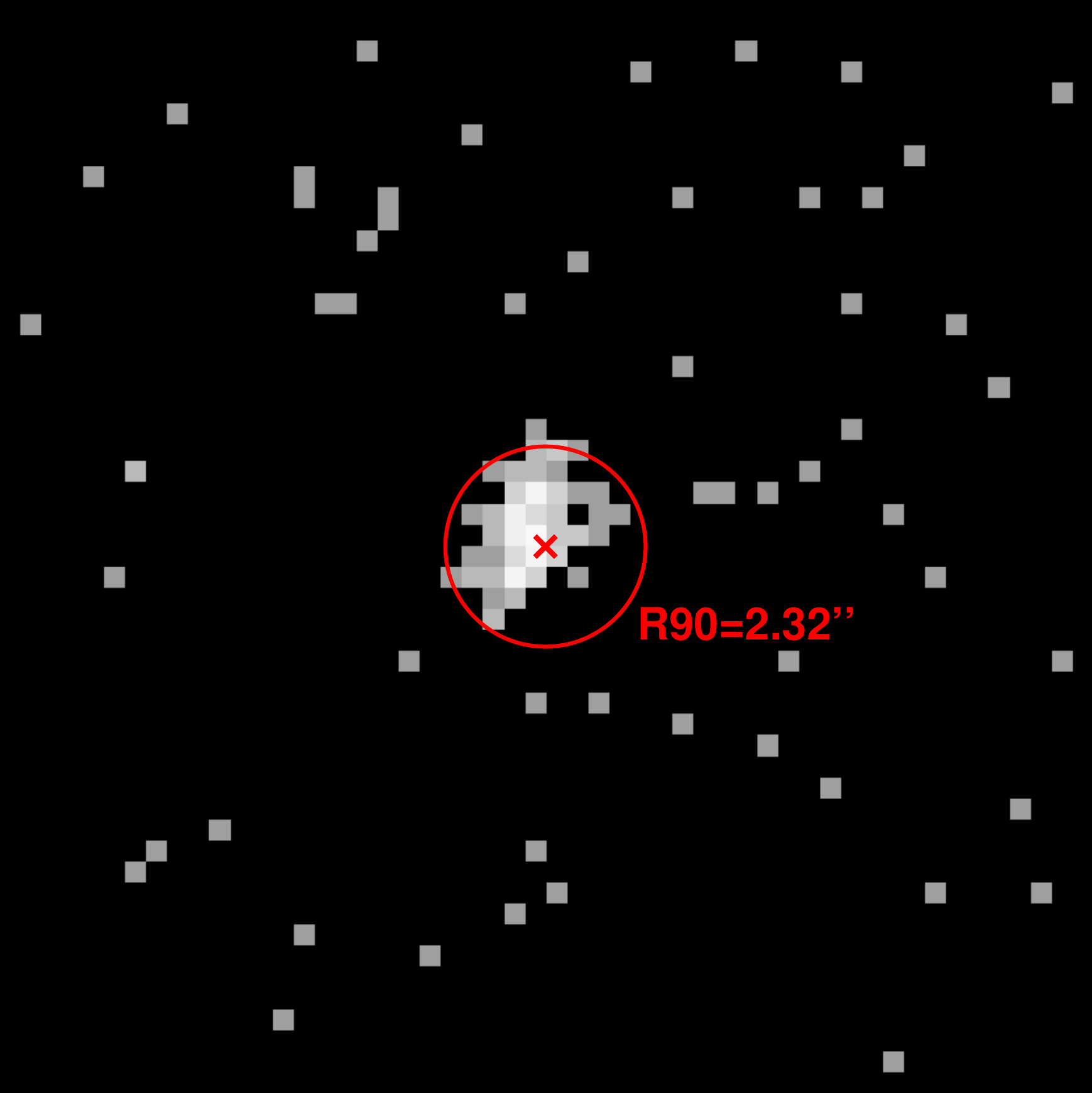}
\end{minipage}
\begin{minipage}{0.3\linewidth}
\centering
\raisebox{-0.2cm}{
\includegraphics[height=1.15\linewidth,angle=270]{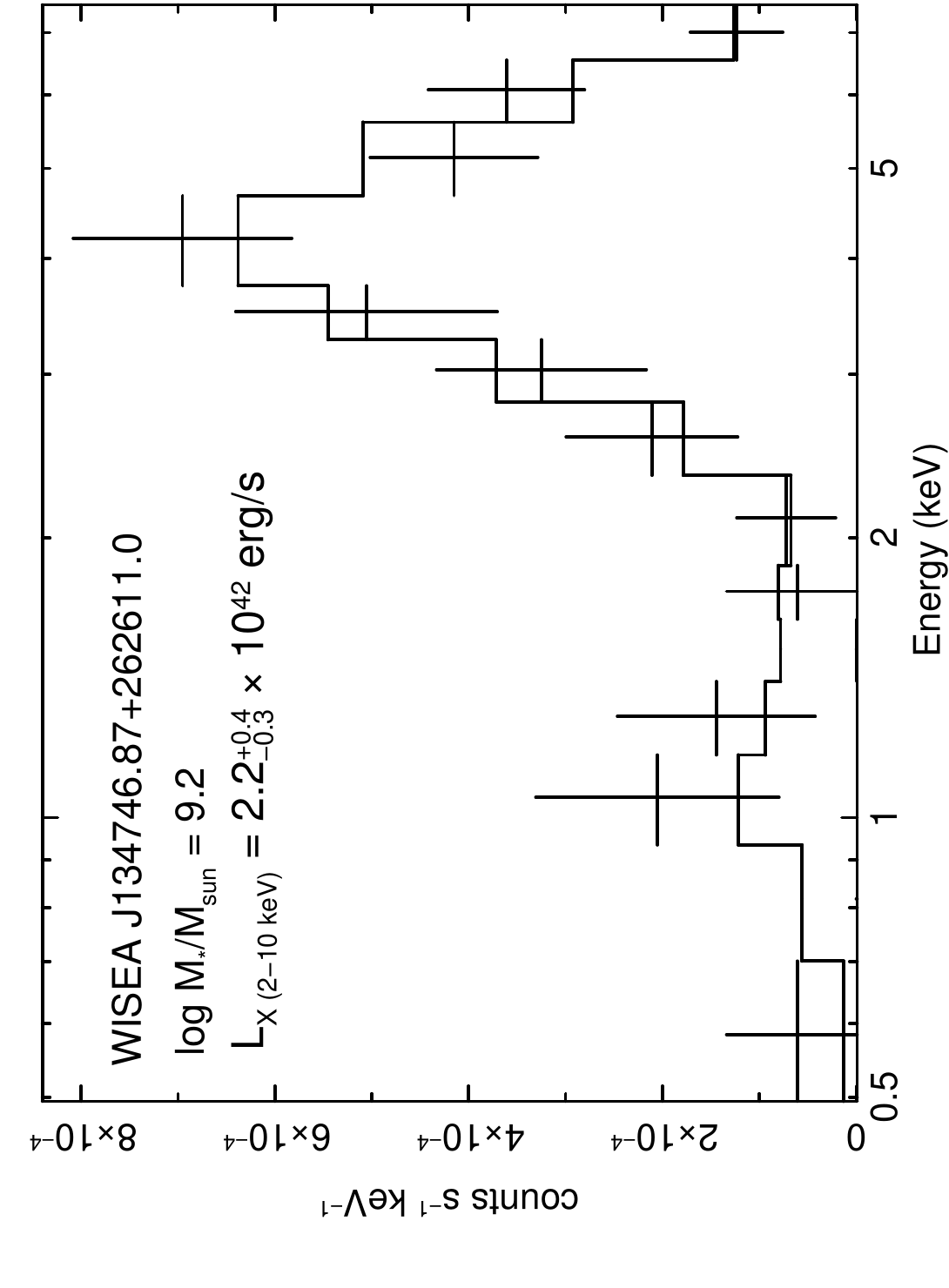}}
\end{minipage}
\\[0.4cm]
\centering
\begin{minipage}{0.3\linewidth}
\centering
\includegraphics[width=\linewidth]{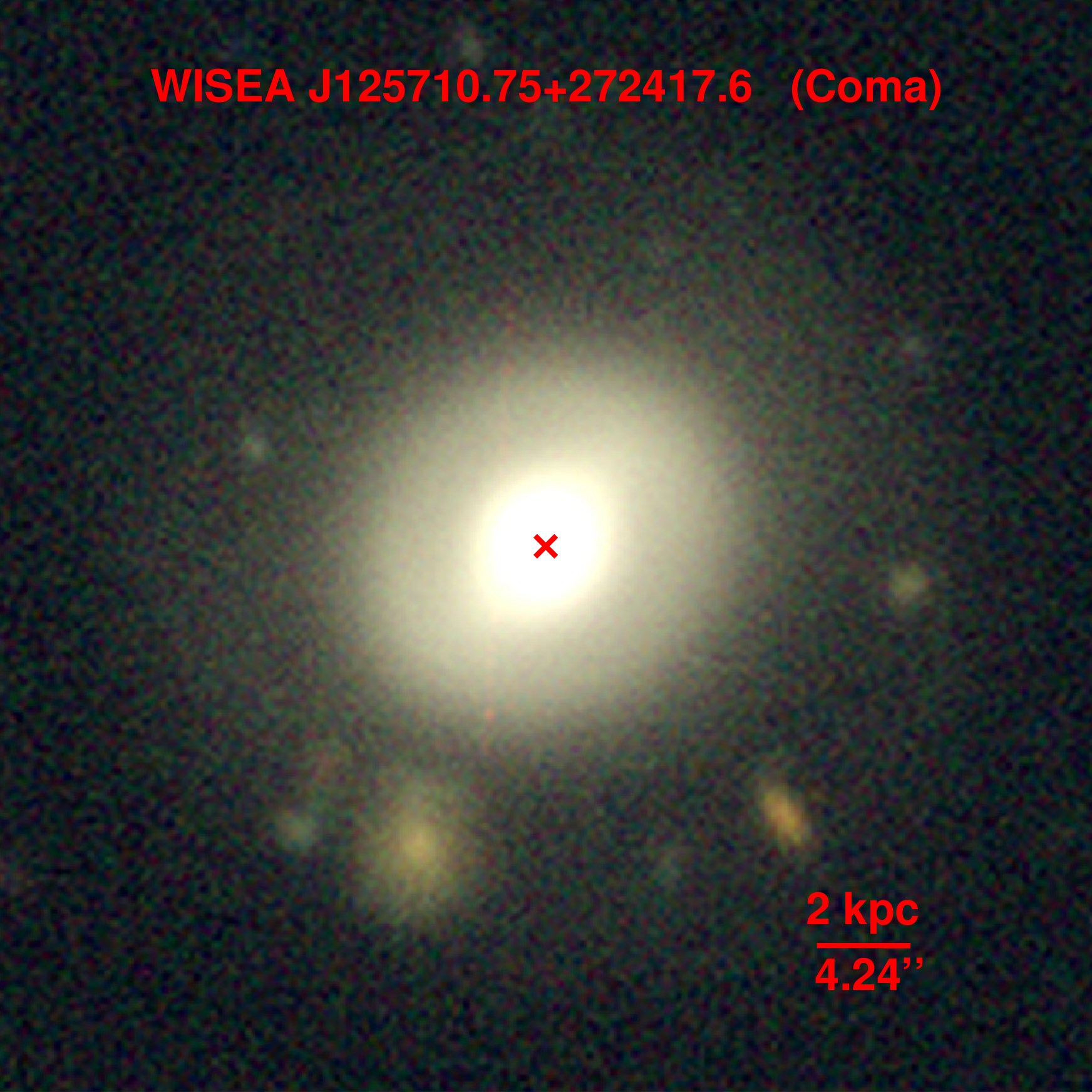}
\end{minipage}
\hspace{0.01\linewidth}
\begin{minipage}{0.3\linewidth}
\centering
\includegraphics[width=\linewidth]{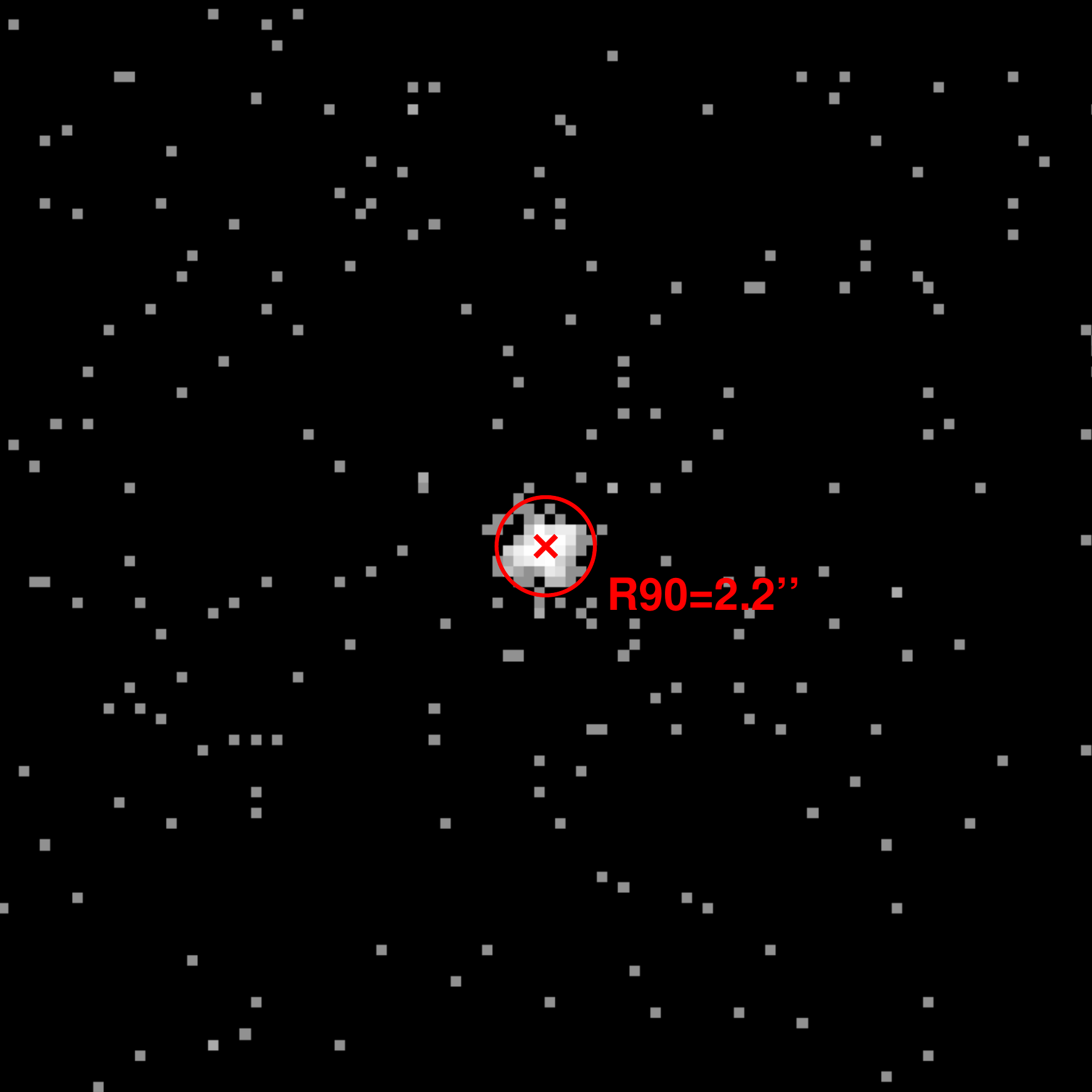}
\end{minipage}
\begin{minipage}{0.3\linewidth}
\centering
\raisebox{-0.2cm}{
\includegraphics[height=1.22\linewidth,angle=270]{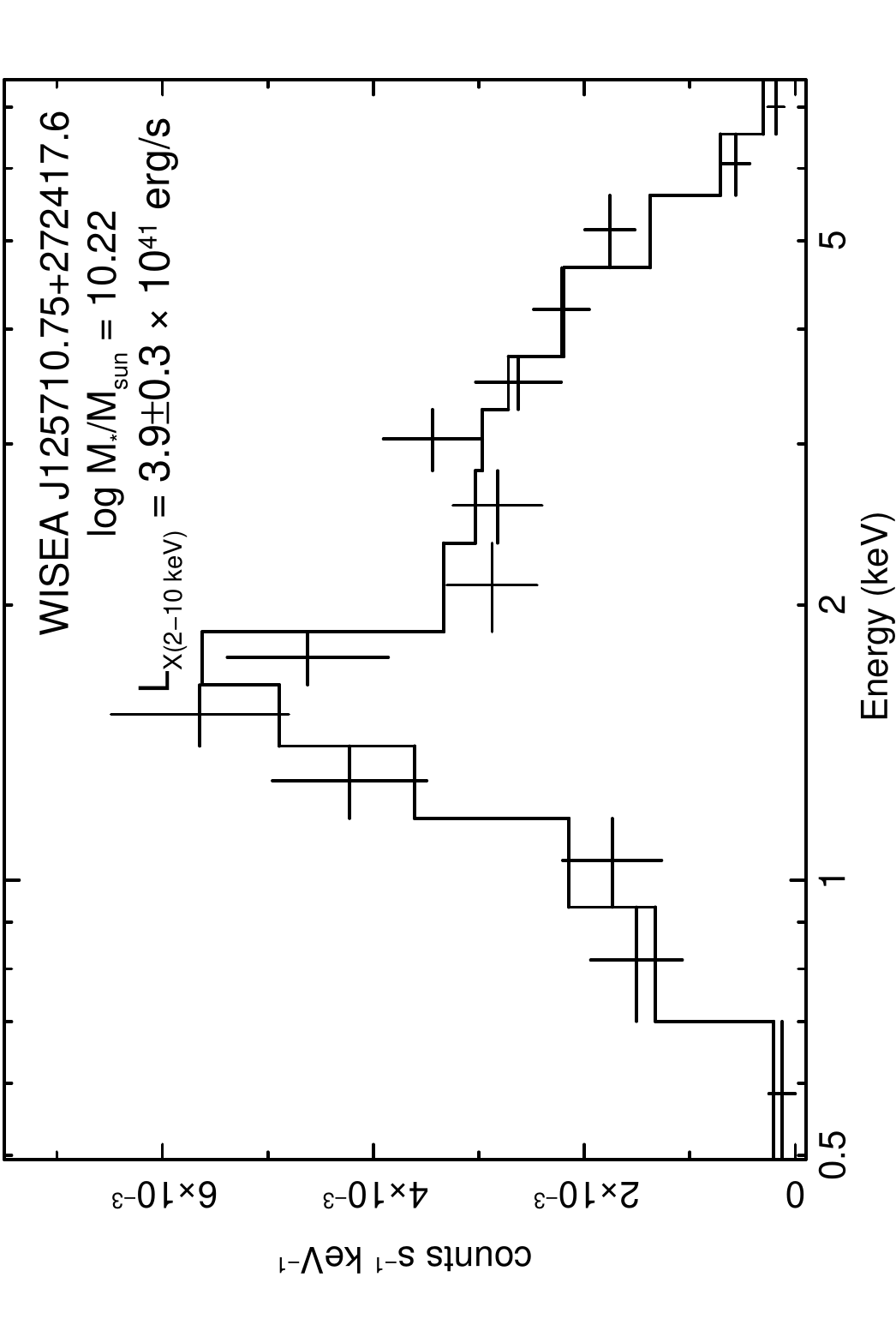}}
\end{minipage}
\\[0.4cm]
\begin{minipage}{0.3\linewidth}
\centering
\includegraphics[width=\linewidth]{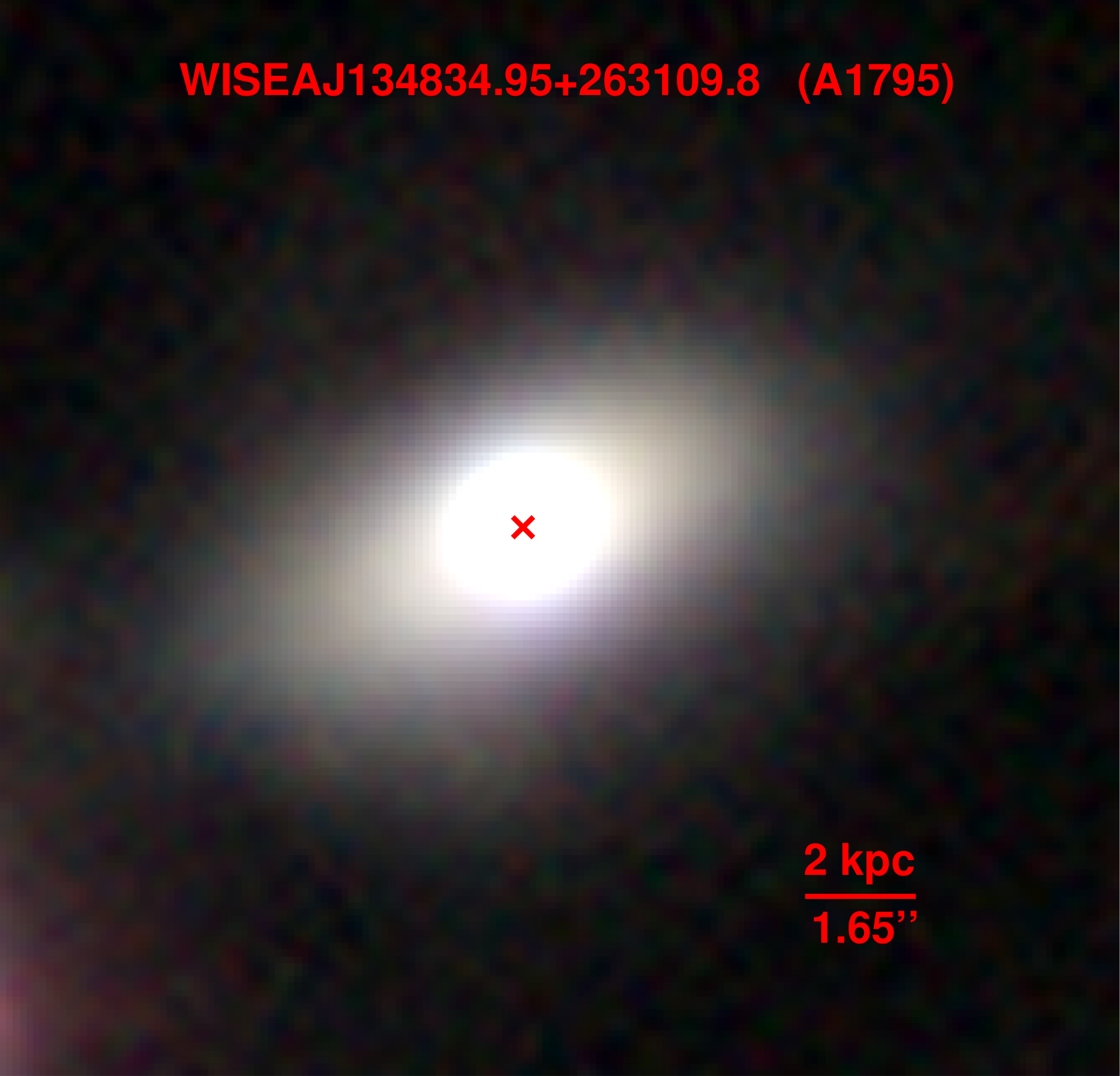}
\end{minipage}
\hspace{0.01\linewidth}
\begin{minipage}{0.3\linewidth}
\centering
\includegraphics[width=\linewidth]{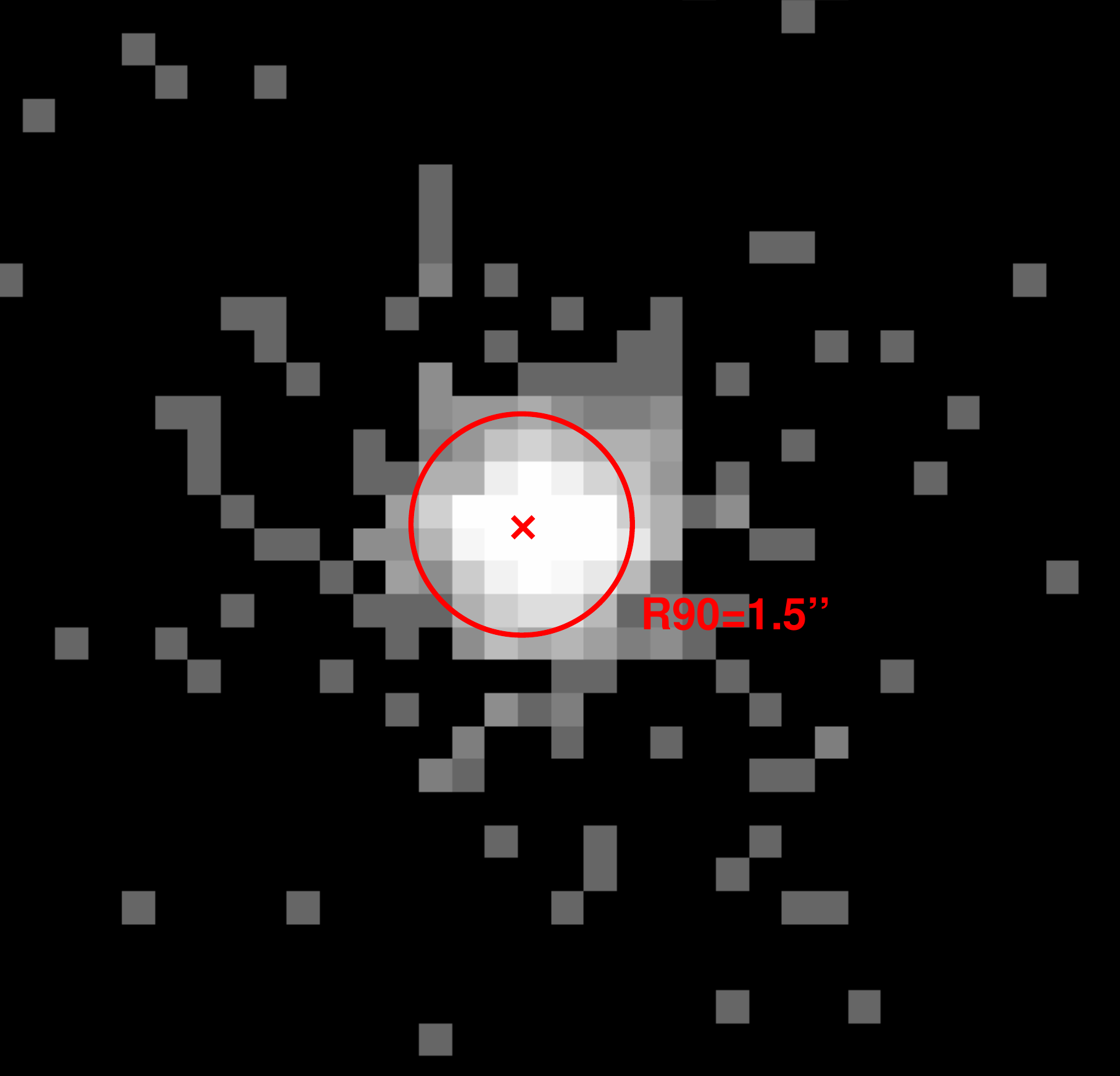}
\end{minipage}
\begin{minipage}{0.3\linewidth}
\centering
\raisebox{-0.2cm}{
\includegraphics[height=1.24\linewidth,angle=270]{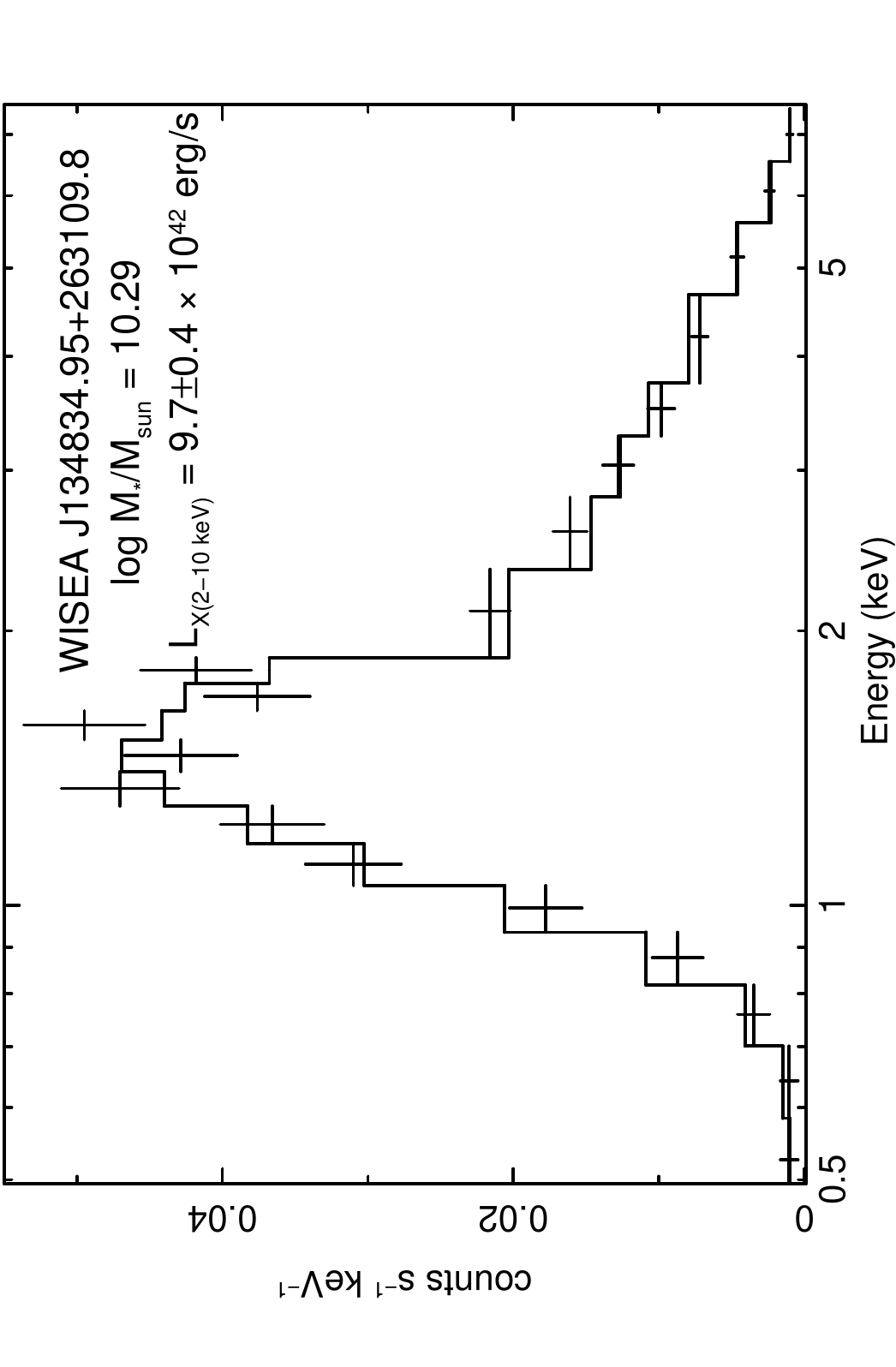}}
\end{minipage}
\\[0.4cm]
\begin{minipage}{0.3\linewidth}
\centering
\includegraphics[width=\linewidth]{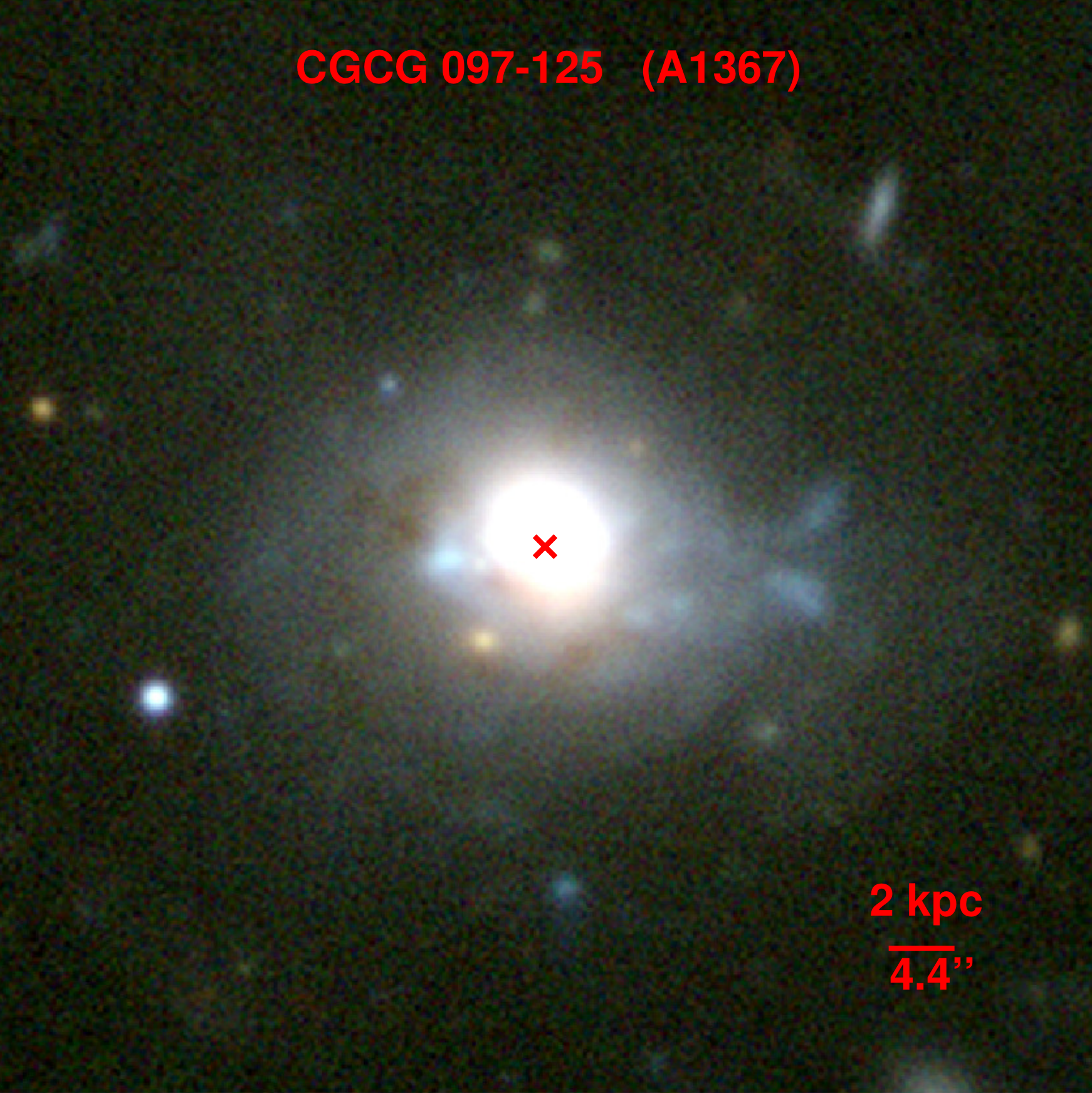}
\end{minipage}
\hspace{0.01\linewidth}
\begin{minipage}{0.3\linewidth}
\centering
\includegraphics[width=\linewidth]{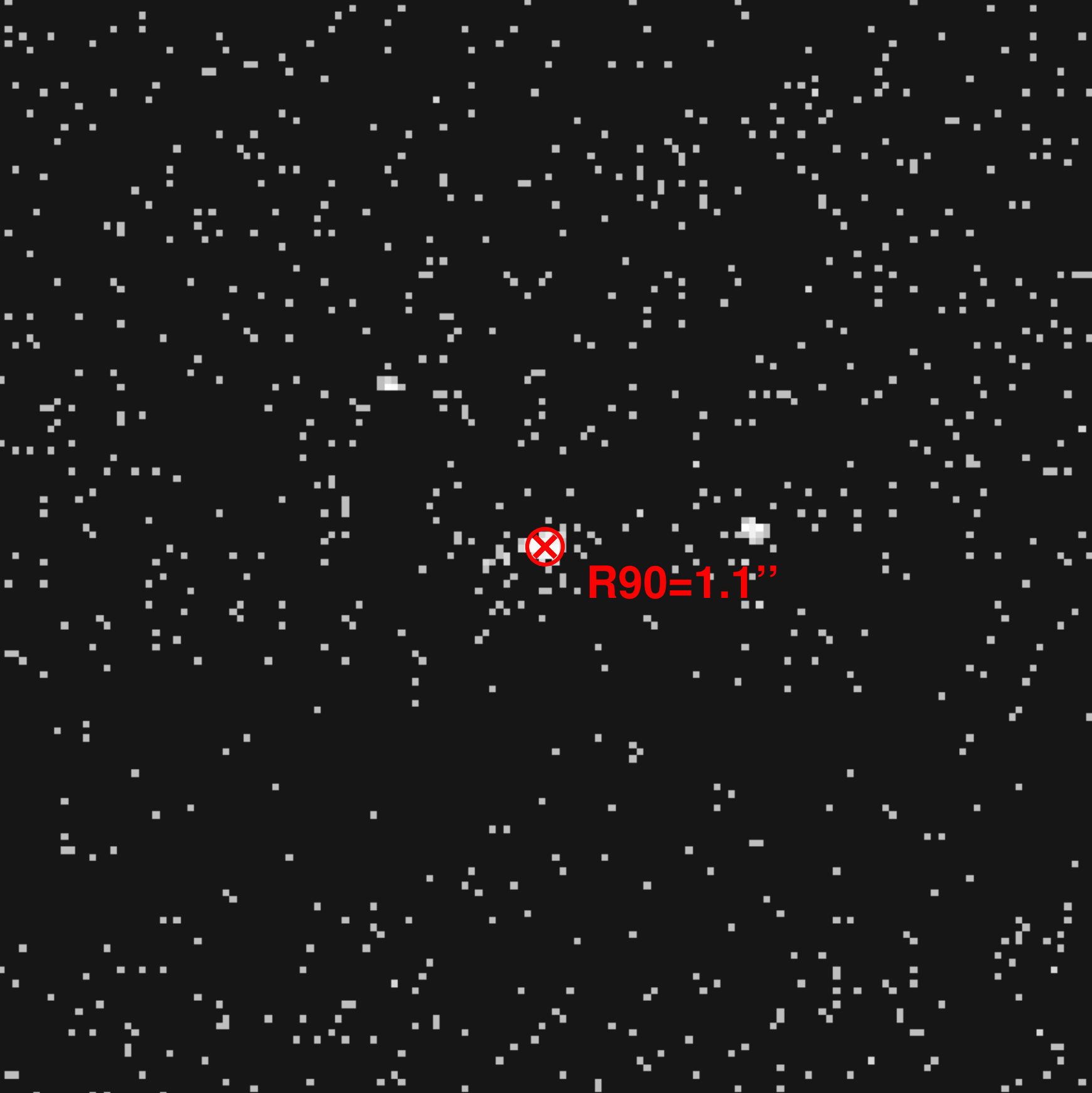}
\end{minipage}
\begin{minipage}{0.3\linewidth}
\centering
\raisebox{-0.2cm}{
\includegraphics[height=1.15\linewidth,angle=270]{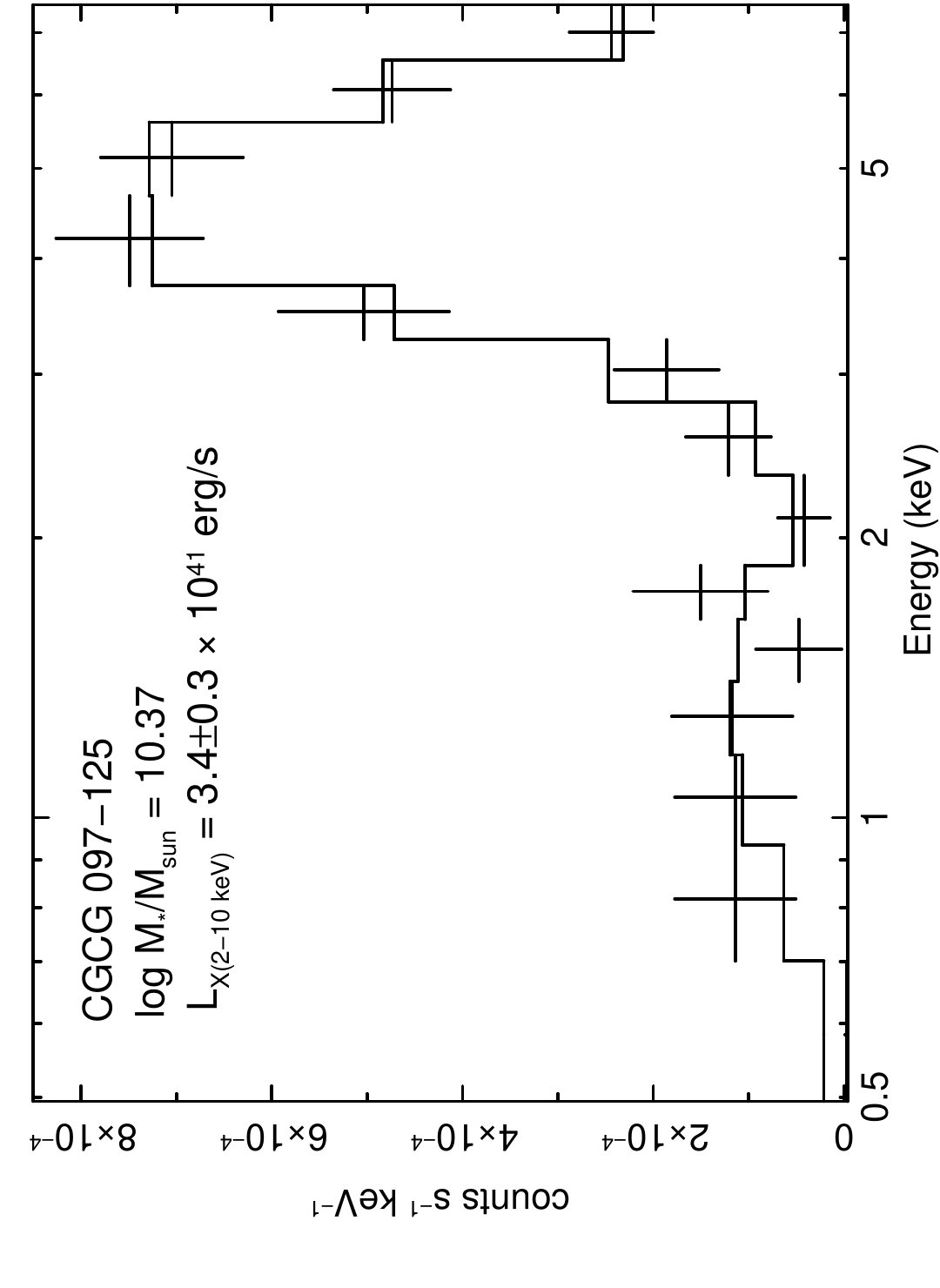}}
\end{minipage}
\\[0.4cm]
\caption{Optical $g,r,z$ colour-composite images, X-ray images and X-ray spectra of dwarf and intermediate-mass cluster members hosting X-ray AGN with $L_{X (2-10\text{ keV})} > 10^{41}$ erg s$^{-1}$.}
\label{fig:agn}
\end{figure*}
\begin{figure*}
\ContinuedFloat
\begin{minipage}{0.3\linewidth}
\centering
\includegraphics[width=\linewidth]{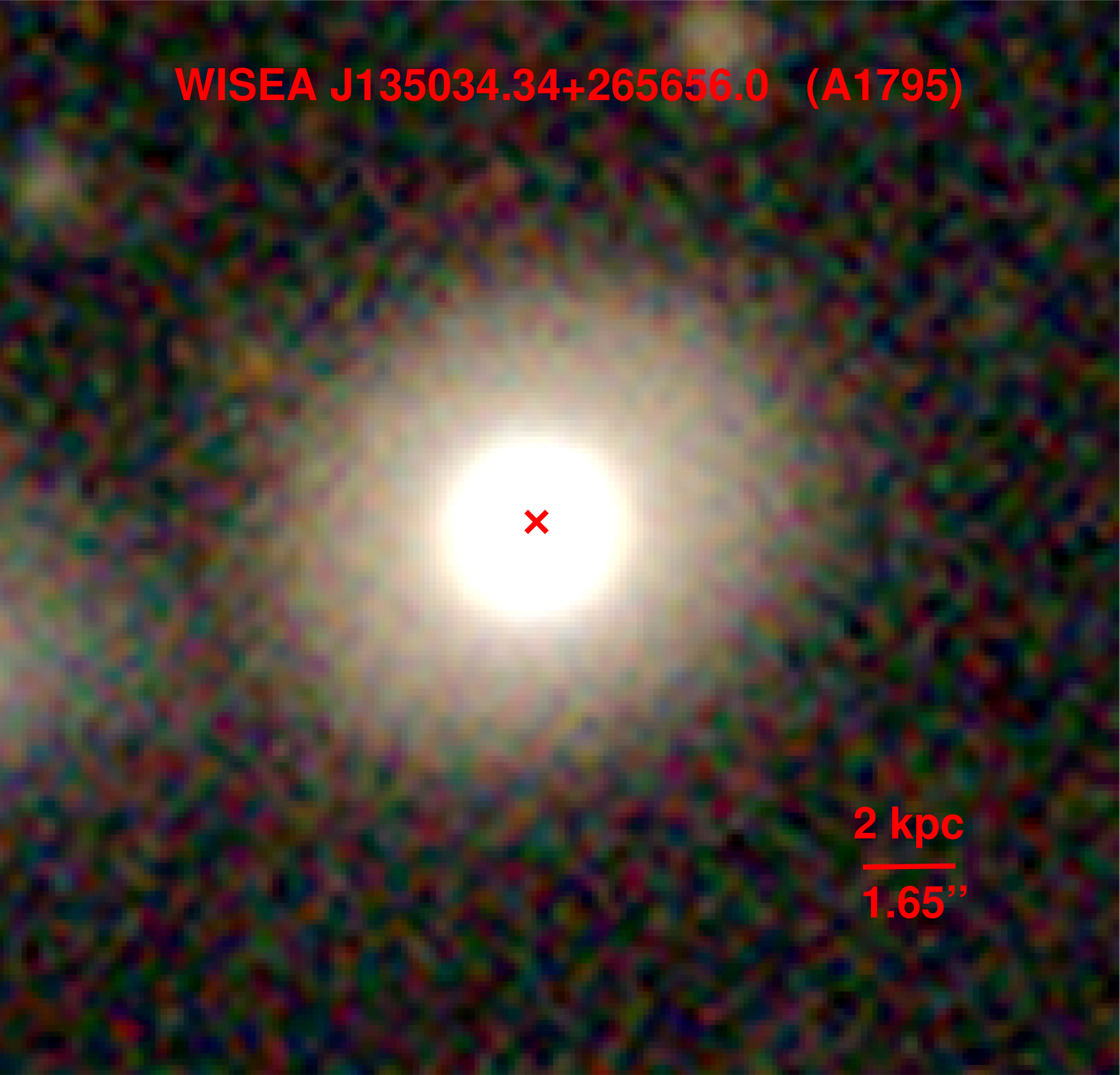}
\end{minipage}
\hspace{0.01\linewidth}
\begin{minipage}{0.3\linewidth}
\centering
\includegraphics[width=\linewidth]{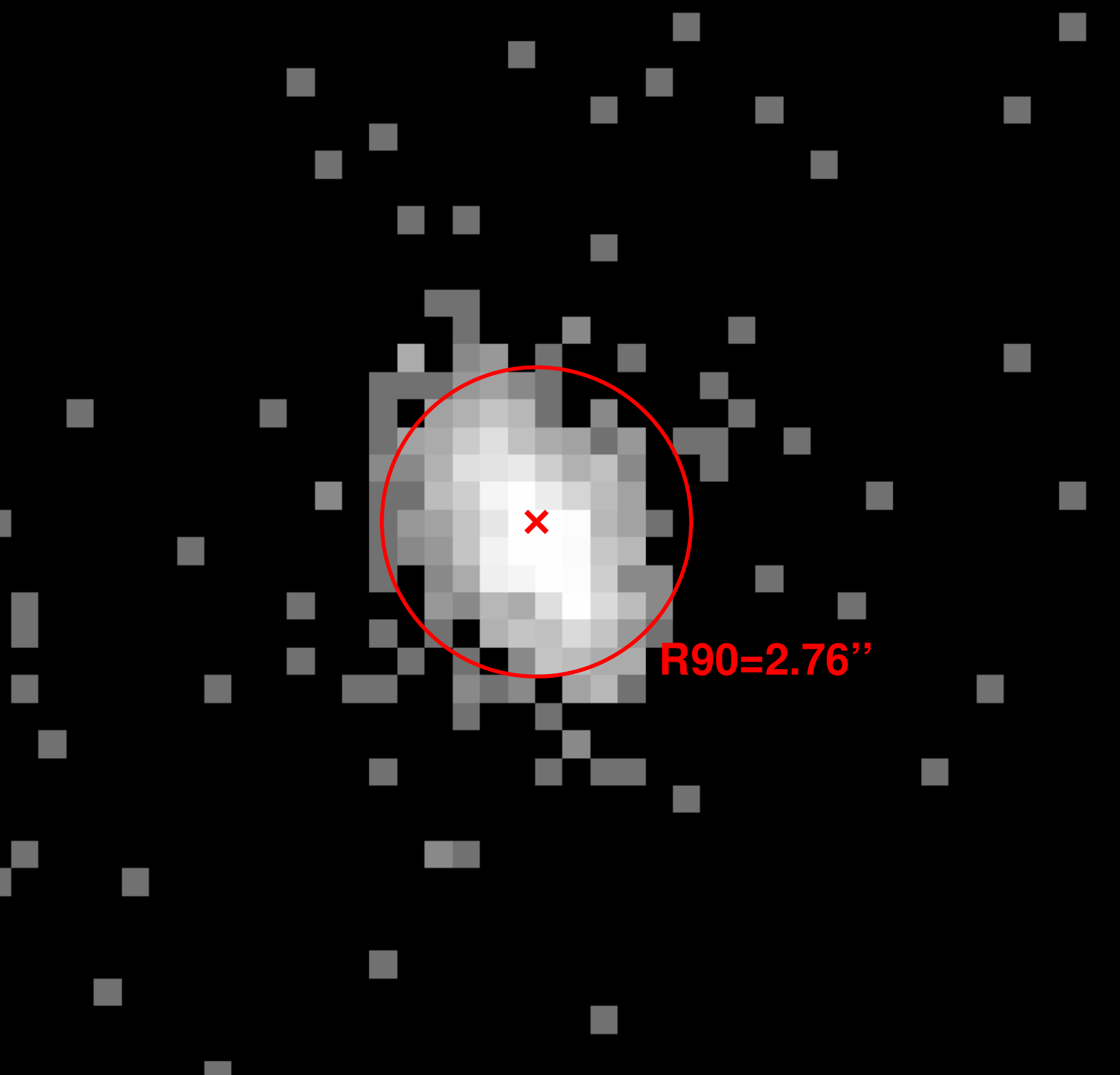}
\end{minipage}
\begin{minipage}{0.3\linewidth}
\centering
\raisebox{-0.2cm}{
\includegraphics[height=1.2\linewidth,angle=270]{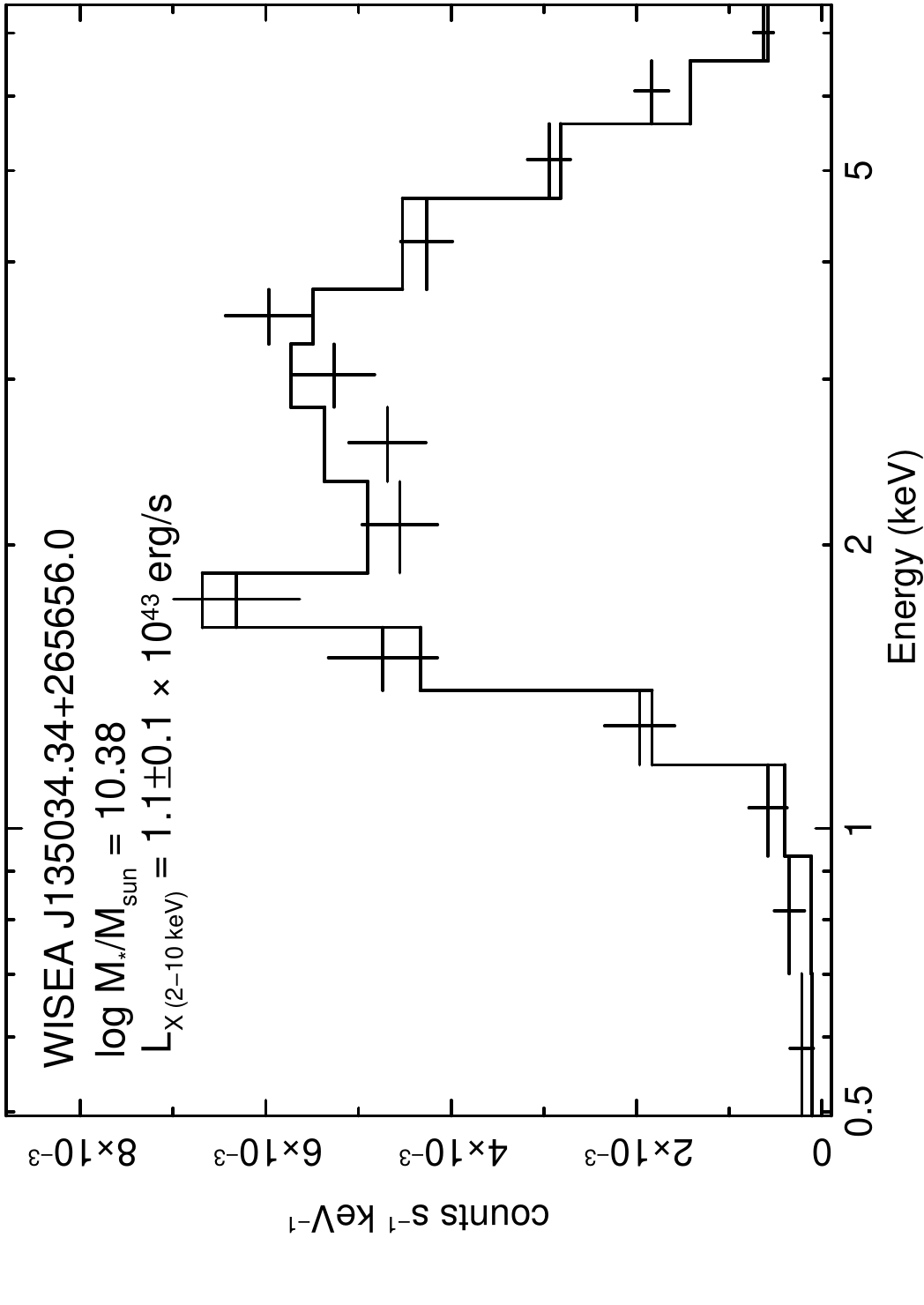}}
\end{minipage}
\\[0.4cm]
\begin{minipage}{0.3\linewidth}
\centering
\includegraphics[width=\linewidth]{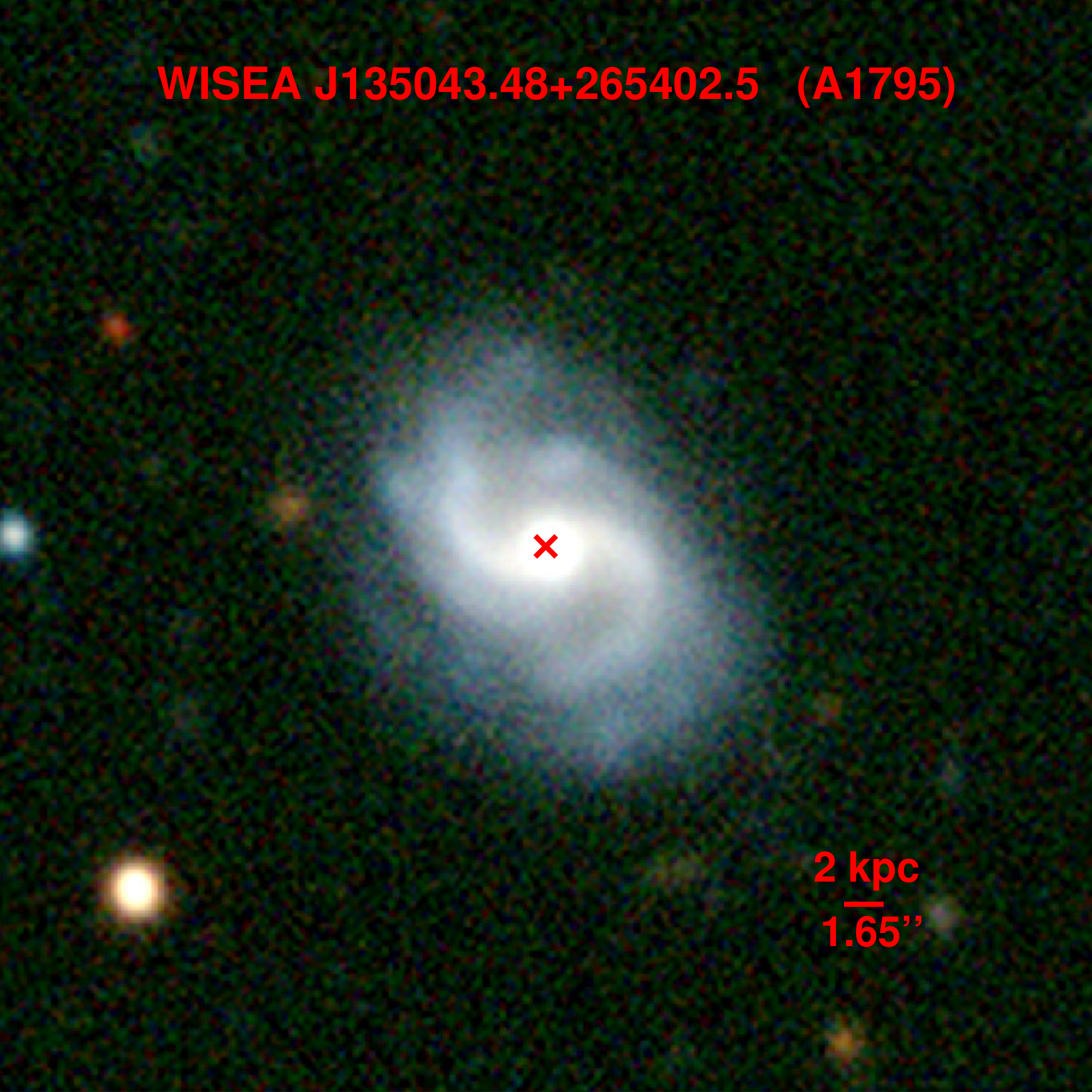}
\end{minipage}
\hspace{0.01\linewidth}
\begin{minipage}{0.3\linewidth}
\centering
\includegraphics[width=\linewidth]{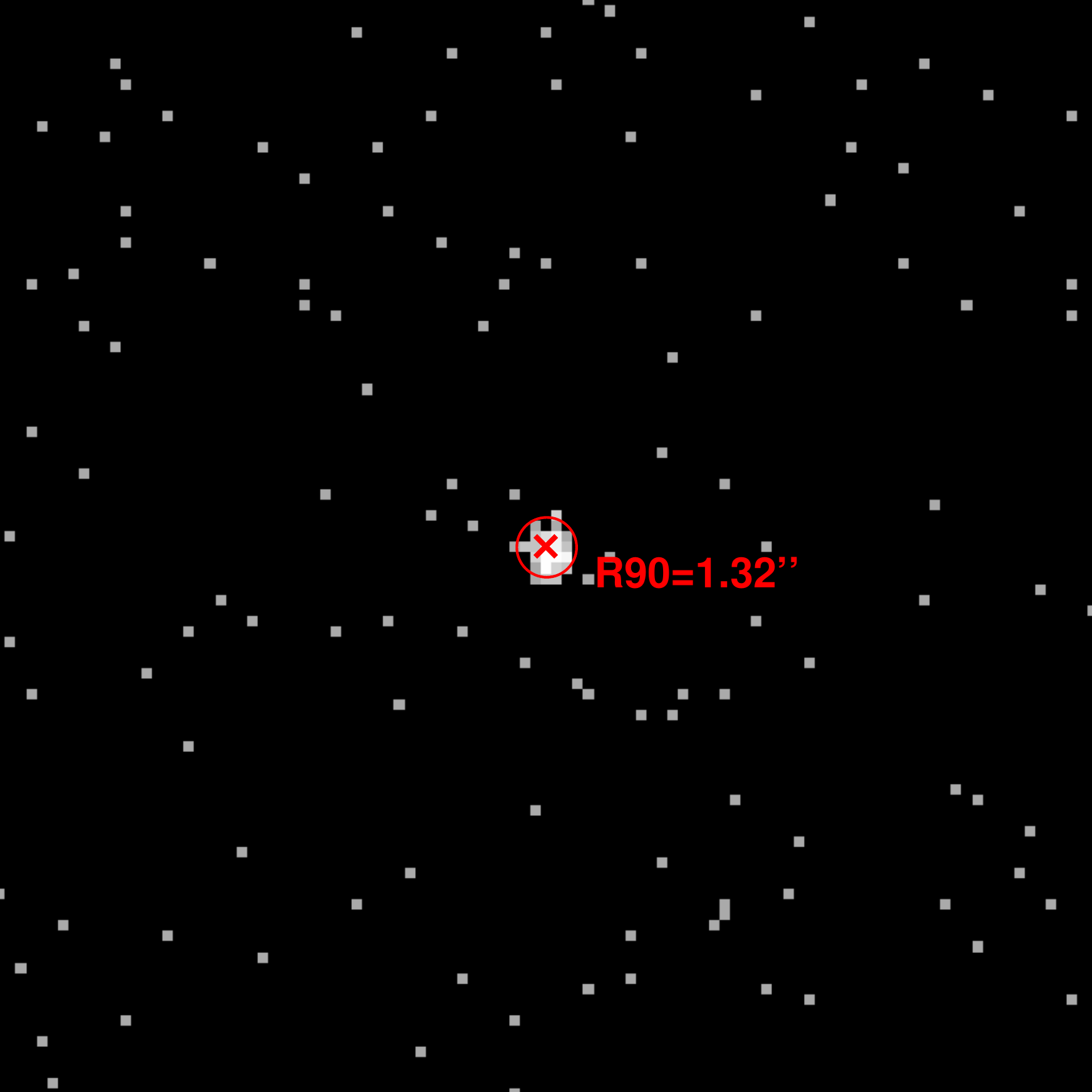}
\end{minipage}
\begin{minipage}{0.3\linewidth}
\centering
\raisebox{-0.2cm}{
\includegraphics[height=1.2\linewidth,angle=270]{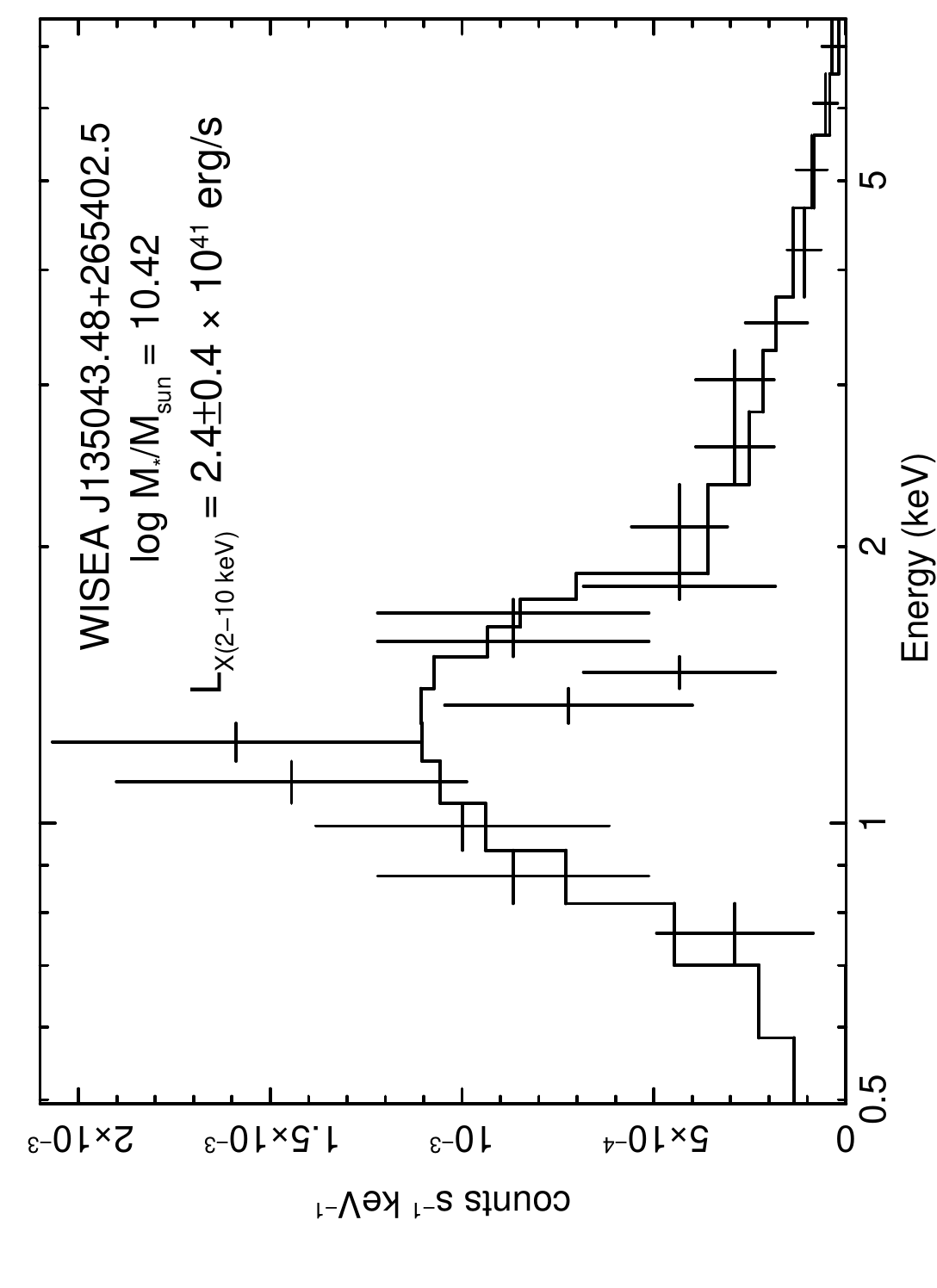}}
\end{minipage}
\caption{--Continued.}
\label{}
\end{figure*}

\clearpage
\onecolumn
\small
\centering
\setlength{\tabcolsep}{10pt}
\begin{xltabular}{\textwidth}{l X l l}
 \caption{Chandra observations used in this work, with the total exposure and total cleaned exposure times.}
\label{tab:obsidlist}\\
\hline
            Cluster&Chandra ObsIDs&Total Time&Total Clean Time\\ 
            &&(ks)&(ks)\\
			\hline\\
            Centaurus&504, 505, 3741, 4190, 4191, 4954, 4955, 5310, 5768, 8179, 11005, 12292, 12957, 15182, 15183, 15184, 15185, 15720, 16223, 16224, 16225, 16534, 16607, 16608, 16609, 16610, 18894, 18902, 20057&1038.67&1034.40\\\vspace{0.2cm}
            A3627 &4956, 4957, 4958, 8178, 9518, 12950  , 26789, 27649&390.18&390.18\\\vspace{0.2cm}
            AWM7&908, 11717, 12016, 12017, 12018&181.67&181.67\\\vspace{0.2cm}
            Perseus&3209, 3237, 3404, 4289, 4946, 4947, 4948, 4949, 4950, 4951, 4952, 4953, 5596, 5597, 6139, 6145, 6146, 8473, 9097, 11713, 11714, 11716, 12025, 12033, 12036, 12037, 13989, 13990, 13991, 13992, 17065, 17258, 17259, 17260, 17261, 17262, 17263, 17264, 17265, 17266, 17267, 17268, 17269, 17270, 17271, 17272, 17273, 17274, 17275, 17276, 17277, 17278, 17279, 17280, 17281, 17282, 17284, 17285, 17286, 18172, 19564, 19565, 19568, 19913, 19914, 19915, 19938, 20520, 20521, 20522, 20523, 20524, 20525, 20526, 20527, 20880, 20881, 21446, 21528, 21529, 21956, 21957, 21958, 21969, 21972, 21973, 22643, 22644, 22888, 22894, 23064, 23084, 23086, 23103, 23702, 23815, 26219, 27864, 27865, 27866, 27867, 27868, 27869, 28474, 28475, 28476, 28477, 28478, 28479, 29633, 30336, 29635&2813.40 &2737.95\\\vspace{0.2cm}
            A1367&514, 4189, 4916, 12999, 17199, 17200, 17201, 17589, 17590, 17591, 17592, 18704, 18705, 18755, 26948, 27219, 27220, 27746, 27747, 29289, 29328&578.56&527.17\\\vspace{0.2cm}
            Coma &555, 556, 1086, 1112, 1113, 1114, 2941, 4724, 4774, 4775, 8188, 9544, 9714, 10672, 10921, 11391, 12887, 13993, 13994, 13995, 13996, 14406, 14410, 14411, 14415, 15023, 15024, 15065, 16599, 16600, 18201, 18234, 18235, 18236, 18237, 18271, 18272, 18273, 18274, 18275, 18276, 18761, 18791, 18792, 18793, 18794, 18795, 18796, 18797, 18798, 19909, 19910, 19911, 19912, 19998, 20010, 20011, 20027, 20028, 20029, 20030, 20031, 20037, 20038, 20039, 20049, 20050, 20051, 20052, 20612, 21068, 21069, 22648, 22649, 22930, 23182, 23361, 23779, 24853, 24854&2590.89&2563.69\\\vspace{0.2cm}
            MKW8&3290, 4942, 7707, 17035, 18266, 18850&167.97&166.0\\\vspace{0.2cm}
           Ophiuchus&3200, 8218, 8219, 8220, 8959, 13734, 15797, 16142, 16143, 16163, 16464, 16626, 16627, 16633, 16634, 16635, 16645, 25293&419.49&412.10\\\vspace{0.2cm}
            A2199&497, 498, 4140, 5618, 10748, 10803, 10804, 10805, 12816, 21425&217.45&207.05\\\vspace{0.2cm}
            A2052&890, 5807, 10477, 10478, 10479, 10480, 10879, 10914, 10915, 10916, 10917&654.04&654.04\\\vspace{0.2cm}
            Hydra A&575, 576, 4969, 4970, 19570, 19571, 19997, 20007, 20968, 20969, 20970, 20971&435.49&413.17\\\vspace{0.2cm}
            A754 &577, 6793, 6794, 6795, 6796, 6797, 6798, 6799, 6800, 10743&217.42&200.99\\\vspace{0.2cm}
            A3667 &513, 889, 4995, 5751, 5752, 5753, 6292, 6295, 6296, 7686, 17202, 17203, 17698&749.42&723.24\\\vspace{0.2cm}
            A85&904, 4881, 4882, 4883, 4884, 4885, 4886, 4887, 4888, 15173, 15174, 16263, 16264&272.92&272.92\\\vspace{0.2cm}
            A2319&3231, 15187, 23846, 24401, 24402, 24403, 24404, 24406, 24408, 24409, 24410, 24411, 24412, 24413, 24414, 24415, 24416, 24417, 27708, 27734, 27801, 27808, 27809&645.92&640.07\\\vspace{0.2cm}
            A133&2203, 3183, 3710, 4103, 9897, 12177, 12178, 12179, 13391, 13392, 13442, 13443, 13444, 13445, 13446, 13447, 13448, 13449, 13450, 13451, 13452, 13453, 13454, 13455, 13456, 13457, 13518, 14333, 14338, 14343, 14345, 14346, 14347, 14354, 20786, 20787, 21864, 21865, 21872, 21873&2653.05&2642.17\\\vspace{0.2cm}
            A1795&493, 494, 3666, 5286, 5287, 5288, 5289, 5290, 6159, 6160, 6161, 6162, 6163, 10432, 10433, 10898, 10899, 10900, 10901, 12026, 12027, 12028, 12029, 13106, 13107, 13108, 13109, 13110, 13111, 13112, 13113, 13412, 13413, 13414, 13415, 13416, 13417, 14268, 14269, 14270, 14271, 14272, 14273, 14274, 14275, 15485, 15486, 15487, 15488, 15489, 15490, 15491, 15492, 16432, 16433, 16434, 16435, 16436, 16437, 16438, 16439, 16465, 16466, 16467, 16468, 16469, 16470, 16471, 16472, 17221, 17222, 17223, 17224, 17225, 17226, 17227, 17228, 17229, 17230, 17231, 17232, 17397, 17398, 17399, 17400, 17401, 17402, 17403, 17404, 17405, 17406, 17407, 17408, 17409, 17410, 17411, 17478, 17479, 17597, 17598, 17599, 17608, 17609, 17610, 17611, 17612, 17613, 17614, 17615, 17616, 17617, 17618, 17619, 17620, 17621, 17622, 17623, 17624, 17625, 17626, 17627, 17628, 17629, 17632, 17656, 17683, 17684, 17685, 17686, 18423, 18424, 18425, 18426, 18427, 18428, 18429, 18430, 18431, 18432, 18433, 18434, 18435, 18436, 18437, 18438, 18439, 18822, 18835, 18877, 19868, 19869, 19870, 19871, 19872, 19873, 19874, 19875, 19876, 19877, 19878, 19879, 19880, 19881, 19939, 19944, 19946, 19964, 19968, 19969, 19972, 20642, 20643, 20644, 20645, 20646, 20647, 20648, 20649, 20650, 20651, 20652, 20653, 21830, 21831, 21832, 21833, 21834, 21835, 21836, 21837, 21838, 21839, 21840, 21841, 22829, 22830, 22831, 22832, 22833, 22834, 22835, 22836, 22837, 22838, 22839, 22840, 24600, 24601, 24602, 24603, 24604, 24605, 24606, 24607, 24608, 24609, 24610, 24611, 25049, 25668, 25669, 25670, 25671, 25672, 25673, 25674, 25675, 25676, 25677, 25678, 25679, 26381, 26382, 27018, 27019, 27020, 27021, 27022, 27023, 27024, 27025, 27026, 27027, 27028, 27029, 27806, 28464, 28465, 28466, 28467, 28468, 28470, 28469, 28471, 28472, 28473, 29453&5846.0&5806.41\\\vspace{0.2cm}
            Cygnus A &360, 1707, 5830, 5831, 6225, 6226, 6228, 6229, 6250, 6252, 17133, 17134, 17135, 17136, 17137, 17138, 17139, 17140, 17141, 17142, 17143, 17144, 17145, 17146, 17505, 17506, 17507, 17508, 17509, 17510, 17511, 17512, 17513, 17514, 17515, 17516, 17517, 17518, 17519, 17520, 17521, 17522, 17523, 17524, 17525, 17526, 17527, 17528, 17529, 17530, 17650, 17710, 18441, 18641, 18682, 18683, 18688, 18846, 18861, 18871, 18886, 19888, 19956, 19989, 19996, 20043, 20044, 20048, 20059, 20063, 20077, 20079&2208.66&2201.53\\
\hline
\end{xltabular}
\twocolumn

\end{document}